\documentclass[12pt,preprint]{aastex}
\usepackage{color}

\slugcomment{VERSION: 6 Jun 2018}

\shorttitle{Atlas of Infrared FUor Spectra}
\shortauthors{Connelley \& Reipurth}

\begin{document}


\title{A Near-IR Spectroscopic Survey of FU Orionis Objects}

\author{Michael S. Connelley\altaffilmark{1,2}
        and 
        Bo Reipurth\altaffilmark{1}}

\affil{University of Hawaii at Manoa, Institute for Astronomy, \\
640 North Aohoku Place, Hilo, HI 96720, USA}

\altaffiltext{2}{Staff Astronomer at the Infrared Telescope Facility, which is operated by the University of Hawaii under contract NNH14CK55B with the National Aeronautics and Space Administration.  }

\begin{abstract}

   We have conducted a homogenous near-IR spectroscopic survey of 33 objects with
varying degrees of similarity to FU~Orionis.

Common spectroscopic features that are characteristic of the
three classical FUors FU Ori, V1057 Cyg, and V1515 Cyg are: strong CO
absorption, weak metal absorption, strong water bands, low gravity,
strong blue shifted He I absorption, and few (if any) emission lines.
Based on these criteria, we classify the 33 objects as either
bona fide FUors (eruption observed), FUor-like objects
(eruption not observed), or peculiar objects with some
FUor-like characteristics, and present a spectral atlas of 14
bona-fide FUors, 10 FUor-like objects, and 9 peculiar
objects. All objects that we classify as FUors or FUor-like have very
similar near-IR spectra.  We use this spectral similarity to determine
the extinction to each source, and correlate the extinction to the
depth of the 3~$\mu$m ice band.  All bona fide FUors still today
maintain the spectrum of a FUor, despite the eruption
occurring up to 80 years ago.
 Most FUors and FUor-like objects occupy a unique space on
a plot of Na+Ca vs. CO equivalent widths, whereas the peculiar objects
tend to be found mostly elsewhere.  Since most FUors show a reflection nebula, we also present an atlas
of K-band images of each target.  We found that the near-IR spectra of FUors and young brown dwarfs can 
be extremely similar, a distinguishing feature being the Paschen $\beta$
absorption in the spectra of FUors.  Although V1647 Ori, AR 6a, and V346 Normae had been previously classified as
candidate FUors, we classify them as peculiar objects with some
FUor-like properties since their spectra now differ significantly from
bona fide FUors.  We confirm two new FUor-like
objects that were initially identified as candidates based on their
near-IR morphology.

\end{abstract}


\section{Introduction}

The major eruption of FU~Ori in 1936 marked an important point in the
study of early stellar evolution, although nobody realized it at the
time (Wachmann 1954). It was Herbig (1966) who argued that FU~Ori
represented some phenomenon of stellar youth.  In 1969
another very similar eruption took place in V1057 Cyg (Welin 1971a,b)
indicating that FU~Ori was not a pathological case but a member of a
new class of objects, for which Ambartsumian (1971) coined the term
FUor. In a classical study, Herbig (1977) presented detailed optical
spectra of three FUors, showing that they all possessed cool
supergiant spectra, with the peculiarity that the spectra become
increasingly late when observed at longer wavelengths, and displaying
a strong lithium line. Of special importance, Herbig demonstrated that
V1057 Cyg prior to its eruption was a T~Tauri star. Reviews about
FUors include Reipurth (1990), Hartmann \& Kenyon (1996), Reipurth \&
Aspin (2010), and Audard et al. (2014). Historical details about
Herbig's work on FUors are discussed in Reipurth (2016).

Hartmann \& Kenyon (1985) suggested that the FUor phenomenon
represents a short-lived major increase in disk accretion, and in a
series of papers they were able to explain many of the observed
features of FUors with this model. The spectral energy distributions
of FUors are reproduced well by modern disk models with accretion
rates of about 10$^{-5}$ M$_{\odot}$~yr$^{-1}$ (e.g., Zhu et al.
2008). Accretion rates of FUors are thus several orders of magnitude
larger than for typical T~Tauri stars, and the FUor phenomenon may
therefore play a significant role in the mass assembly of young stars,
and may also explain the significant luminosity spread observed among
embedded stars (e.g., Baraffe et al. 2012).

Various ideas have been proposed to explain the triggering of FUor
outbursts.  The first is a throttle mechanism that controls the
passage of gas through the inner disk (e.g., Zhu et al.  2009). The
disk may receive gas from an infalling envelope at a different rate
than it transfers through the disk, sometimes causing material to pile
up, which may later be released and causing an eruption.  The second
also deals with a disk that receives gas from an infalling envelope
and assumes that if the accretion rate through a disk increases, and
thus the disk temperature, then the opacity of the gas may rise so
fast that heat becomes trapped in the disk and a runaway situation
develops during which the inner disk may be emptied out (e.g., Bell \&
Lin 1994, Armitage et al. 2001).  The third mechanism proposes that a
companion in an eccentric orbit perturbs the disk at periastron
(Bonnell \& Bastien 1992, Reipurth \& Aspin 2004a).  Perturbations by
a planet (Clarke et al. 2005) or with another member of a dense
cluster (Pfalzner 2008) have also been considered.  Finally, the
fourth idea assumes accretion of a large body, like a planet (Larson
1980) or a large 'gas blob' in a circumstellar disk (Vorobyov \& Basu
2015).

George Herbig and Peter Petrov proposed an alternative to the disk
model in a series of papers (Herbig et al. 2003, Petrov \& Herbig
1992, 2008), and concluded that the {\em optical} spectra are caused
by a rapidly rotating bloated star near the edge of instability, as
originally proposed by Larson (1980). Interestingly, the bloated star
scenario and the disk accretion concept are not mutually exclusive:
Recent calculations of episodic accretion onto low-mass protostars
suggest that, for sufficiently high accretion rates, the central star
may expand significantly, thus increasing its luminosity far beyond
the expectation based on its mass and age (e.g., Hosokawa et al. 2011,
Baraffe et al. 2012). Thus, optical spectra may primarily refer to the
central object -- extended due to rotation near breakup speed (Larson
1980) -- while infrared observations are dominated by disk emission
further out.

No FUor has been observed to undergo more than one eruption (although
the peculiar object V1647~Ori, which has certain FUor characteristics,
had a prior outburst), but through a statistical analysis Herbig
(1977) showed that the phenomenon must be recurrent with an outburst
rate of roughly 10$^{-4}$ per year per star in a star forming
region.  Subsequent studies have supported this conclusion, albeit
with somewhat different outburst rates (e.g., Hartmann \& Kenyon 1996,
Hillenbrand \& Findeisen 2015). Currently only about two dozen FUors
and FUor-like objects are known, and better statistics will emerge as
ongoing wide field surveys identify new cases.

Spectroscopically FUors share certain broad characteristics: the
significant P~Cygni profiles at specific lines, the gradually cooler
spectra at longer wavelengths, and the low-gravity spectral
appearance.  But photometrically FUors form a rather heterogeneous
group, with a significant difference in rise-times for different
objects, ranging from approximately a year for FU~Ori and V1057~Cyg to
more than 15 years for V1515~Cyg, and also major differences in decay
times: V1057~Cyg has already faded to only about 2 magnitudes brighter than its
original brightness, whereas for example V883~Ori and BBW~76 are still
bright more than a century after their presumed eruptions. On much
shorter timescales, EXors can attain almost the same amplitudes as
FUors, but they decay on timescales of a year and typically have
strong emission-line spectra (Herbig 2007, 2008). And some eruptive
variables defy classification (e.g., Herczeg et al. 2016).  As
more and more young objects are found to erupt (e.g., Contreras Pe\~na
et al.  2016), an ever wider range of photometric behaviors may well
emerge. It thus appears likely that spectroscopic criteria will
form a firmer basis for classification of eruptive variables. And
since an increasing number of eruptions are found among embedded
objects, infrared spectroscopy may emerge as a critical tool to
distinguish between different categories of objects.

In this paper, we present an atlas of near-infrared spectra of all
currently accepted or probable FUors and FUor-like objects, as well as
of a few eruptive variables which are sometimes suggested as FUors. We
attempt to identify common spectroscopic characteristics of the FUors,
which may prove useful for future classifications.

\section{Observations}

Our observations were carried out with the 3.0~m NASA Infrared
Telescope Facility (IRTF) on Maunakea, Hawaii with the recently
upgraded SpeX instrument in the short cross-dispersed mode, which
covers 0.7~\micron~ to 2.5~\micron~ in each exposure (Rayner et al.
2003).  We used the 0\farcs5 wide slit, which gives a resolving power
of R=1200.  Stars were nodded along the slit, with two exposures taken
at each nod position.  The total exposure time was driven by our goal
to get a high S/N spectrum for each target on each night.  The
individual exposure times were limited to three minutes to ensure that
the telluric emission lines would cancel when consecutive images taken
at alternate nod positions were differenced.  An A0 telluric standard
star was observed after at least every other science target for
telluric correction, usually within 0.1 air masses of the target.  Our
observing log is shown in Table~1.

An argon lamp was observed for wavelength calibration and a quartz
lamp for flat fielding.  An arc/flat calibration set was observed for
each target/standard pair.  The SpeX data were flat fielded,
extracted, and wavelength calibrated using \emph{Spextool} (Cushing et
al. 2004).  After extraction and wavelength calibration, the
individual extracted spectra were co-added with \emph{xcombspec}.
\emph{Xtellcor} was then used to construct a telluric correction model
using the observed A0 standard, after which the observed spectrum of
the target was divided by the telluric model.  Finally,
\emph{xmergeorders} was used to combine the spectra in the separate
orders into one continuous spectrum.  However, wavelength regions
where the J, H and K orders overlap are noisy due to atmospheric
opacity, and thus the scaling of the orders can adversely affect the
flux ratio of lines in different orders.  As such, we did not scale
the orders prior to merging so as to best preserve the line flux
ratios, which is particularly important in determining the extinction.
These codes are all IDL routines written by Cushing et al. (2004) to
reduce SpeX data.  Spectral line flux, equivalent width, and FWHM were
measured using the SPLOT routine in IRAF.

When the observing
conditions were photometric, we took K-band images for flux calibration using the
SpeX guider camera. 
For these imaging data, the telescope was dithered
using a nine point 10\arcsec\ pattern to allow us to make "sky flats''
from the data.  A dark frame was made by averaging together 10
individual dark frames of the same exposure time as the science data.
This dark was then subtracted from each target frame.  Sky flats were
made by scaling each dark subtracted image to have the same median
value, then averaging them together using a min-max rejection.  The
resulting sky flat was then normalized using the median value of the
pixel counts.  Each dark subtracted (but not scaled) target frame was
divided by this normalized sky flat.  The median sky value for each
frame was subtracted from each frame to set the average background
counts in each frame to 0 to account for changes in the brightness of
the sky.  The images were then aligned and averaged together using an
average sigma clipping rejection.  Standard stars that have been
observed by UKIRT through the MKO filter set (Simons \& Tokunaga 2002,
Tokunaga \& Simons 2002) were selected from the UKIRT faint standard
star list, and were observed for photometric calibration.

Table~2 lists details of the 33 objects discussed in this paper.

\section{What is a FUor?}

Attention was originally drawn to FU~Ori because of its major and
rapid brightening. Similarly, the next two FUors to be recognized,
V1057~Cyg and V1515~Cyg, were discovered because of their
brightenings, albeit not as rapid as that of FU~Ori. In his pioneering
study, Herbig (1977) identified the optical spectroscopic
characteristics of these three objects, which are now universally
accepted as 'classical' FUors. 

Since then a wealth of objects have been classified as FUors, with
justifications that range from solid to at best tenuous. It is
noteworthy that there are now many objects that have been classified
as FUors based on their spectra rather than on the basis of an
observed outburst.  Evidently this easily leads to a rather
heterogeneous group of objects.

While Herbig's analysis was based on optical high-resolution spectra,
many of the more recent discoveries of FUors are sufficiently embedded
that they can only be observed in the infrared. We therefore 
attempt to define here the {\em near-infrared} spectral characteristics of
FUors, based on the three classical FUors discussed in Herbig (1977).
In the following we examine the near-infrared spectra of
FU~Ori, V1057~Cyg, and V1515~Cyg.

\subsection{Near-IR Spectral Characteristics of FU~Ori, V1057
  Cyg, and V1515 Cyg}

Figure~\ref{classicFUors} shows the near-infrared spectra in the J, H,              
and K bands of the three classical FUors FU~Ori, V1515 Cyg, and V1057
Cyg. All three stars have only modest extinction (see Appendix A and
Table~2) and their spectra in Figure~\ref{classicFUors} are not
corrected for this minor reddening. In the following we describe the
spectral characteristics that these three stars have in common.

\textbullet~ {\bf CO}.
~Strong absorption at the CO bands, starting at 2.29~$\mu$m, is
perhaps the most visible characteristic of the three FUors, and in the
past it has been used as a spectroscopic characteristic to identify
candidate FUors whose eruptions were not observed. However, we caution
that \emph{strong CO absorption is necessary but alone is not
  sufficient} to characterize a YSO as a FUor.  Several other YSOs
have spectra that show strong CO absorption (e.g., IRAS 06393+0913),
but lack other important spectroscopic characteristics of FUors
(Connelley \& Greene 2010). The equivalent width of the bluest CO band
of FU Ori is slightly greater than for a late-M dwarf, but less than a late-M giant (per the
SpeX library, Rayner et al.  2009).

It is noticeable that our spectrum of V1057~Cyg does not show the same
pronounced CO absorption as the other two FUors.  The first
near-infrared spectra of FU~Ori and V1057~Cyg were obtained 40 years
ago in 1977 by Mould et al. (1978), only eight years after the
eruption of V1057~Cyg. Visual comparison of the CO bands in these
spectra of FU~Ori and V1057~Cyg show them to be of equal strength (see
Fig.~1 of Mould et al. 1978), as they also appear to be in the spectra
obtained in 1986 by Hartmann \& Kenyon (1987).  Their much weakened
appearance in our new spectra of V1057~Cyg is thus a more recent
development.  There seem to be two possible explanations. First, we
note that V1057~Cyg has faded considerably since its eruption, more
than any other known FUor, and it is now about 2/3 of the way down to
its pre-outburst range (e.g., Kopatskaya et al. 2013). Since most of
the light from a FUor emanates from the disk, the star (or central
bloated object) may therefore now contribute a larger fraction of the
light from V1057~Cyg. Second, we speculate that the disk could have
undergone a structural change, somehow losing the outer disk regions
which mainly contribute the CO bands (see Figure~5 of Calvet et al.
1991), possibly as the result of dynamical interactions with a binary
companion.



\textbullet~ {\bf Water vapor bands}.  ~The continuum profiles of the
three classical FUors superficially appear similar to a mid-M type
star.  Strong water vapor bands on each end of the H-band window shape
the H-band continuum into a characteristic triangular shape. All three
classical FUors have a 'peaky' or 'triangular' H-band continuum that
is often interpreted as a signature of low gravity in young stars, in
contrast to the more rounded appearance seen in evolved stars.  There
are also strong breaks in the continuum at 1.33~$\mu$m and 2.29~$\mu$m
due to the onset of water vapor absorption.

    
\textbullet~ {\bf Other molecular bands}.
~The J-band spectra show rather wide absorption from Vanadium oxide
bands around 1.05~$\mu$m and 1.19~$\mu$m. These bands are highly
sensitive to gravity, and are not visible in M-dwarfs, but begin to
appear at spectral type M7 for higher luminosity stars (e.g., Rayner
et al. 2009). The TiO bands that so dominate the red end of optical
spectra of late-type stars reach slightly into the J-band, and weak
bands are discernible in V1515~Cyg at 0.88, 0.92, and 1.11~$\mu$m, but only at 1.11~$\mu$m for the other two sources.

\textbullet~ {\bf Hydrogen lines}. ~The Paschen-$\alpha$, $\beta$,
$\gamma$, $\delta$ lines are pronounced absorption lines in the J and H-bands.
  The Brackett series of lines are generally not observed.
Brackett $\gamma$, a line frequently observed in young stars and often
associated with mass accretion, is very weakly seen at K-band with a possible
P~Cygni profile.


\textbullet~ {\bf Absence of emission lines}. 
~In contrast to other young erupting stars like the EXors, FUors tend
to have few, if any,  emission lines. As in the optical wavelength
range, the three classical FUors show emission in the J, H, K bands
exclusively as strong P~Cygni profiles, notably at the Ca~II triplet
at 0.850, 0.854, and 0.866~$\mu$m. In particular, we note that
Pa$\beta$ and H$_2$ are {\em not} seen in emission, while Br$\gamma$ appears very weak with a possible P Cygni profile.


\textbullet~ {\bf Metallic lines}.  
Absorption from Na (2.208~$\mu$m) and Ca (2.256~$\mu$m) are weakly observed in the three
FUor spectra.  The EW of these lines is less than half that of a late-M giant.  
For 9 photospheric lines from several species, the median of the ratio of EWs measured in the spectrum of FU
Ori compared to an M7 giant is 0.55.

\textbullet~ {\bf Helium I}.  ~The He I line at 1.083~$\mu$m is a
pronounced absorption line in the spectra of all three FUors.  The blue edge 
of the line has a velocity of --300~kms$^{-1}$, --400~kms$^{-1}$, and --400~kms$^{-1}$
for FU Ori, V1515 Cygni, and V1057 Cygni, respectively.  



\textbullet~ {\bf Gradual change of spectral type with wavelength}.
~We note that the main optical spectroscopic peculiarity recognized by Herbig
(1977) in these three stars, namely the gradual change towards later
spectral types at longer wavelengths, is not obvious from the J to the
K band, presumably due to the rather low spectral resolution. But the
overall spectral type of the three stars is late M, which is much
later than the spectral types derived by Herbig in the optical range.

In summary, each of the above spectral characteristics of the three
classical FUors is seen in other stars. But if one focuses only on
young stars, the {\em combination} of these spectral characteristics
is unique to the FUors. Hence the first task when faced with a
potential new FUor is to establish -- or at least make plausible --
the youth of an erupting object. While lithium offers one such
assurance in optical spectra, no similar line is available in the
near-infrared. Hence, for embedded objects for which lithium is unobservable,     
an infrared excess, an association with a reflection nebula, or at                                                             
least association with a star forming region or molecular cloud, are
important additional criteria.  \footnote{\textbf{A similar condition was required by Herbig (1960)               
when he defined what is now known as Herbig Ae/Be stars. When later
researchers abandoned the requirement of a reflection nebula, the
catalogs of Herbig Ae/Be stars became polluted with many stars that are not young.}}


  In consequence, attaching the label of FUor or FUor-like to an object 
will necessarily be done with varying degrees of confidence, depending on how 
many of the above characteristics are observed.  In many cases, the firm classification 
of an object as a FUor or FUor-like object will require the gradual accumulation over time 
of the necessary information. 

\section{Comments on Individual Objects}   

In the following we evaluate the spectra and the
photometric histories together with other pertinent information for each
of 33 objects, all of which have at some point been suggested as
FUors. Table~2 lists basic information about each of the 33 objects as
well as our evaluation of their classification.   Table~3 presents the basis
for our classification of each object.  

Not all of these objects have been observed to undergo an eruption.
We therefore divide the 33 objects into
three categories: (1) bona fide FUors with spectra displaying the
near-infrared spectral characteristics recognized in Section~3 and
with witnessed eruptions, (2) FUor-like objects that similarly have
spectra like FUors but for which an eruption was not witnessed, and
(3) objects that have (or have had) some spectral similarities to
FUors, but enough differences that a FUor classification is not
warranted.  In Appendix~C, we additionally describe a few objects that in the past
have been erroneously classified as FUors.

\subsection{Bona Fide FUors}

We here discuss 14 objects that have been observed to erupt and whose
spectra are very similar to those of the classical FUors discussed in
Section~3, and which we therefore deem to be bona fide FUors. Their
near-infrared spectra are shown in the spectral atlas in
Figure~\ref{page1}. Since some of these 14 objects suffer considerable             
extinction, we have dereddened their spectra and plotted them next to
each other in Figure~\ref{FUor_outburst}, which facilitates a direct                    
comparison of their spectra with those of the three classical FUors.
The He I absorption line and the CO absorption bands are important characteristics of FUors, and
these features are shown in detail in Figures \ref{FUor_outburst_HeI} and \ref{FUor_outburst_CO}, respectively.  

\textbf{RNO 1b:} This object, also known as V710 Cas, was identified
as a new FUor by Staude \& Neckel (1991), who noticed that it had
brightened by at least 3 magnitudes between 1978 and 1990.  Their
optical spectrum is consistent with a FUor, showing H$\alpha$ and Li
in absorption, with a blueshifted absorption trough at H$\alpha$.  The
near-IR spectrum in Greene \& Lada (1996) and the one shown here show strong CO absorption,
'triangular' H band, and a clear break in the continuum at 1.33~$\mu$m
from water vapor.

\textbf{V582 Aurigae:} The amateur astronomer Anton Khruslov
discovered the eruption of this star (Samus 2009).  It is located in
the L1516 cloud at a distance of 1.3~kpc (Kun et al. 2017).
Semkov et al. (2013) found that the eruption happened between late 1984 and
early 1985 based on archival photographic plates, and showed that the
optical spectrum displays typical strong P~Cyg profiles at H$\alpha$
and the Sodium doublet.  Our spectrum is the first IR spectrum of this target,
and it shows the typical FUor characteristics: CO absorption, a
'triangular' H band, and a clear water vapor break in the continuum at
1.33~$\mu$m.

\textbf{V883 Orionis:} This source is optically very faint but at
infrared wavelengths it is one of the most luminous sources in the
L1641 cloud in Orion. The source was suggested as a FUor by Strom \&
Strom (1993) based on an optical spectrum of its reflection nebula
which shows a blueshifted absorption profile of H$\alpha$.  The K-band
spectrum presented in Reipurth \& Aspin (1997) showing strong CO
absorption and a featureless continuum is consistent with our
spectrum, which additionally shows the characteristic triangular
H-band continuum, a strong break at 1.33~$\mu$m, VO bands, and no
emission lines.  The source is associated with the little HH object
HH~183, and illuminates the small optical reflection nebula IC~430.
The IC~430 nebula was significantly larger and brighter in 1888 when
-- as discussed by Strom \& Strom (1993) -- Pickering (1890)
photographed the region. This is reminiscent of the bright reflection
nebulae that appeared when V1057 Cyg and V1647~Ori erupted, and we
accept the bright state of IC~430 around 1888 as evidence of a prior
eruption, and accordingly classify the object as a bona fide FUor.
Cieza et al. (2016) used ALMA observations to determine a dynamical
mass of 1.3$\pm$0.1~M$_\odot$ for V883~Ori, and found evidence that
the water-line was moved outwards in the disk during the eruption, although
Schoonenberg et al. (2017) offer a different interpretation.


\textbf{V2775 Orionis:} The outburst of this object occurred between
March 2007 and November 2008, with an increase in brightness of 3.8
magnitudes at K-band, as documented by Caratti o Garatti et al.
(2011).  Their near-IR spectrum shows several of the features that
characterize FUor spectra: strong water and CO absorption, a
'triangular' H-band continuum, VO bands and no emission lines.
Fischer et al. (2012) observed He I absorption blue-shifted by $\sim$
300~kms$^{-1}$.  Whereas our spectrum also shows He I absorption, it
is much weaker and is centered at the rest wavelength. This object has only 
been detected at infrared wavelengths, and we argue in Appendix A 
that this object is on the back side of its cloud.

\textbf{FU Orionis:} The eruption of FU~Ori was observed in 1936
(Wachmann 1954, Herbig 1966).  The light curve shows a rapid brightening
by $\sim$5.5 magnitudes over 8 months, followed by a slow decline of $\sim$1/2
magnitude over the next 40 years with small-scale faster variability (Herbig
1977).  Wang et al. (2004) noted a highly reddened faint companion
0\farcs5 south of FU Ori, which was demonstrated to be a young star by
Reipurth \& Aspin (2004a), and Beck \& Aspin (2012) and Pueyo et al.
(2012) determined that this companion is the more massive component of
the FU~Ori binary. FU~Ori is the prototype of the FUors, and since it
suffers very little extinction we use its spectrum as the reference
for all other potential FUors.

\textbf{V900 Monocerotis:} The outburst of this target was first
noticed in 2009 by Thommes et al. (2011).  Optical and IR spectra
taken by Reipurth et al. (2012) confirmed several FUor-like spectral
features. The object is a partly embedded Class~I object with a
considerable extinction of about A$_V$$\sim$13 magnitudes.  Infrared
photometry by Varricatt et al. (2017) suggests that the object is
still brightening. Our spectrum of V900 Mon appears to be consistent
with the spectrum in Reipurth et al. (2012) and displays all the FUor
characteristics discussed in Sect.~3.1.

\textbf{V960 Monocerotis:} This object is also known as 2MASS
J06593158-0405277.  It is a recent FUor whose eruption was first
noticed on Nov 3, 2014 by Maehara et al.  (2014).  The outburst
happened between January and October 2014.  Hackstein et al. (2015)
presented the light-curve for this target back to Nov 2010, showing
the target getting brighter in the R-band and I-band by $\sim$2.6
magnitudes. After the outburst was discovered, Hackstein et al. (2015)
found that the R-band brightness declined by $\sim$0.5 magnitudes in
the following 6 months while showing a low-amplitude 17-day
oscillation, which they interpreted as evidence for a (unseen)
companion. K\'osp\'al et al. (2015) found V960 Mon to be associated
with 8 likely T~Tauri stars and an embedded Class~0
source. Jurdana-Sepic \& Munari (2016) found no other eruptions in a
historical light curve spanning from 1899 to 1989. Caratti o Garatti
et al. (2015) imaged a faint companion at a 0$\farcs$2 separation, and
possibly another even closer companion. We observed it on Dec 22,
2014, very shortly after the eruption was announced, when the object
was at K=7.42, whereas the 2MASS magnitude is K=9.45. The spectrum
displays all the FUor characteristics.

\textbf{V1515 Cygni:} This object is located near the NGC 6914 region
in Cygnus at an approximate distance of 1050~pc according to Racine
(1968). It was first identified as a FUor by Herbig (1977), and the
FUor classification was further supported by the observations
presented by Kenyon et al. (1991).  The light curve in Herbig (1977)
shows a slow and gradual brightening over $\sim$10 years, starting in
about 1950, different from the very rapid rise of FU~Orionis.  Clarke
et al. (2005) extend the light curve to 2003, showing a slow gradual
decline with $\sim$0.3 magnitudes of short-term variability.
Optical spectra taken around 1974 show a high luminosity G type star,
with P~Cygni profile at H$\alpha$, and strong lithium at 671 nm.  The
K-band spectrum presented by Reipurth \& Aspin (1997) shows the strong
CO absorption characteristic of FUors. Our spectrum is discussed in
Sect.~3.1.

\textbf{HBC 722:} Also known as V2493 Cygni. A faint visible star
erupted in July 2010 and brightened by R$\sim$4.2 magnitudes in
$\sim$2 months (Semkov et al.  2010, 2012).  It is
associated with the small cluster around LkH$\alpha$~188 that is
located in the active star forming cloud L935 in the North America
Nebula at a distance of 550~pc (e.g., Armond et al. 2011).  Shortly
after the outburst was noticed, the optical spectrum of this object
was observed to be similar to FUors (e.g., Munari et al. 2010).
Importantly, a pre-outburst optical spectrum was fortuitously recorded
and found to be consistent with a classical T Tauri star showing
hydrogen emission with a K7-M0 spectral type (Cohen \& Kuhi 1979,
Miller et al. 2011).  Despite undergoing an outburst of accretion, the
non-detection of this source in the sub-millimeter by SMA puts an
upper limit on the total circumstellar material at 0.02~M$_\odot$
(Dunham et al. 2012). After the initial eruption, HBC~722 faded
slightly, then brightened to a relatively stable plateau (e.g.,
K\'osp\'al et al. 2016). Our spectrum displays all the FUor
characteristics discussed in Sect.~3.1.


\textbf{V2494 Cygni:} Also known as IRAS 20568+5217 and HH 381 IRS.
The L1003 cloud in Cygnus is a region of active star formation (e.g.,
Magakian et al.  2010) with numerous HH objects, including HH~381
associated with IRAS~20568+5217 which illuminates a bright reflection
nebula (Devine et al. 1997, Connelley et al. 2007). Reipurth \& Aspin
(1997) suggested the FUor nature of the object based on near-infrared
spectra, a classification further supported by Greene et al. (2008)
and Aspin et al. (2009b).  The star brightened by R$\sim$2.5
magnitudes sometime between 1952 and 1983 after which it seems to have
reached a plateau (Magakian et al.  2013).  The optical spectrum shows
H$\alpha$ and forbidden emission lines originating from the HH~381
flow near the star, which has a projected length on the sky of 4.7~pc
(Magakian et al. 2013, Khanzadyan et al. 2012). Our spectrum displays
all the FUor characteristics discussed in Sect.~3.1.



\textbf{V1057 Cygni:} The star, which lies in a dense cloud within the
North America Nebula (NGC 7000), brightened by 6 magnitudes in 1969
(Welin 1971a,b).  A fortuitous pre-eruption low resolution optical
spectrum taken in 1957 shows an "advanced T Tauri emission spectrum
with no detectable absorption lines" (Herbig \& Harlan 1971). The
outbursting star was studied in detail photometrically and
spectroscopically by Herbig (1977). The light curve from 1969 to 2012
has been compiled by Kopatskaya et al. (2013), who showed that the
rapid decline in brightness can be fitted with an exponential
curve. They also found that the I-band photometry has a strong 523-day
periodic variation with $\sim$0.5 mag amplitude. V1057~Cyg has a
high-velocity long-lasting wind which manifests itself through strong
blueshifted absorption troughs at several prominent optical lines
(Herbig 2009).  Hartmann \& Kenyon (1987) presented high resolution
near-infrared spectra, which are consistent with the disk accretion
model for FUors.  Most recently, Green et al. (2016) have detected a
faint companion to V1057~Cyg with a projected separation of 58 mas and
a brightness difference of $\Delta$K$\sim$3.3 mag. Our spectrum is
discussed in Sect.~3.1.


\textbf{V2495 Cygni:} Also sometimes called the Braid Nebula Star.  A
bright near-infrared reflection nebula appeared between 1999 and 2001,
brightening at K-band by at least 4 magnitudes (Movsessian et
al. 2006).  Their near-IR spectrum shows many characteristics of FUors
(e.g.  'triangular' H-band continuum, strong water and CO absorption),
and is essentially identical to the spectrum we show here.  The source
is associated with several HH objects, HH 629/635, with a projected
length on the sky of 0.8 pc (Magakian et al. 2010, Khanzadyan et
al. 2012).  The object is highly extincted, so our H-band spectrum is
noisy and the J-band spectrum is missing. The CO absorption is very
deep and the H-band continuum has a clear triangular shape.

\textbf{V1735 Cygni:} Also known as Elias~1-12, IRAS 21454+4718. and HBC 733.  Elias (1978) showed
that this object, located in the dark molecular filaments stretching
$\sim$2$^\circ$ west of IC~5146, had brightened by at least 5 visual
magnitudes between 1952 and 1965, and identified it as a new FUor.
Sandell \& Weintraub (2001) found a bright submillimeter source,
V1735~Cyg~SM1, about 24~arcsec northeast of V1735~Cyg. SM1 is almost
coincident with the IRAS source 21454+4718. Both sources were detected
in X-rays with XMM-Newton by Skinner et al. (2009).  Six out of seven
of our spectra of this object are consistent with the spectrum shown
in Greene \& Lada (1996), demonstrating that the object is quite
stable.  However, one spectrum differs, and this
spectroscopic variability is further discussed in Sect.~5.3. Our
spectrum shows the FUor characteristics discussed in Sect.~3.1, with
exceptionally strong water vapor features.

\textbf{V733 Cephei:} Also known as Persson's Star.  The outburst of
this object was first noticed in 2004 (Persson 2004).  Peneva et al. (2010) 
found that the earliest photographic plate showing the
outburst was taken in 1971, and the object slowly rose in brightness
until 1993.  From 2007 to 2009, they observed this object to slowly
fade.  Optical and near-IR spectra (Reipurth et al. 2007) show great
similarity to FU Ori, securing this object's classification as a FUor.
Our spectrum of V773 Cep is consistent with the spectrum
shown in Reipurth et al. (2007).

\subsection{FUor-like Objects}

Wide-field and all-sky surveys are becoming increasingly common, and
as a result it is more and more likely that a new FUor eruption will
be quickly detected. Until recently, however, this was not the case,
and hence we can be certain that in the past many FUor eruptions have been
overlooked. Since some FUors retain their spectroscopic
characteristics for up to a century or even more (e.g., FU~Ori 1936
eruption, V883~Ori $<$1888 eruption), we expect to find a number of
objects that have all the FUor characteristics, but whose outbursts
were missed. A very similar situation exists for novae, for which the
terms nova and nova-like object have been introduced, depending on
whether an eruption has been witnessed or not. In analogy, Reipurth et
al. (2002) suggested to use the nomenclature FUors and FUor-like
objects. In this section, we list 10 objects which have clear spectroscopic FUor
characteristics, but for which an eruption was missed. In Section~5
we compare the properties of bona fide FUors and the FUor-like
objects.  The de-reddened spectra of the FUor-like objects
are presented in  Figure~\ref{FUor_like_flat}.                          

\textbf{RNO 1c:} This heavily extincted object is located
$\sim$6~arcsec from the FUor RNO~1b in the L1287 cloud. Unlike RNO 1b,
however, no outburst has been observed.  Based on a 2.291-2.298~$\mu$m
spectrum, Kenyon et al. (1993) found that RNO~1c has the same deep CO
absorption characteristic as FUors and, combined with its energy
distribution, suggested its FUor nature.  Our complete 1-4~$\mu$m
spectrum mostly supports this conclusion. The spectrum shows strong, distinct
CO band heads, although the 'triangular' H-band continuum resulting
from water absorption that is a characteristic of FUors is quite weak.


\textbf{PP 13S:} Also known as IRAS 04073+3800.  PP 13 is a small
compact reflection nebula in the L1473 cloud in Perseus (Parsamian \&
Petrossian 1979).  Cohen et al. (1983) found the nebula to be
associated with two sources, PP~13N, a typical T~Tauri star, and
PP~13S, a cometary nebula with no H$\alpha$ emission and illuminated
by an embedded infrared source.  Further near-infrared imaging was
presented by Smith (1993), who resolved PP~13N into a close binary.
Sandell \& Aspin (1998) and Aspin \& Sandell (2001) found PP~13S to be
driving a molecular outflow, and detected deep, broadened 2.3~$\mu$m
CO absorption bands in a 2.0-2.5~$\mu$m spectrum, and on this basis
suggested that PP~13S could be in an elevated FUor state. Aspin \&
Reipurth (2000) found a small Herbig-Haro jet, HH~463, emanating from
PP~13S along the axis of the molecular outflow. Perez et al. (2010)
presented CARMA 0\farcs15 resolution 227~GHz observations of the
circumstellar disk of PP~13S.  Since 1998, the K-band magnitude has
faded from 9.2 to 10.8.  However, our data show that the spectrum in
the K-band is little changed from 1998.  There is still weak H$_2$
emission, presumably from HH~463, and the CO band heads are indistinct.
At first glance the red featureless continuum is unlike any bona fide
FUor, however, this is primarily due to the high extinction towards
the object.  The water absorption bands that form the characteristic
'triangular' shape of the continuum become evident when the spectrum
is dereddened by an A$_V$ of about 56 mag.

\textbf{L1551 IRS 5:} This is a deeply embedded object (Fridlund et
al. 1980), that was classified as a FUor on the basis of its
optical spectrum seen via a reflection nebula (Mundt et al. 1985,
Stocke et al.  1988).  The source is a binary with projected
separation of 0$\farcs$36 (Bieging \& Cohen 1985, Lim et al. 2016), with
each component driving a jet (Fridlund \& Liseau 1998, Rodr\'\i guez et
al. 2003).  The near-infrared spectrum in Reipurth \& Aspin (1997) shows H$_2$
emission and CO band absorption.  Strong [FeII] emission seen in
our spectrum at 1.644~$\mu$m is unusual for FUors, but - as the H$_2$
emission - it likely derives from the associated jets. Greene et al.
(2008) presented high-resolution near-infrared spectroscopy, and used
cross-correlation with FU~Ori to demonstrate their similarities: deep
CO absorption, triangular H-band continuum, and strong break at
1.33~$\mu$m.


\textbf{Haro 5a IRS:} This source, also known as IRAS 05329-0505, is
one of the more luminous embedded sources in Orion (e.g., Wolstencroft
et al.  1986).  Haro 5a and Haro 6a designate the two lobes of an
optically visible reflection nebula surrounding the IRAS source (Haro
1953).  This is the source for the giant flow HH~41, 42, 128, 129,
294, and 295 (Reipurth et al. 1997a).  The object was identified as a
FUor-like object based on its near-IR spectrum (Reipurth \& Aspin
1997).  Our spectrum shows an extremely red continuum with strong CO
absorption, and without photospheric absorption lines.  The water
absorption bands creating the triangular shape in the H-band appears
to be weakly present, but when dereddened by an extinction of
A$_V$$\sim$57~mag, their presence become evident (see
Figure~\ref{FUor_like_flat}).

\textbf{IRAS 05450+0019:} This is an embedded source also known as
NGC~2071~MM3 which illuminates a large bipolar reflection nebula in
the NGC~2071 region at a distance of 400~pc (Connelley et al. 2007). It
was identified as a FUor-like object by Connelley \& Greene (2010)
based on its near-infrared spectrum, which shows very strong CO
absorption and a triangular continuum shape in the H-band. Our present
spectrum is identical to the earlier spectrum.

\textbf{Z Canis Majoris:} Herbig (1960) identified Z~CMa as a young
star, which he classified as a HAeBe star. Later, Hartmann et al.
(1989) suggested that Z~CMa is a FUor on the basis of weak CO
absorption and doubled absorption lines from Li I and Ca I near 671 nm
which they interpret as originating from a rotating accretion disk.
The star also displays the wide blueshifted absorption troughs at the
H$\alpha$ and Sodium doublet characteristic of FUors (e.g., Hessman et
al. 1991, Welty et al. 1992). Both the HAeBe and the FUor
classification turn out to be correct, since Z~CMa has been found to
be a 0$\farcs$1 binary (e.g., Haas et al. 1993, Velazquez \& Rodr\'\i guez
2001), where the SE component is a FUor and the NW component is a
Herbig Be star (e.g., Bonnefoy et al. 2016). Z~CMa drives a giant HH
flow with a well collimated jet (Poetzel et al. 1989). Whelan et al.
(2010) demonstrated that this large-scale jet is driven by the Herbig
Be star, but that also the FUor drives a small $\sim$0$\farcs$4 jet,
see also Antoniucci et al. (2016). Z~CMa shows major optical
variability, Covino et al. (1984) collected the photometric history
from 1923 to 1983, revealing an amplitude of 3 magnitudes, but no
outburst, implying that it must have occurred prior to 1923. This
variability has been traced to the Herbig Be component (e.g., Bonnefoy
et al. 2016).  Our near-infrared spectrum shows little resemblance to
a FUor spectrum, which evidently is due to the superposition of the
FUor spectrum with the (brighter) Herbig~Be spectrum (both of which
are separated in the integral field spectra of Bonnefoy et al. 2016).
We note that the Br$\gamma$ line is in emission with broad blue and
red shifted wings extending out to $\pm \sim$1000 kms$^{-1}$, whereas
the He I line only has a blue shifted absorption wing extending out to
1000 kms$^{-1}$.

\textbf{BBW 76:} Also known as V646 Puppis.  This object was
identified as a FUor based on its optical spectrum and near-IR
photometry (Reipurth 1985b, Reipurth 1990).  Photographic plates
from 1900 show it as bright as it is today (Reipurth et al. 2002), rather
than the usual slow decline observed from most FUors.  Evans et al.
(1994) did not detect a CO outflow at 346 GHz, and it is
not associated with an HH object (Reipurth \& Aspin 1997).  Although the
optical line profiles changed in observations obtained over a period of 10 years starting in
1985 (Reipurth et al. 2002), our spectrum of this target is consistent with
the K-band spectrum 20 years earlier presented in Reipurth \& Aspin (1997).

\textbf{Parsamian 21:} Also known as IRAS 19266+0932. The object,
which displays a prominent reflection nebula (e.g., Connelley et al.
2007), was identified as a FUor by Staude \& Neckel (1992) based on
its optical spectrum and spectral energy distribution.  The optical
spectrum showed H$\alpha$ absorption blueshifted by
$\sim$600~kms$^{-1}$, Li absorption, and an F5 giant spectral type.
The near-IR spectra presented by Greene et al. (2008) and Connelley \&
Greene (2010), as well as our spectrum shown here, show a remarkable
similarity to FU~Ori and support the conclusion that this is a
FUor-like object.  The visible light brightness has not appreciably
changed for at least 57 years (Semkov
\& Peneva 2010).  SED analysis (Gramajo et al. 2014) and near-IR
polarimetry (K\'osp\'al et al. 2008) show that this target is seen
nearly edge-on.  We do not see a far-side (southern) reflection nebula
for Parsamian~21, despite background stars being visible to the south.  
Parsamian~21 appears to be at the
southern edge of its cloud, so it is likely that there is simply no
dust to the south of Parsamian 21 to scatter light.

\textbf{CB 230 IRS1:} Also known as IRAS 21169+6804 or
CB230~A.  This is a deeply embedded Class~0/I protostar in the L1177
cloud. A companion, IRS2, is located $\sim$10$''$ to the east (e.g.,
Massi et al. 2008), corresponding to about 3000~AU at the 300~pc
distance (e.g., Das et al. 2015). High-resolution VLA observations
reveal another close companion at 0$\farcs$3 separation, making it a
triple system (Tobin et al. 2013). No outburst has been witnessed.
Massi et al. (2008) mention the possibility that some of the
spectral features could be from a circumstellar disk, as in a FUor,
but do not classify this object as a FUor.  Our spectrum shows that
the source shares prominent characteristics of FUors: strong CO
absorption, 'triangular' H-band continuum, and a dearth of
photospheric metal lines.  Hence, we classify CB230~IRS1 as a new
FUor-like object.

\textbf{HH 354 IRS:} Also known as IRAS 22051+5848 or L1165~SMM1. This
is the source of a giant HH outflow (Reipurth et al. 1997a) and it
illuminates a reflection nebula visible in the optical and near-IR
(e.g., Connelley et al. 2007).  HH~354 has a projected separation of
2.4~pc from IRAS 22051+5848 at the $\sim$750~pc distance of the L1165
cloud (Reipurth et al. 1997a). The source is an embedded Class~0/I
protostar, and Reipurth \& Aspin (1997) showed that the K-band
spectrum has a red featureless continuum and strong CO absorption, and
noted that the K-band spectrum is very similar to L1551~IRS5.  No
outburst has been detected. The K-band brightness of this object has
varied from K=13.3 in 1987 (Persi et al. 1988) to K=9.7 in 1999
(2MASS) to K=10.8 in this paper.  Additional near-infrared
spectroscopy supports the classification as a FUor-like object (Greene
et al. 2008, Connelley \& Greene 2010), and the spectrum we present
here is consistent with previous spectra.  High-resolution
radio continuum VLA observations have revealed a 0$\farcs$3 companion
(Tobin et al. 2013).

\subsection{Peculiar Objects with Some FUor Characteristics}

A number of objects have been identified over the years as FUors,
which either through closer comparison with the spectra of the three
classical FUors (Section~3) or through spectral variability are no
longer convincingly classifiable as FUors. In the following we comment
on 8 such objects.

\textbf{V1647 Orionis:} The eruption of this object was discovered by
the amateur astronomer Ian McNeil, when a bright reflection nebula,
now known as McNeil's Nebula, appeared in 2004 in the Lynds 1630 cloud
(McNeil et al. 2004, Brice\~no et al. 2004). Near-IR spectra, taken
shortly after the outburst was noticed, showed emission from the CO
band heads near 2.3~$\mu$m, atomic H, Na I, Ca I, and Ca II, and strong
absorption from He I at 1.083~$\mu$m, water ice, and solid state CO at
4.6~$\mu$m (Reipurth \& Aspin 2004b, Vacca et al. 2004).  After
photometrically declining to pre-outburst brightness, V1647 Ori
unexpectedly brightened again in August 2008 to be as bright as it was
in 2004.  During this second brightening, the near-IR spectrum was
quite different than during its 2004 eruption.  The spectrum showed
strong water vapor absorption and 'triangular' H-band continuum common
among FUors, Br$\gamma$ emission, and possible CO band head absorption
(Aspin et al. 2009a). The Harvard plate collection reveals that this
object previously erupted in 1966-67 (Aspin et al. 2006). Our spectrum
shows that atomic hydrogen and the CO bands are now back in emission, but other than that
the overall spectrum is remarkably FUor-like. However, the presence of
CO emission is such a major deviation from 'normal' spectroscopic FUor
appearance that we do not include the object among the FUors. One
possible, but unproved, explanation could be that V1647~Ori is a close
binary with a FUor plus an active companion that causes the CO
emission, a scenario reminiscent of Z~CMa. Principe et al. (2017) mapped
two distinct, misaligned molecular outflows, but do not find evidence
for any companion at distances larger than 40~AU. If a companion is
responsible for the recent outburst, however, it would still be close
to periastron, and thus difficult to detect. 

\textbf{IRAS 06297+1021W:} This IRAS source, located in the NGC 2264
clouds, is resolved at near-infrared wavelengths into two sources with
a separation of 70$''$ (Connelley et al. 2008a,b). A near-infrared
spectrum of the western component was presented by Connelley \& Greene
(2010), who noted certain spectroscopic features characteristic of
FUors, but also strong discrepancies. The continuum looks like a FUor,
with a triangular H-band profile and He absorption.  However, in
common with V1647 Ori and Z CMa, it has CO in emission.  Also similar
to Z CMa, this object shows emission from sodium (2.208~$\mu$m),
Br$\gamma$, Pa$\beta$, and the Ca IR triplet (0.85~$\mu$m).

\textbf{AR 6a:} Also known as V912 Mon. This is a bright near-infrared
source in NGC 2264.  While an outburst has not been observed in this
object, Aspin \& Reipurth (2003) suggested, based on a K-band spectrum
showing strong CO and Br$\gamma$ absorption, a FUor-nature for AR~6a.
However, our present spectrum, taken more than 10 years later, now
shows only weak CO absorption and the Br series in absorption, giving
the appearance of a highly reddened early type star.  Notably there is
no triangular shape of the continuum in the H-band.  Since 2003, the
spectrum of this object has evidently changed and it is no longer
similar to a FUor.

\textbf{AR 6b:} This is a faint companion to AR~6a, at a separation of
about 3$''$.  An outburst has not been observed.  The K-band spectrum
of Aspin \& Reipurth (2003) shows stronger CO absorption and more IR excess
than AR~6a.  Aspin \& Reipurth (2003) argue that this should be classified as a
FUor due to the CO absorption and overall spectroscopic
similarity to PP 13S.  Our 1-2.5~$\mu$m spectrum shows high veiling,
and possibly water absorption from 1.7 to 2~$\mu$m, but the spectrum
is too noisy to make definite statements on the presence of water.
The individual CO bands are not distinct, possibly due to high
rotation.  The spectrum is certainly not one typical of T~Tauri stars,
and a FUor-like classification is still possible, but we feel it safer
to withhold judgment until better data are in hand.

\textbf{IRAS 06393+0913:} Connelley \& Greene (2010) obtained a SpeX
near-infrared spectrum of this object in NGC 2264, and noted certain
similarities to FUors, especially the very deep CO band head
absorption, but due to the lack of any water absorption creating the
triangular continuum shape in the H-band they stepped back from
classifying the object as a FUor-like object. {Moreover,
as discussed in Section~7, the luminosity of this object is as low as
0.9~L$_\odot$, which opens the question whether it might be a young
brown dwarf, see Section~6.


\textbf{V346 Normae:} Also known as HH~57~IRS. The eruption was
discovered in 1983 and the star is visible on plates taken as early as 1980
(Graham \& Frogel 1985, Reipurth 1985a).  Optical spectra taken at that time
showed H$\alpha$ absorption blue-shifted by $\sim$440~kms$^{-1}$ and
Li absorption. V346~Nor drives the little HH flow HH~57 (Prusti et al.
1993, Reipurth et al. 1997b). \'Abrah\'am et al. (2004) compiled the
light curve up to that time, and most recently Kraus et al. (2016) and
K\'osp\'al et al. (2017) noted that V346 Nor had declined significantly in
brightness by 2010, indicating a drop in accretion rate by three
orders of magnitude, followed by another brightening. Our spectrum is
taken in July 2015 during this second brightening, but with the source
still about 4 magnitudes fainter than during its previous outburst. It
has been assumed up to now that V346~Nor is a bona fide FUor, mainly
because of its brightening during the early 1980's, but also due to
its strong P~Cyg profile at H$\alpha$ (e.g., Graham \& Frogel 1985).
However, our spectrum as well as one by Graham \& Frogel (1985) show
virtually no CO absorption, while another spectrum from the mid-1990's (Reipurth et al. 1997b)
shows pronounced CO emission and Br$\gamma$ emission (which also has
disappeared again). Moreover, H$_2$ emission is now present in the
spectrum. The unusual light curve and the spectroscopic variability
makes V346 Nor more similar to V1647~Ori, and we similarly consider it
to belong to the group of peculiar stars with some FUor characteristics. We
speculate that also V346~Nor could be a binary with a FUor and an active
companion, as the FUor-like object Z~CMa.

\textbf{IRAS 18270-0153W:} This object was identified as a possible
FUor-like object by Connelley \& Greene (2010), showing a 'triangular'
H-band continuum, strong CO absorption, and no photospheric absorption
lines, with an appearance similar to other FUor-like objects.
However, our more recent spectrum, although rather noisy, no longer
looks like a FUor, showing weaker CO absorption and a smoother H-band
continuum.

\textbf{V371 Serpentis:} Originally listed as EC 53 by Eiroa \& Casali
(1992), who noted a small cometary nebula in their infrared imaging
survey of the Serpens core. Hodapp (1999) found the nebula to be
variable, and Hodapp et al. (2012) determined a period of 543 days
with an amplitude of $\Delta$K$\sim$2 mag, possibly due to a very
close companion, and also noted that the source has a resolved
0\farcs3 companion.  The high resolution spectra from Doppmann et
al. (2005) of three segments in the K-band are consistent with our
spectrum, which shows CO absorption, triangular H-band continuum, and
a dearth of identifiable photospheric absorption lines. Several
emission lines of [FeII] and H$_2$ are seen which
we ascribe to shocks in a small jet. The object is extremely red, so
the H-band spectrum is noisy, and the J-band spectrum is absent
despite 4 hours of integration. It is conceivable that V371~Ser could
be a FUor-like object, but the data are not good enough to confirm
that. Moreover, as we discuss in Section~7, there is the concern that
V371~Ser has a luminosity of only 1.6~L$_\odot$, which should be
divided between two or possibly three components. In Section~6 we
discuss the possibility that V371~Ori is a newly born brown dwarf.


\textbf{IRAS 18341-0113S:} Similar to the previous object, this one
was also identified as a possible FUor-like object by Connelley \&
Greene (2010), showing a 'triangular' H-band continuum, strong break
in the continuum slope at 2.3~$\mu$m, but barely detectable CO
absorption.  Our new spectrum, however, no longer looks like a FUor,
showing no appreciable CO absorption and a smoother red H-band
continuum. We therefore include also this object in the category of
peculiar objects with some FUor characteristics. We also
note that IRAS~18341-0113S has a very low luminosity of only
0.8~L$_\odot$, in fact it is the lowest luminosity object of all our
33 targets, and in Section~6 we discuss the possibility that it may be
a very young brown dwarf.

\subsection{Spectral Variability of the Bona Fide FUors}

FUors are much less spectroscopically variable than other YSOs, and
their spectra maintain their FUor characteristics decades after the
eruption.  The only major spectroscopic change seen in a bona fide
FUor is the strong weakening of the CO band in V1057~Cyg, which has
accompanied its significant fading (Section~3.1).  Connelley \& Greene
(2014) monitored two FUors (V1735 Cyg and V2494 Cyg, each observed 5 times) among 19 YSOs.
The He~I line at 1.08~$\mu$m (one of the few strong lines in a FUor
spectrum) varied on average by 18\% for the two monitored FUors, with
a maximum variability of just under 60\%.  However, this line varied
on average by 57\% for the other YSOs, with a maximum variability of a
factor of 12.  In the course of this survey, 11 of the 33 targets were observed more than once
in SXD mode.  V1735 Cyg was the only target to show a significant change in this time.

Combining the data in Connelley \& Greene (2014) with our new data,
V1735 Cyg has been monitored with SpeX eight times over the past nine
years.  The spectrum showed little change (other than slight changes
in the He~I line profile) until July 8, 2016.  The spectrum taken on
that night was significantly redder (Figure~\ref{V1735Cyg_var}), and    
the He~I line showed a P~Cygni profile for the first time
(Figure~\ref{V1735Cyg_HeI}).  The additional flux appears to be a        
featureless red continuum, but with added CO absorption.  The change
in the CO and He~I lines shows that this was not a simple data
processing error, or a change in the line-of-sight extinction.  Before
this event, the He~I line showed blueshifted emission with a maximum velocity
of --500~kms$^{-1}$ to --800~kms$^{-1}$.  This event accelerated the wind velocity  up
to --950~kms$^{-1}$.  An observation taken the following month shows
that V1735 Cyg had returned to its previous state.  The He~I and CO
lines, as well as the continuum, were again consistent with previous
spectra.

Despite the obvious spectroscopic changes during this unusual
event, the K-band brightness of V1735 Cygni appears to have changed
very little, if at all.  The K-band flux on July 8, 2016 is consistent
with our photometry from Aug 28, 2015, a year before this event.  On
the month after the event, the photometry on Aug 20, 2016 is only 0.2
magnitudes fainter than on July 8, 2016.  Considering the very slight
photometric change, it is unlikely that this event represents a new
outburst or significant change in the state of this FUor.  We seem to
have caught a rare, isolated, and transient event that demonstrates
that FUors {\em can} exhibit significant spectroscopic variability
without much photometric variability.

\section{FUors as Low-Gravity Objects}  

Figure~\ref{CO_vs_metals2} shows a comparison of the equivalent widths        
of the bluest CO band (measured from 2.292 to 2.320~$\mu$m) versus the sodium
plus calcium lines of FUors, of some Class~I sources, and of late
field dwarfs, giants, and supergiants.  It is readily seen that this
diagram effectively differentiates FUors and FUor-like objects from
the Class~I sources.  This diagnostic diagram has also been used by
Greene
\& Lada (1996) and later by Connelley \& Greene (2010).  The Na
(2.208~$\mu$m) and Ca (2.265~$\mu$m) lines were used, as they are near
the CO band heads so they are all equally affected by veiling and
extinction.  In this figure, adding veiling tends to push a star
closer to (0,0).  Both bona fide FUors and FUor-like objects mostly
lie in a tight group near the giant stars (and along the veiling
vector from the super giants) in Figure~\ref{CO_vs_metals2}.  This is
consistent with the interpretation of FUors being low gravity objects.
If we let {\em m} be the average equivalent width of the Na + Ca
lines, then to be on the FUor side of the figure the CO equivalent
width should be greater than 3{\em m}+5 (shown by the gray dashed
line).  We note that the peculiar objects tend to mostly be located closer to
(0,0) in a group separate from the majority of the bona fide FUors and
FUor-like objects.

The separation into two distinct groups of FUors plus FUor-like
objects vs. peculiar objects is broken by four of the 33
objects in Figure~\ref{CO_vs_metals_crop}, and we discuss these four        
objects in the following:

1) The only bona fide FUor that clusters with the peculiar objects is
V1057~Cyg. This is unexpected and somewhat disturbing, since V1057~Cyg
is one of the classical FUors which helped to define FUors as a class.
As discussed in Section~3.1, V1057~Cyg has undergone a significant
evolution in luminosity, with a concomitant major decrease in its CO
band strength, which accounts for its location further left in
Figure~\ref{CO_vs_metals_crop} than any other FUor or FUor-like
object. We note that Calvet et al. (1991) show (their Fig.5) that CO
absorption is only dominant in the outer annuli of the circumstellar
disk. Hence as the disk cools and fades, the outer annuli may become
undetectable. Or if the outer annuli are not continuously replenished
from an envelope, they will spiral inwards. The explanation for its
weaker sodium and calcium lines is even less clear, we speculate that
hot spots due to accretion onto the stellar surface  begins to dominate
as the disk emission fades, thus increasing veiling. Whatever the
specific reason, we believe that the location of V1057~Cyg outside the
clump of FUors and FUor-like objects in Figure~\ref{CO_vs_metals_crop}
is due to the recent rapid decay from its high FUor state.


2) The FUor-like object V371~Ser is also located among the peculiar
objects and is close to V1057~Cyg in the diagram. The object is deeply
embedded and is only seen as a reflection nebula, even at
near-infrared wavelengths. This is in itself unlikely to be the cause
of its location among the peculiar objects, since seven other objects
are also so embedded that they are only seen as compact reflection
nebulae. We have no explanation for the location of V371~Ser in the
diagram.

3) Z~CMa is located in the extreme lower left corner of
Figure~\ref{CO_vs_metals_crop}, but this is readily understood, since
the object is a close binary combining a FUor and a Herbig Ae/Be
star, where the latter dominates the light in the near-infrared.

4) IRAS 06393+0913 is located in the middle of the group of FUors and
FUor-like objects in Figure~\ref{CO_vs_metals_crop}. This is perhaps
not surprising, since its main deviation from a FUor spectrum is the
absence of broad water bands and those would not affect the line
measurements used for the diagram.

Figure~\ref{FUor_outburst} shows that all of the bona fide FUors have
a 'triangular' H-band continuum profile.  As stated by Barman et al.
(2011), the absorption bands of water naturally meet in the middle of
the H-band, shaping the continuum into the well known triangular
profile.  As gravity increases, collisionally induced absorption (CIA)
from H$_{2}$ increases, and CIA can be the dominant source of opacity
at the higher gravity of dwarf stars (Borysow et al. 1997).  CIA has a
very broad wavelength dependence, effectively smoothing out the
otherwise triangular profile into the more rounded profile seen in
late type dwarfs.  The ubiquity of a triangular H-band profile among
bona fide FUors is further evidence in support of FUors being very low
gravity objects.




\section{Brown Dwarfs as FUor Impostors}  

Here we address the question of whether there are other objects in
star forming regions that may mimic the spectra of FUors. Broadly
speaking, FUors in the near-infrared have mid- to late-type
M-spectra. The only other objects found in star forming regions with
such spectra are young brown dwarfs. At an age of 1 Myr, the brown
dwarf mass limit corresponds to a spectral type around
M5.5. Figure~\ref{FUOri+BD} shows the near-infrared spectrum of FU~Ori    
together with that of HBC~341 (Dahm \& Hillenbrand 2017), a partly
embedded M5 star bordering the brown dwarf regime with an age assumed
to be roughly 1 Myr, and that of TWA-8B (Allers \& Liu 2013), a brown
dwarf in the 10-Myr old TWA association. 

The similarity of the spectra is striking, and perhaps even
surprising, given that the FUor spectrum is emitted by a disk, while
the BD spectra presumably comes from a photosphere. The
spectra are very similar even in most details, but there are a few
differences. The first thing to notice is that the BD spectra show
somewhat weaker CO bands, however, when a larger sample of young BDs
and FUors are compared, the two distributions of CO strength do have
some overlap, which makes it unusable as a criterion for distinction.

Paschen-$\alpha$ is very pronounced in FU~Ori, and much weaker in
HBC~341, but its location in a region of limited atmospheric
transmission makes it a poor criterion. Paschen-$\beta$, however, is
better placed, and shows much higher strength compared to
BDs. Figure~\ref{FUOri+BD-EQW} shows a histogram of equivalent widths
of Paschen-$\beta$ for the FUors and FUor-like objects vs a number of
young BDs in the TW Hya association.  In this figure, the number of
objects at a given EW is the sum of the objects whose EW range of
uncertainty included that EW value.
These EW distributions just overlap due to a single FUor-like object, BBW 76.  Z~CMa shows strong emission at the Paschen-$\beta$ line, and is off to the far left in the figure.

A set of weaker lines in the J-band also show certain differences. The
Na doublet at 1.138/1.141 is pronounced in BDs, but is absent in
FU~Ori. Other doublets of potassium (1.169/1.178 and 1.243/1.253) and
aluminum (1.313/1.315) are stronger in BDs than in FU~Ori, as are some
iron lines (1.169/1.177/1.189/1.198). However, the fact that many
FUors are highly extincted makes the J-band often difficult to observe. 

In summary, for a purely spectroscopic classification, the
Paschen-$\beta$ strength is the best, although not
completely unique, way to distinguish between a FUor and a BD.

In this context, two interesting questions arise: can
brown dwarfs undergo FUor eruptions? And if so, how can we distinguish
between a substellar object with and without major accretion?

Regarding the first question, we note that brown dwarfs are formed
like stars, except that they fail to reach the hydrogen-burning limit
because either their mass reservoirs are not large enough (Padoan \&
Nordlund 2004) or they are prematurely expelled from their cloud core
by companions (Reipurth \& Clarke 2001) or their mass reservoir is
destroyed by a nearby massive star (Whitworth \& Zinnecker 2004). All
three mechanisms are likely to operate, although identifying which
mechanism is responsible for a given object is most likely not
possible. The key is that in all three cases there is a smooth
transition of properties across the stellar/substellar limit. We know
that stars with a mass as low as an early M-type star can erupt as a FUor
(Reipurth, in prep.), so there seems little reason why a slightly less
massive brown dwarf also cannot undergo a FUor outburst. Additionally,
we know that brown dwarfs have circumstellar disks, albeit smaller
than those of higher-mass T~Tauri stars (e.g. Testi et al. 2016). We
conclude that there is no obvious reason why brown dwarfs should not
undergo FUor eruptions like their higher-mass brethren, although they
may perhaps not reach as high luminosities. 

Regarding the second question, as we have shown above, the
photospheric spectra of brown dwarfs and the disk spectra of FUors are
almost indistinguishable at the resolution employed here. Future
higher resolution spectroscopy may be key to separate the two types of
objects, e.g. by looking for the double-lined profiles characteristic
of rotating disks in FUors (e.g., Hartmann \& Kenyon 1996). Meanwhile
it appears that luminosity is a necessary parameter to separate brown
dwarfs from brown dwarfs undergoing FUor outbursts. However, setting a
limit in luminosity between brown dwarfs and brown dwarfs in a FUor
state may not be straightforward, since young brown dwarfs are likely
to experience a continuum in accretion luminosity.

In Section~7 we determine the luminosities of all 33 objects of this
study.  The three lowest-luminosity objects are IRAS 06393+0913
(0.9~L$_\odot$), V371 Ser (1.6~L$_\odot$), and IRAS 18341-0113S
(0.8~L$_\odot$). We have classified all three as peculiar, and they
are individually described in Section~4.3.  With such low luminosities, 
and their spectra having some features reminiscent of FUors, these spectra 
may be from a photosphere of a very low mass object with at most a modest 
contribution from accretion.

\section{Bolometric Luminosity Distribution}  

We now discuss the bolometric luminosities of
FUors, and to that end their distances are important. The distances we
use are listed in Table~2 together with references to the distance
determinations.  Although a few FUors have well determined distances,
generally they are not well known, and in some cases are little more
than wishful thinking. We have critically reviewed each distance,
taking into account the many new accurate distance determinations of
star forming regions in recent years. See Appendix B for a discussion on the adopted distance to each star forming region.

We have estimated the bolometric luminosity for each target using the
distances listed in Table~2.  In an attempt to make a
homogeneous set of luminosity estimates, we used publicly available
photometry from 2MASS (JHK), WISE (3.4 to 22~$\mu$m), and Akari (65 to
160~$\mu$m).  For the longer wavelengths of the spectral energy
distributions (SED), we appended a 20~K blackbody for each object,
scaling the blackbody curve to match the Akari 160~$\mu$m photometry.
The 20~K temperature was chosen as that temperature provides a smooth
transition from the measured SED to the extrapolated blackbody.  The
20~K blackbody 'tail' on average adds 9\% to the bolometric luminosity
of the objects.  We have not attempted to correct the photometry for
extinction.

The bolometric luminosity distribution is shown in
Figure~\ref{Lbol_dist}.  FUors are shown in red, FUor-like objects in
The median bolometric luminosity for bona fide FUors is 99 ~L$_\odot$
and for FUor-like objects it is 35~L$_\odot$. If taken at face
value, the smaller value for the FUor-like objects could be understood
under the assumption that FUors decline in brightness, and that the
eruptions of some FUor-like objects may have taken place long ago,
before the sky was well patrolled. However, a 2-sample K-S test shows
that the bolometric luminosity distributions cannot be differentiated
at the 90\% confidence level, and we thus cannot show that these two
distributions are statistically different.

We note that, despite being a spectroscopically rather homogenous
group, FUors and FUor-like objects span a wide range of bolometric
luminosities.  Their luminosities span 3 orders of magnitude, from
$\approx6$~L$_{\odot}$ to $\approx3500$~L$_{\odot}$. It has been
widely assumed that FUors have considerable luminosities, but
evidently some objects have luminosities that are only slightly
brighter than typically found for classical T~Tauri stars, despite
their dramatically different spectra.






\section{Conclusions}

We have conducted a homogeneous near-IR spectroscopic
survey of all currently known FUor and FUor-like YSOs as well as 
some peculiar objects with certain FUor characteristics, in total 33 objects with
varying degrees of evidence that they are indeed FUors. Our goal was
to decide the nature of these objects by more stringent means.  We
have obtained the following results:

1. First we have determined the near-infrared spectroscopic
   characteristics of the three classical FUors FU~Ori, V1057~Cyg, and
   V1515~Cyg in order to establish a set of reference criteria. The
   main near-infrared spectral characteristics of these objects is
   deep CO band absorption, weak metal absorption lines, pronounced
   water vapor at the short and long sides of the H-band window giving
   rise to a characteristic triangular H-band continuum, a strong
   break at 1.32~$\mu$m due to water vapor, strong He~I absorption
   (frequently blueshifted) at 1.083~$\mu$m, and few if any emission
   lines.

2. We then applied these criteria to the 33 objects, and divided them
   into three categories: bona fide FUors (objects that display the
   above characteristics and for which an eruption was observed),
   FUor-like objects (which display the above characteristics but for
   which an eruption was not witnessed), and peculiar objects which
   either miss some FUor characteristics or display additional unusual
   features.

3. We examined multiple spectra of three FUors (V1057~Cyg,
   V1735~Cyg, and V2494~Cyg). V1057~Cyg has in recent years seen a
   significant weakening of the CO bands, possibly related to a steep
   fading in brightness. Of the eight spectra of V1735~Cyg obtained
   over the past 9 years, only one deviates from an otherwise stable
   appearance, one night showing an additional red continuum and
   stronger CO band absorption, with no significant change in
   brightness. V2494~Cyg showed no spectral variability.

4. We have plotted the equivalent widths of the $\sim$2.2~$\mu$m lines
   of Na and Ca versus the equivalent width of CO for all observed
   objects. FUors and FUor-like objects do not fall in the same area
   as Class~I sources, but are found together with giant and
   supergiant stars, supporting other evidence that FUors are
   low-gravity objects. With only a few exceptions, the FUors and
   FUor-like objects are located together in a group, whereas the
   peculiar objects are mostly displaced to another nearby region of the
   diagram.

5. Whereas the first FUors to be discovered were bright optically
   visible stars, in recent years an increasing number of FUors are
   found that are embedded. Extinction is therefore an important
   parameter in the study of FUors, and we have estimated the amount
   of extinction for all FUors and FUor-like objects. We assume that
   any spectral differences between FUors are dominated by extinction
   rather than spectral differences and/or veiling. We then matched
   the slope of all FUors and FUor-like objects to FU~Orionis, which
   has a low extinction of A$_V$$\sim$1.5$\pm$0.2. Many of the objects
   were also observed in the L-band, and often show a pronounced
   3~$\mu$m ice band. By measuring the depth of the ice band and
   comparing to the estimated extinction we have calibrated the
   relationship between the ice band optical depth $\tau$ and the visual
   extinction: $\tau~=~0.048*A_V - 0.13$.

6. Most FUors display reflection nebulae, and we provide an atlas of
   K-band images of the surroundings of all of the objects studied. We
   note that three objects have no optical counterpart and yet do not
   have pronounced reflection nebulae at K-band. We speculate that these objects
   may not be deeply embedded, but merely located on or just beyond the back side of
   their clouds.

7. The near-infrared spectra of FUors and the more massive brown
   dwarfs display great similarity, and at low resolution can be
   indistinguishable. Close examination of their spectra show some
   discrepancies. Paschen-$\alpha$ is very pronounced in FU~Ori, but
   is located in a region with poor atmospheric
   transmission. Paschen-$\beta$ on the other hand is much stronger in
   FUors than in brown dwarfs, where it is often absent. Certain weak
   metallic lines in brown dwarfs are much stronger than in FUors, but
   both Paschen-$\beta$ and these lines are in the J-band, which makes
   them hard to detect in highly extincted objects. It is
   likely that also brown dwarfs can undergo FUor eruptions, and this
   may add to the confusion between brown dwarfs and FUors.

8. We have estimated the bolometric luminosity for each target using a
   homogeneous data set consisting of photometry from 2MASS (JHK),
   WISE (3.4 to 22~$\mu$m), and Akari (65 to 160~$\mu$m), and from the
   160~$\mu$m data point we appended a 20~K blackbody. In view of the
   major improvement in distances to star forming regions in recent
   years, we review for each object its most accurate distance. The
   median bolometric luminosity for bona fide FUors is 99~L$_\odot$
   and 35~L$_\odot$ for the FUor-like objects. Given that FUors
   decline with time and that FUor eruptions are more likely to be
   witnessed today than at earlier times, the difference is not
   unexpected. However, a 2-sample K-S test shows that the
   distributions cannot be distinguished at the 90\% level.

9. We have confirmed two new FUor-like objects, V371~Ser and
   CB230~IRS1, that initially attracted attention due to their bright
   near-infrared cometary reflection nebulae. We classify V1647~Ori,
   frequently considered a FUor, as a peculiar object with some FUor
   characteristics since its spectrum differs significantly from that
   of FUors. We speculate that the object is a close binary with one
   component being a FUor and the other being responsible for the
   emission line spectrum. Similarly, V346~Nor has long been
   classified as a FUor, but its unusual light curve, its almost
   complete lack of CO absorption and presence of emission lines
   points to a peculiar nature, possibly another case of a close
   binary. Finally both AR6a/6b are deemed peculiar, in the case of
   AR6a because its spectrum has changed and is no longer
   FUor-like.


\vspace{0.5cm}

%



\acknowledgments  

We are grateful to Nuria Calvet for a critical reading of an early
version of the manuscript, to Will Best for providing spectra of brown
dwarfs, to Scott Dahm for the spectrum of HBC~341, to Maria Kun
for information on the distance of the L1287 cloud, to Adwin Boogert for 
information on water ice absorption, and to the anonymous referee 
for a helpful report.  We acknowledge
the support of the NASA Infrared Telescope Facility, which is operated
by the University of Hawaii under contract NNH14CK55B with the
National Aeronautics and Space Administration, and we are grateful for
the professional assistance from Dave Griep, Eric Volquardsen, Brian
Cabreira, Tony Matulonis, and Greg Osterman.  This research has made
use of the SIMBAD database, operated at CDS, Strasbourg, France, and
NASA's Astrophysics Data System.  This publication makes use of data
products from the Two Micron All Sky Survey, which is a joint project
of the University of Massachusetts and the Infrared Processing and
Analysis Center/California Institute of Technology, funded by the
National Aeronautics and Space Administration and the National Science
Foundation.  This research has made use of NASA's Astrophysics Data
System.



Facilities: \facility{IRTF}.

\clearpage

\appendix
\section{Extinction and Reflection Nebulae}
FUors are generally associated with interstellar material, and
especially among the more recently discovered ones there are a number
of embedded cases. It is of interest to estimate the amount of
extinction that each FUor suffers along our line of sight, and we use
our spectral library to make such an attempt. As can be seen from
Figure~\ref{FUor_outburst}, FUor spectra have fundamental
similarities. On the assumption that their difference in spectral
slope is dominated by extinction rather than intrinsic spectral
differences, we have determined the amount of extinction that must be
added to the spectrum of FU~Ori to make the best fit to the observed
spectra of the other 32 objects studied here.  We adapted the spectral fitting code used
by Connelley \& Greene (2010), using FU Ori as the reference star, and only considering extinction and not adding any veiling.
This code adds extinction to the spectrum of FU Ori in an effort to
minimize the RMS error between that model and the spectrum of the FUor
in question.  The estimated uncertainty in the extinction is                             
the amount of change in extinction required to double the RMS fitting
error between the FUor in question and the extincted spectrum of FU
Ori.

Herbig (1977) concluded that FU~Orionis has very little extinction,
and Hartmann \& Kenyon (1985) and Kenyon et al. (1988) suggested an
A$_V$ of 1.55 and 2.2, respectively, by comparison with disk models.
More recently, Zhu et al. (2007) determined an A$_V$ of 1.5$\pm$0.2
from a detailed comparison of observations with models, a value which
we adopt here.  Table~2 lists the derived extinctions for each object after adding FU Ori's 1.5 magnitudes of extinction, and Figures~\ref{FUor_outburst} and
\ref{FUor_like_flat} show the dereddened spectra. Two comments are
pertinent: (1) Most of the objects in this study have a rather strong
spectral similarity to FU~Ori, thus giving some confidence in the
estimated extinction values, but evidently the more discrepancies to
FU~Ori that a spectrum shows, the more uncertain the value becomes;
(2) the extinction estimates refer to the light that reaches us, which
may not always be dominantly from the source itself, but could be from
reflected light that escapes through a cavity in those cases where the
source is deeply embedded.  Thus, the listed extinctions may be lower limits.

In order to evaluate whether we see a FUor directly, or we observe a
compact reflection nebula, we have examined the K-band images that
were obtained at the same time as each spectrum; Table~2 lists the
corresponding K-magnitudes. Figure~\ref{Images1} shows panels with     
these K-band images. It is immediately clear that not all objects have
stellar point spread functions.  The following
eight sources are not seen directly even in the K-band, but are
compact reflection nebulae: L1551~IRS5, Haro~5a~IRS, IRAS 05450+0019,
V371~Ser, Parsamian~21, V2495~Cyg, CB~230~IRS1, and HH~354~IRS.  All of
these, with the puzzling exception of Parsamian~21, are among the
objects with highest extinctions, which demonstrates that the path of
light from each source passes copious amount of dust.

It is well known that FUors in optical light are surrounded by often
prominent reflection nebulae (e.g., Herbig 1977, Goodrich 1987).
Indeed, CB230~IRS1 and V371 Ser were initially suspected of being
FUor-like on account of their nebula morphology alone; their true
nature was confirmed by subsequent spectroscopy. The reflection
nebulae on larger scales are obviously more pronounced at short
wavelengths, but these near-infrared images offer insight into the
immediate surroundings of the objects. We note that ten objects do not
show evidence for reflection nebulae in the K-band, these are:
V883~Ori, V2775~Ori, FU~Ori, IRAS~06297+1021W, IRAS~06393+0913, Z~CMa,
BBW~76, V1515~Cyg, V1735~Cyg, and V733~Cep. All of these objects are
among the ones with lowest extinctions, although in three cases it is
not negligible.  The remainder show various degrees of reflection
nebulosity. In eight cases we clearly see cometary nebulae with the
source embedded at the apex, indicating that it is illuminating an
outflow cavity. Perhaps not surprisingly, those eight are the same
sources listed above that are not seen directly.

Only three sources do not show any sign in the visible of either a
star or a reflection nebula.  This is evidence of how the detection of
FUors until recently has been primarily done in the optical. Since all
of the sources are associated with dark clouds, it follows that almost
all of these objects are located on the side of the clouds facing
towards the observer.  We note that \emph{all} optically visible FUors
and FUor-like objects are associated with an optical reflection
nebula.  A roughly equal number of objects should be located on the
far side of dark clouds. Three sources without any optical
counterpart, V2775~Ori, IRAS~06297+1021W, and IRAS~06393+0913, do {\em
not} show evidence of reflection nebulae, suggesting that they are not
deeply embedded, and these three objects may thus be part of the
missing population of FUors that are located near the backside of
clouds (we note that a weak reflection nebula around V2775~Ori was
detected in an HST image, see Fig.~2 of Fischer et al. 2012).


Our extinction estimates along with our 2 to 4~$\mu$m spectra allow us
to calibrate the relationship between the 3.0~$\mu$m ice band depth
and the visual extinction for our sample of YSOs.  The relationship between the depth of the
3.0~$\mu$m ice band and extinction is well established.  Beck et al.
(2001) showed that the 3.0~$\mu$m absorption is much greater towards T
Tau S than T Tau N, consistent with T Tau S being seen through greater
extinction.  Beck (2007) fit an ice absorption model to 2 to 4~$\mu$m
spectra of several YSOs.  However, she considered the applicability of the 
relation between ice column density and visual extinction to be questionable.  
Chiar et al. (2011) clearly showed a linear trend
between 3.0~$\mu$m ice band depth and the visual extinction through a
quiescent dark cloud.

   We adopted an empirical approach towards calibrating the
   relationship between the 3.0~$\mu$m ice band depth and the visual
   extinction.  We measured the continuum flux at 2.50, 3.05, and
   3.80~$\mu$m over a bandwidth of 0.05~$\mu$m.  We fit a line between
   the 2.50 and 3.80~$\mu$m flux levels, and calculated the interpolated
   3.05~$\mu$m continuum level (Figure~\ref{Icedefinition}).  The relationship between A$_v$    
   (derived by comparison to FU Ori) and the ice band depth is shown in
   Figure~\ref{ice_depth}.  For this figure, we used all of the FUors     
   and FUor-like objects for which we have 2 to 4~$\mu$m spectra
   except for Z CMa, since its continuum is so unlike other FUor-like
   objects.  The data in red are objects affected by reflection
   nebulae in the near-IR.  Most of the very highly extincted objects
   have reflection nebulae, and are affected by scattered light.  The
   depth of the ice band appears to saturate near $\tau\approx2$,
   corresponding to a visual extinction of about 30.  Despite having
   sources seen through higher extinction, the ice band depth does not
   appreciably deepen for these highly extincted objects.  We fit a
   linear regression line through the data of all but the two highest 
  extinction objects, for which the ice band optical depth appears to have saturated.  
  We derive the following the relationship:

\begin{equation}
\tau = (0.048\pm0.008)*A_V - (0.13\pm0.14)
\end{equation}

where $\tau$ is the ice band optical depth and A$_V$ is the visual
extinction in magnitudes.  $\tau$ was calculated as --ln(flux at
3~\micron/continuum at 3~\micron)

  This method of calculating the ice band optical depth does not go to
  zero at zero extinction, due to a natural curvature in the spectrum
  of a FUor at this wavelength
 range.  This curvature of the continuum is also seen among late M-giants in the 
 SpeX library.  To compensate for this effect, we subtracted the $\tau$ of FU~Ori
 (likely the lowest extinction target) from all of the sources, and this is reflected in
 the equation above.  Some objects strongly deviate from
  this relationship, such as Parsamian 21, which is a low extinction
  object (the reflection nebula is seen in visible light) but has a
  strong ice band absorption feature.  In this case, a large fraction
  of the light from Parsamian 21 may be scattered, and the true
  extinction to the source may be much higher.

\section{Comments on Adopted Distances}

{\bf L1287.} The two sources RNO~1b/c are found in Cassiopeia and are
associated with the L1287 cloud located in the Orion arm of our Galaxy
(Kun 2008). These sources are commonly assumed to be at the kinematic
distance of 800~pc (Persi et al. 1988). However, recently Reid et al.
(2014) used the VLBI to measure the parallax of a maser source in
the L1287 cloud, and determined a distance of 930$\pm$35~pc, which we
henceforth use.

{\bf The California Molecular Cloud.} The FUor-like object PP13S is
located in a dense core named L1473 within the giant California
Molecular Cloud. A variety of distances have been determined for
different components of the California Molecular Cloud, the most
recent is the extinction-based study by Lada et al. (2009) suggesting
a distance of 450$\pm$23~pc, which we adopt here.

{\bf L1551.}  The small cloud L1551 in Taurus is located just south of
the large Taurus cloud complex. Only three degrees from L1551 and also
south of the Taurus complex is a similar small cloud containing the
eponymous T~Tauri. For T~Tauri a VLBA distance of 146.7$\pm$0.6~pc has
been determined by Loinard et al. 2007, and we here assume the same
distance for L1551, but adopt a greater uncertainty, 147$\pm$5~pc.

{\bf Orion.} The structure of the large Orion complex of star forming
regions is increasingly well understood, see Bally (2008) for a
review.  Recently Kounkel et al. (2017) observed 36 YSOs with the Very
Large Baseline Array (VLBA) and determined a distance to the Orion
Nebula Cluster of 388$\pm$5~pc (in excellent correspondence with the
earlier, less accurate measurements of Jeffries 2007, Sandstrom et al.
2007 and Mayne \& Naylor 2008) and found that the L1641 (Orion~A)
cloud is inclined away from us, with a distance of 428$\pm$10~pc
towards its southern portion, while similarly the L1630 (Orion~B)
cloud is increasingly distant towards the north, with NGC~2068 at
388$\pm$10~pc and NGC~2024 at roughly $\sim$420~pc. Consequently we
adopt 428~pc for V2775~Ori, which is located in the southern part of
L1641, and 388~pc for V883~Ori which is located in the northern part
of L1641 near the ONC. For Haro~5a~IRS in OMC~2 we also adopt the ONC
distance of 388~pc, and the same for V1647~Ori just south of NGC~2068
and for IRAS~05450+0019 in NGC~2071 just north of NGC~2068. Due to
plasma radio scattering in the $\lambda$~Ori region, Kounkel et al.
were not able to measure sources in that region, hence for FU~Ori the
best distance estimate remains the 400$\pm$40~pc of Murdin \& Penston
(1977). FU~Ori was measured by Gaia, but the First-Release value of
$353^{+82}_{-56}$~pc is still not a significant improvement, so for
now we remain with the 400~pc value.

{\bf NGC 2264.} Four of the objects in this study are located towards
NGC 2264, these are IRAS 06297+1021W, IRAS 06393+0913, and AR6a/b.
Hence we re-examine the current status of distance estimates to this
region. NGC~2264 is often assumed to be at a distance of about 800~pc,
although various determinations range from 700~pc to 950~pc, see
Table~1 in Dahm (2008) for a full summary. Subsequent distance
determinations of the cluster are by Baxter et al. (2009), who suggest
913$\pm$40~pc based on model fits to the rotational velocity
distribution, Turner (2012) who suggests 777$\pm$12~pc based on main
sequence fitting to 13 B-type stars, and Kamezaki et al. (2014), who
find 738~$^{+57}_{-50}$~pc based on VLBI parallax measurements of two
water maser sources. Reviewing these values and their uncertainties,
we deem that the 738~pc distance for now remains the best value for
NGC~2264.

{\bf Monoceros.} The two FUors V900~Mon and V960~Mon are found in this
region. For V900~Mon we adopt the distance of about 1100~pc suggested
by Reipurth et al. (2012). For V960~Mon there is much uncertainty
about the distance, Caratti o Garatti et al. (2015) and K\'osp\'al et al.
(2015) assume 450~pc.  However, this FUor is associated with the L1649/1650
cloud complex, for which Kim et al. (2004) derive a kinematic
distance of 1100~pc, and we adopt that distance here.

{\bf Serpens.}  The three objects V371 Ser, IRAS 18270-0153W, and IRAS
18341-0113S are all located within the Serpens constellation, but in
three different regions. V371~Ser is embedded in the densest part of
the Serpens Core, which has frequently been assumed to be rather
nearby, e.g. Straizys et al. (1996) determined that extinction rises
at a distance of about 259$\pm$37~pc. However, a more recent VLBA
parallax study by Dzib et al. (2010) of a young binary in the Serpens
Core indicated a distance of 415$\pm$5~pc, later refined to
429$\pm$2~pc (Dzib et al. 2011). This was supported by additional VLBA
observations that yielded a distance of 436$\pm$9~pc (Ortiz-Le\'on et
al. 2016). This suggests that the Serpens Core is not part of the
nearby Aquila Rift, but is located behind it, and the earlier
extinction-based estimates most likely refer to the Aquila cloud
complex. IRAS 18270-0153W is located in the northernmost cloud core in
the filamentary cloud that contains the embedded Serpens-South cluster
(see Figure~1 of Friesen et al. 2016). It has long been assumed that
Serpens-South and the neighboring W40 star forming region are
physically associated due to their similar gas radial velocities.
Several stars in W40 were included in the parallax study of
Ortiz-Le\'on et al. (2016), and it thus follows that Serpens-South is
likely to be at a distance of 436$\pm$9~pc, which we therefore also adopt
as the distance to IRAS~18270-0153W. For the third object in Serpens,
IRAS~18341-0113S, there is, however, no evidence that it is associated
with the more distant clouds, and at its location on the southeastern
edge of the Aquila Rift (see Figure~1 of Ortiz-Le\'on et al. 2016), it
seems more likely that it is part of the Aquila Rift at the
259$\pm$37~pc distance of Straizys et al. (1996), which we thus adopt

{\bf Aquila.} The FUor Parsamian 21 has a very poorly determined
distance, with suggestions varying from 400~pc (e.g., K\'osp\'al et al.
2008) to 1800~pc (Staude \& Neckel 1992). The object is located
towards Cloud~A of Dame \& Thaddeus (1985), who suggest a kinematic
distance of 500~pc. For lack of better determinations, we here adopt a
distance of 500~pc, while recognizing its considerable uncertainty.

{\bf North America Nebula.} The two FUors V1057 Cyg and HBC 722 are
located towards the North America Nebula (NGC 7000), which together
with the Pelican Nebula (IC 5070) form a single large HII region W80
that is bifurcated by the L935 dark cloud. Numerous distance
determinations have been made over time, and we here adopt
550$\pm$50~pc as determined by Laugalys et al. (2006). This distance
is likely to apply to V1057~Cyg, which is seen towards the northern
part of the HII region.  For HBC~722, which is located on the
frontside of the L935 cloud, it is possible that the distance is
slightly closer, but we also retain 550~pc for this object.

{\bf Khavtassi 141.} The two FUors V2494 Cyg and HH~381~IRS are
located in the L1003 cloud in Cygnus. Together with the nearby L988
cloud they form the two highest extinction regions in the large cloud
Khavtassi 141, sometimes called the Northern Coalsack. Kh~141 is
located within the Cyg~OB7 association, with which it is assumed to be
associated.  Numerous distance estimates towards this region have been
made, resulting in values from 500 to 800~pc; for details see the
discussions in Herbig \& Dahm (2006) and Khanzadyan et al. (2012).
Following Herbig \& Dahm (2006) we accept 600~pc as a compromise
value.

{\bf IC 5146.} V1735~Cyg is located towards the western part (L1045)
of a long filamentary cloud which at its eastern extremity harbors the
star forming region IC~5146. Starting with the study of Walker (1959)
almost all studies have suggested a distance between 900 and 1200~pc
(see Table~1 of Harvey et al. 2008) except Lada et al. (1999) who used
extinction measurements to suggest a distance of 460$\pm$60~pc. Upon
reviewing the various methods, we adopt the distance proposed by
Harvey et al. (2008) of 950$\pm$80~pc.

{\bf L1165.}  The FUor HH354~IRS is embedded in the small cloud L1165,
which is located east of the HII region IC~1396, at a separation of
2.6$^\circ$.  It is unclear whether or not the two are physically
associated. The distance to IC~1396 has been determined as
870$\pm$80~pc by Contreras et al.  (2002), consistent with previous
estimates (e.g.  $\sim$1000$\pm$100~pc by Garrison \& Kormendy 1976).
However, the CO local standard of rest velocity is discrepant, and
hence it has been suggested that L1165 may form a foreground cloud,
with a distance variously assumed between 200~pc and 400~pc (for a
discussion, see Kun et al. 2008). HH354~IRS is associated with a
maser, so an accurate distance may one day be measured, but until then
we adopt a distance of 300~pc (e.g. Dobashi et al. 1994).

\section{Objects sometimes considered FUors, but not included here}

We here mention three objects which occasionally appear in lists of
FUors but which we exclude in this study for the reasons stated.

\textbf{Re 50 IRS1:} Strom \& Strom (1993) suggested that this
embedded source is a FUor based on the H$\alpha$ profile seen via
reflected light. Re~50 is a complex reflection nebula that appeared in
the southern part of the L1641 cloud sometime between 1955 and 1979.
It is illuminated by the embedded IRAS source 05380--0728, with a
luminosity of $\sim$250~L$_\odot$, and a spectrum of the associated
reflection nebula shows H$\alpha$ and the Ca triplet in emission
(Reipurth \& Bally 1986). Recently the northern part of the reflection
nebula, Re~50N, closest to the source has brightened considerably,
suggesting the possibility of a FUor eruption, but a near-infrared
spectrum shows only a steeply rising featureless continuum (Chiang et
al. 2015).  While it is possible to imagine scenarios in which heated
dust could swamp a FUor spectrum, we consider the absence of
near-infrared spectroscopic features characteristic of FUors as
disqualifying for a FUor classification.

\textbf{OO Ser:} This is a deeply embedded young star that erupted
before 1994, but which faded almost back to its pre-outburst
brightness in only 5 - 7 years (Hodapp et al. 2012), a timescale that
places this object between the FUors and the EXors. A K-band spectrum
obtained in 2006 showed no CO absorption bands (K\'osp\'al et al. 2007).

\textbf{V1331 Cyg:} This luminous emission-line T~Tauri star
has occasionally been found in lists of FUors. It was Welin (1976) who
first suggested that V1331~Cyg might be in a {\em pre-}FUor state,
based on the similarity of its spectrum to the pre-outburst
low-dispersion objective prism spectrum of V1057~Cyg obtained by
Herbig (1977). While V1331~Cyg indeed has very strong winds and
P~Cygni profiles with deep absorption troughs on certain optical lines
(e.g., Petrov et al.  2014), its rich emission line spectrum,
including the 2-$\mu$m CO bands in emission (Carr 1989), is completely
different from all known FUors, and any speculations on a relation to
FUors remain just speculations.






\clearpage

 \begin{figure}
 \plotone{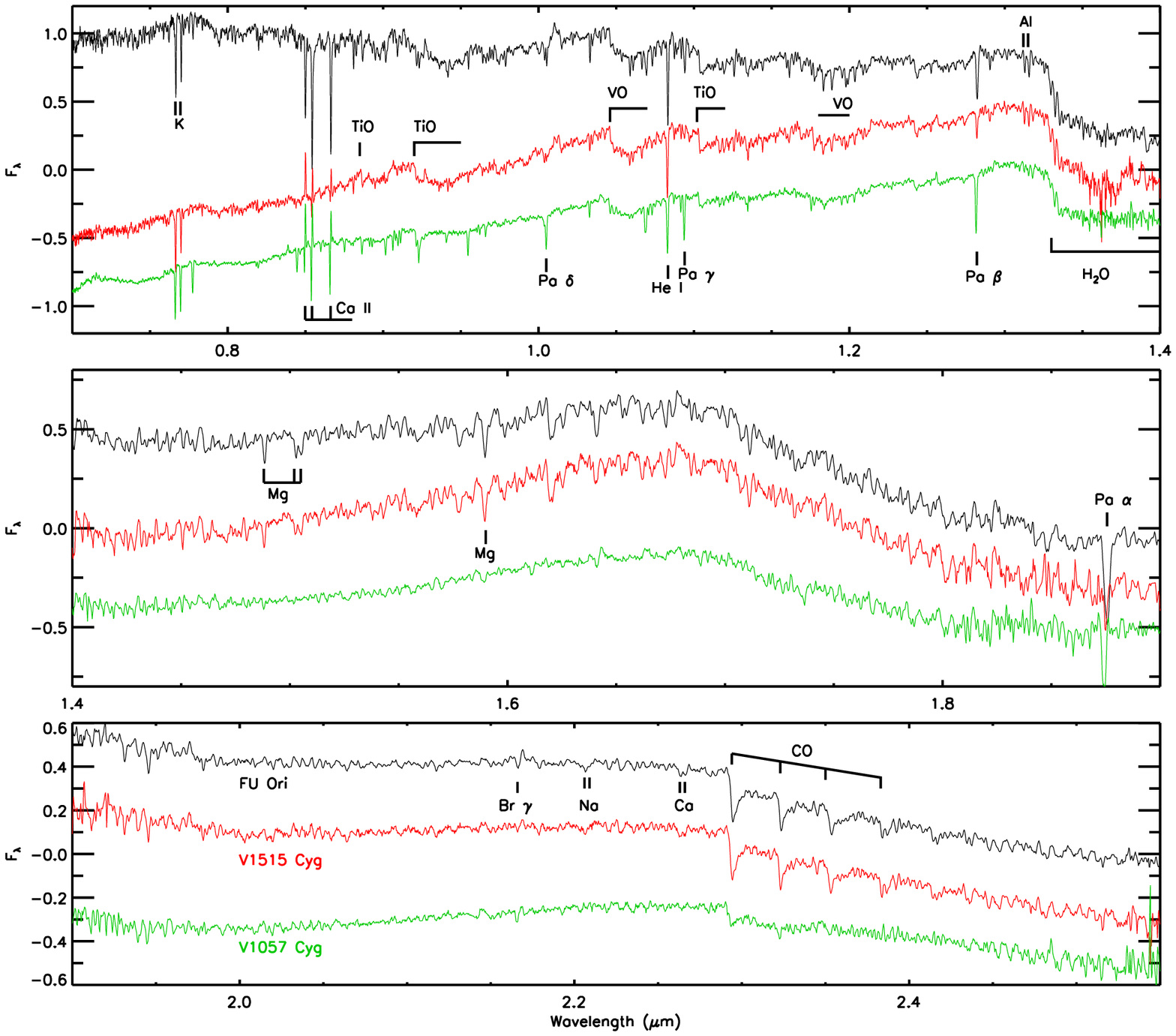}
\caption{J, H, and K-band spectra of the three classical FUors discussed by Herbig (1977): FU~Ori (black), V1515~Cyg (red), and V1057~Cyg (green). 
  \label{classicFUors}}
 \end{figure}
\clearpage

 \begin{figure}
 \plotone{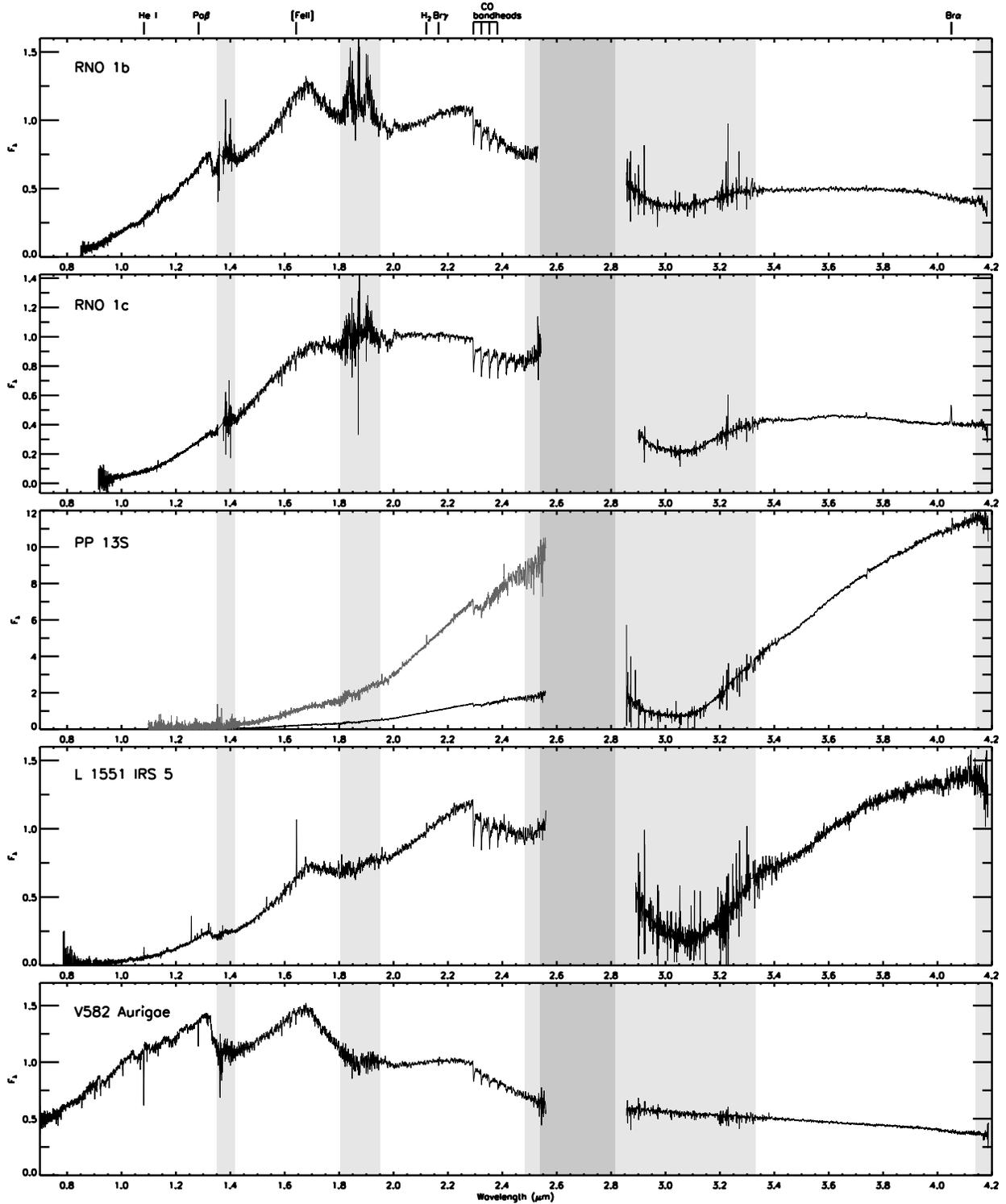}
   \caption{An atlas of 33 objects each of which have been suggested
     in the literature to be related to the FUor phenomenon.  The light
     gray regions show areas of strong telluric absorption, and the
     darker gray area shows where the atmosphere is quite opaque.  For
     PP 13 S, the 1-2.5~$\mu$m spectrum has been overplotted in dark
     gray rescaled by a factor of 5 to
     better show the features in this part of the spectrum.
     \label{page1}}
 \end{figure}
\clearpage

 \begin{figure}
 \plotone{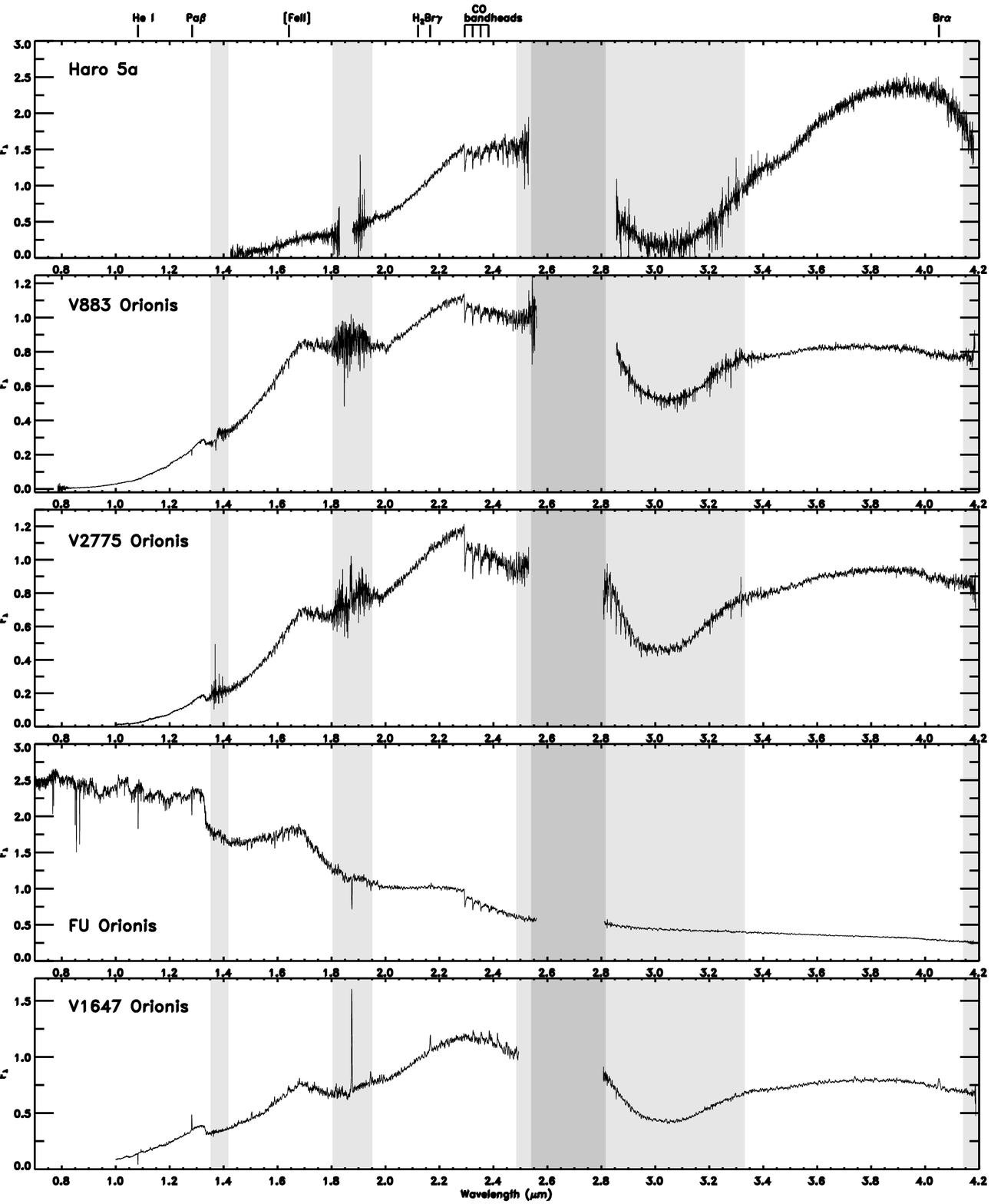}
  \addtocounter{figure}{-1}
\caption{continued
 \label{page2}}
 \end{figure}
\clearpage

 \begin{figure}
  \addtocounter{figure}{-1}
\plotone{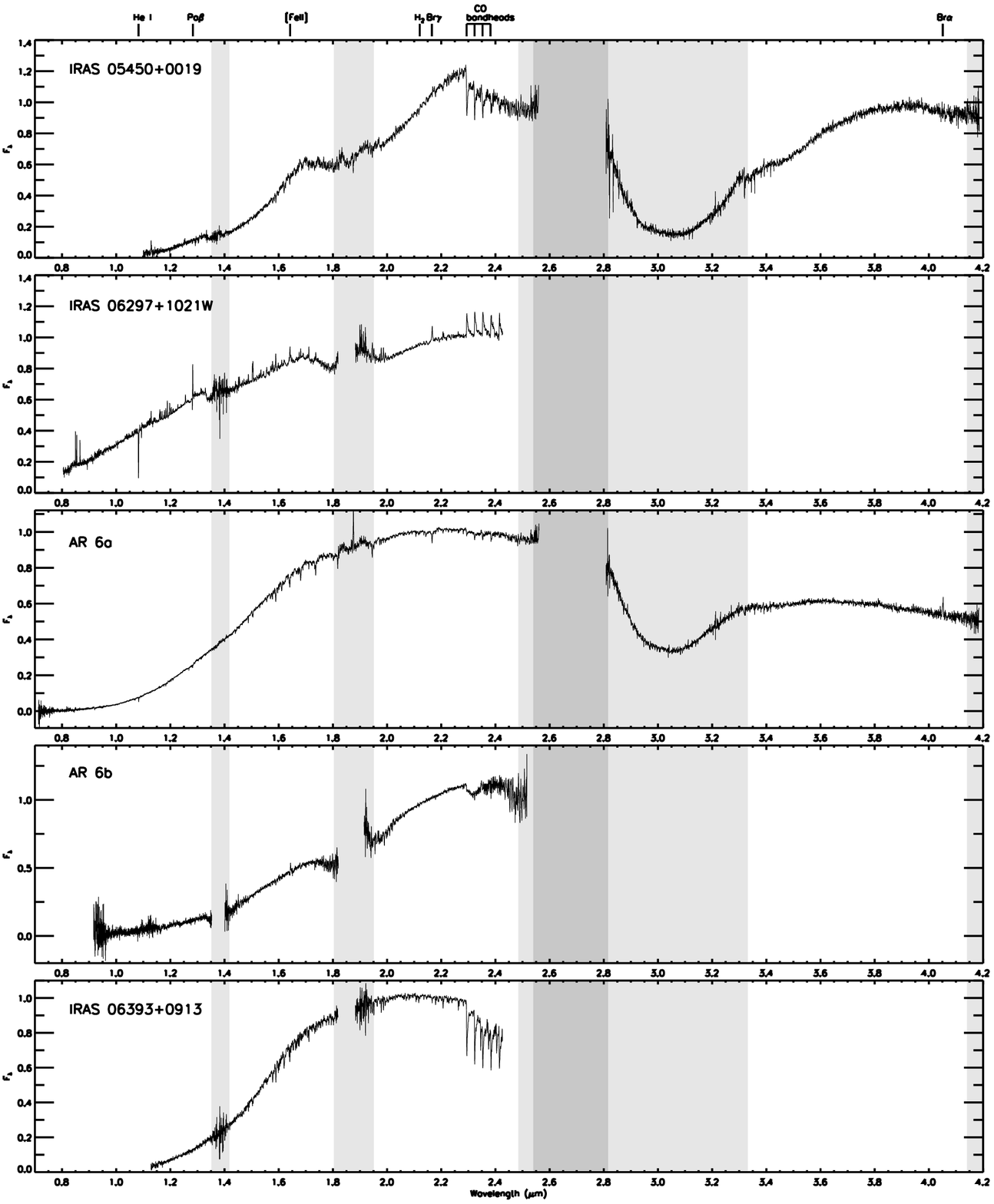}
 \caption{continued
 \label{page3}}
 \end{figure}
\clearpage

 \begin{figure}
  \addtocounter{figure}{-1}
\plotone{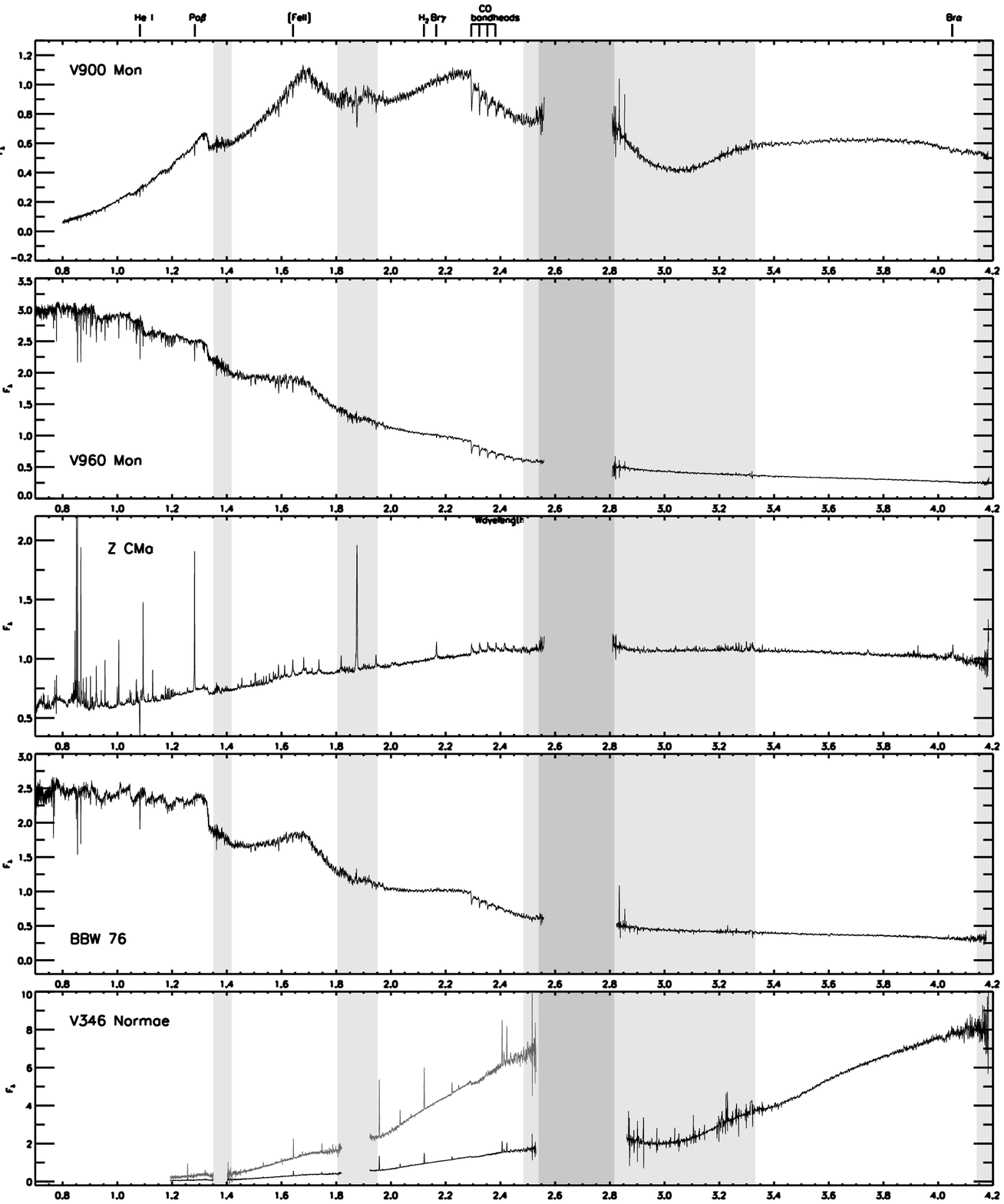}
 \caption{.  For V346 Normae, the
   1-2.5~$\mu$m spectrum that has been overplotted in dark gray was
   scaled by a factor of 4 to better show the features in this part of
   the spectrum.  \label{page4}}
 \end{figure}
\clearpage

 \begin{figure}
  \addtocounter{figure}{-1}
\plotone{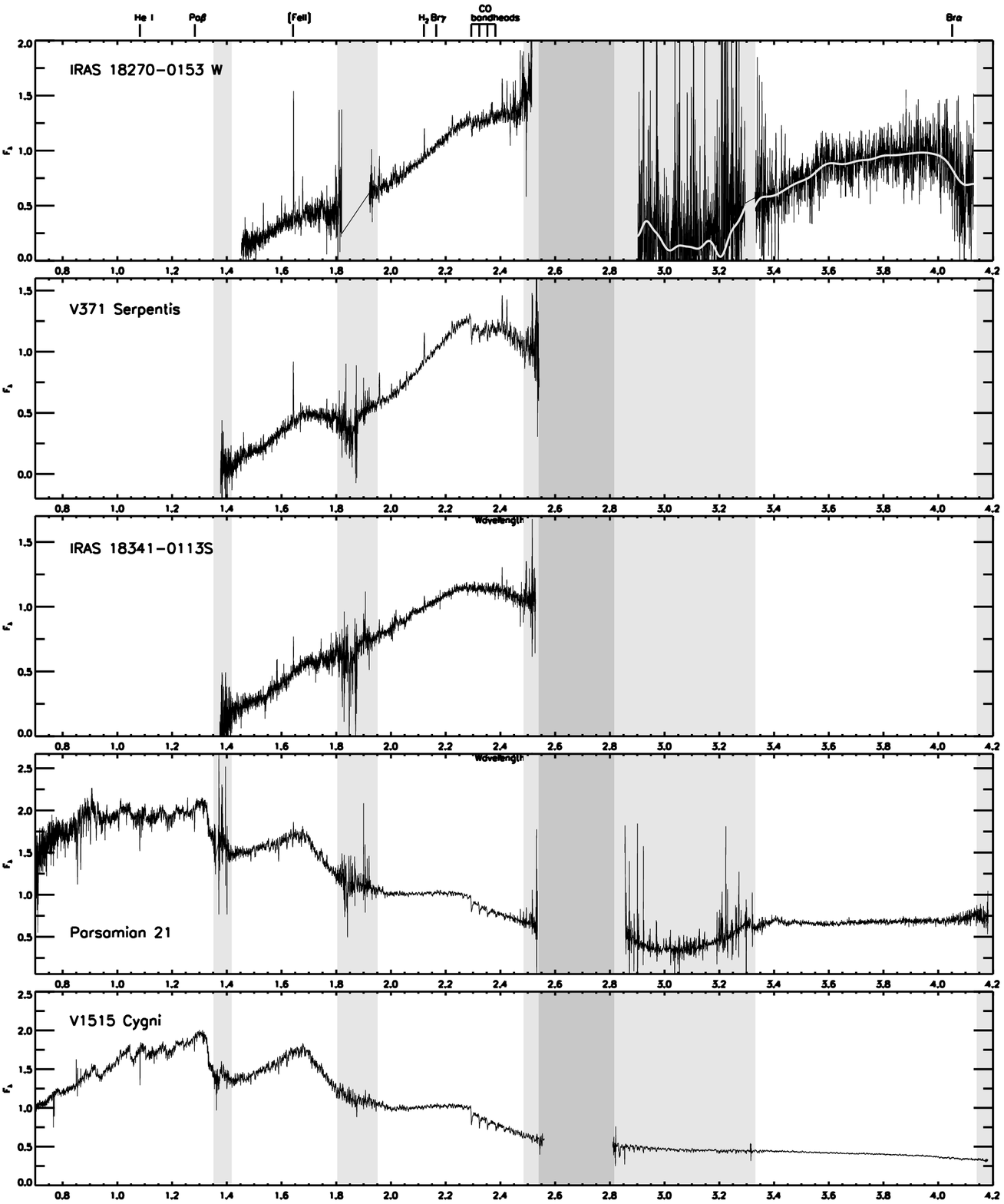}
 \caption{The white line through the L-band portion of the spectum of IRAS 18270-0153W has been boxcar smoothed to better show the overall shape of the spectrum in this region. \label{page5}}
 \end{figure}
\clearpage

 \begin{figure}
 \addtocounter{figure}{-1}
 \plotone{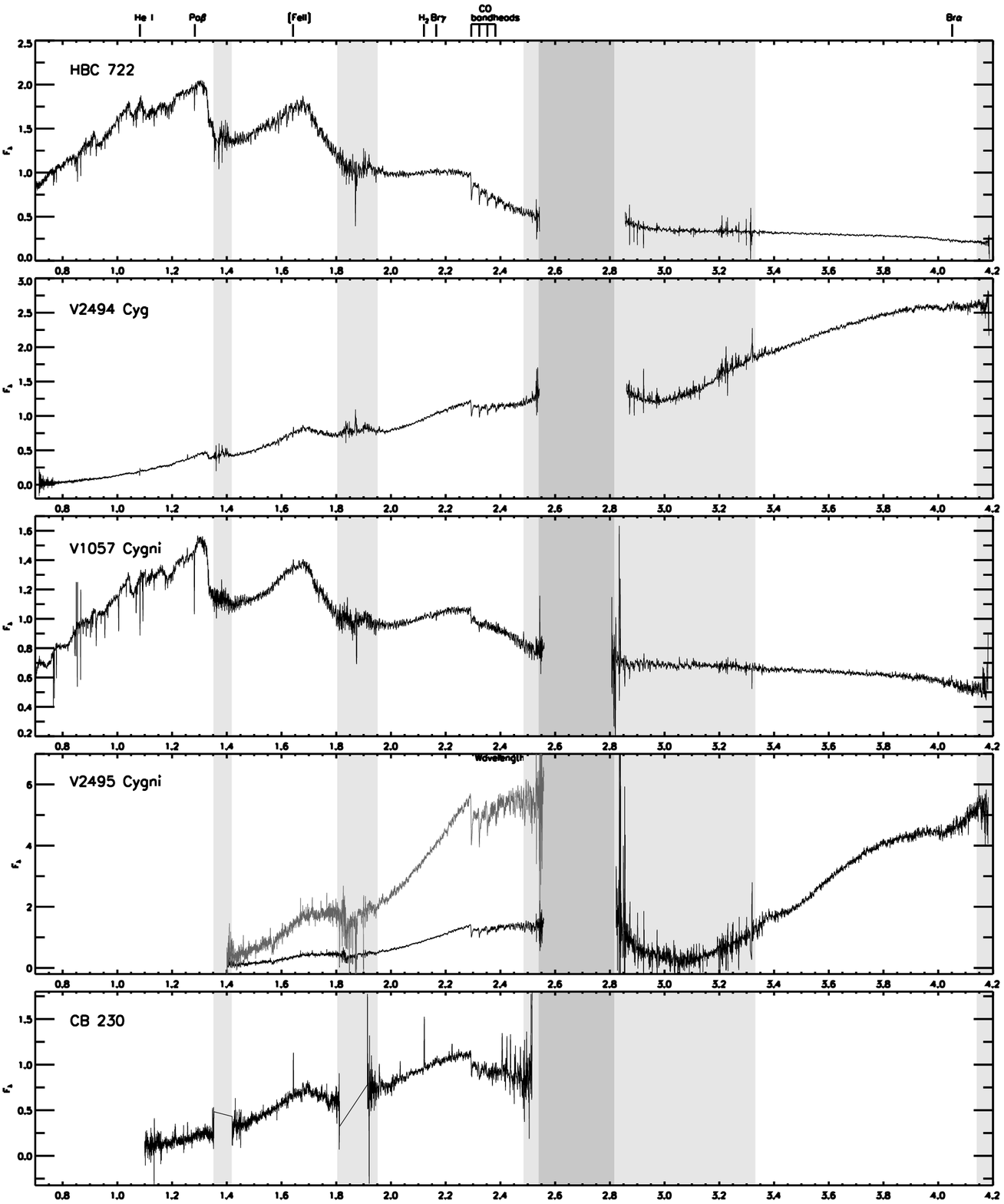}
 \caption{For V2495 Cygni, the
   1-2.5~$\mu$m spectrum that has been overplotted in dark gray was
   scaled by a factor of 4 to better show the features in this part of
   the spectrum.  
  \label{page6}}
 \end{figure}
\clearpage

 \begin{figure}
 \addtocounter{figure}{-1}
 \plotone{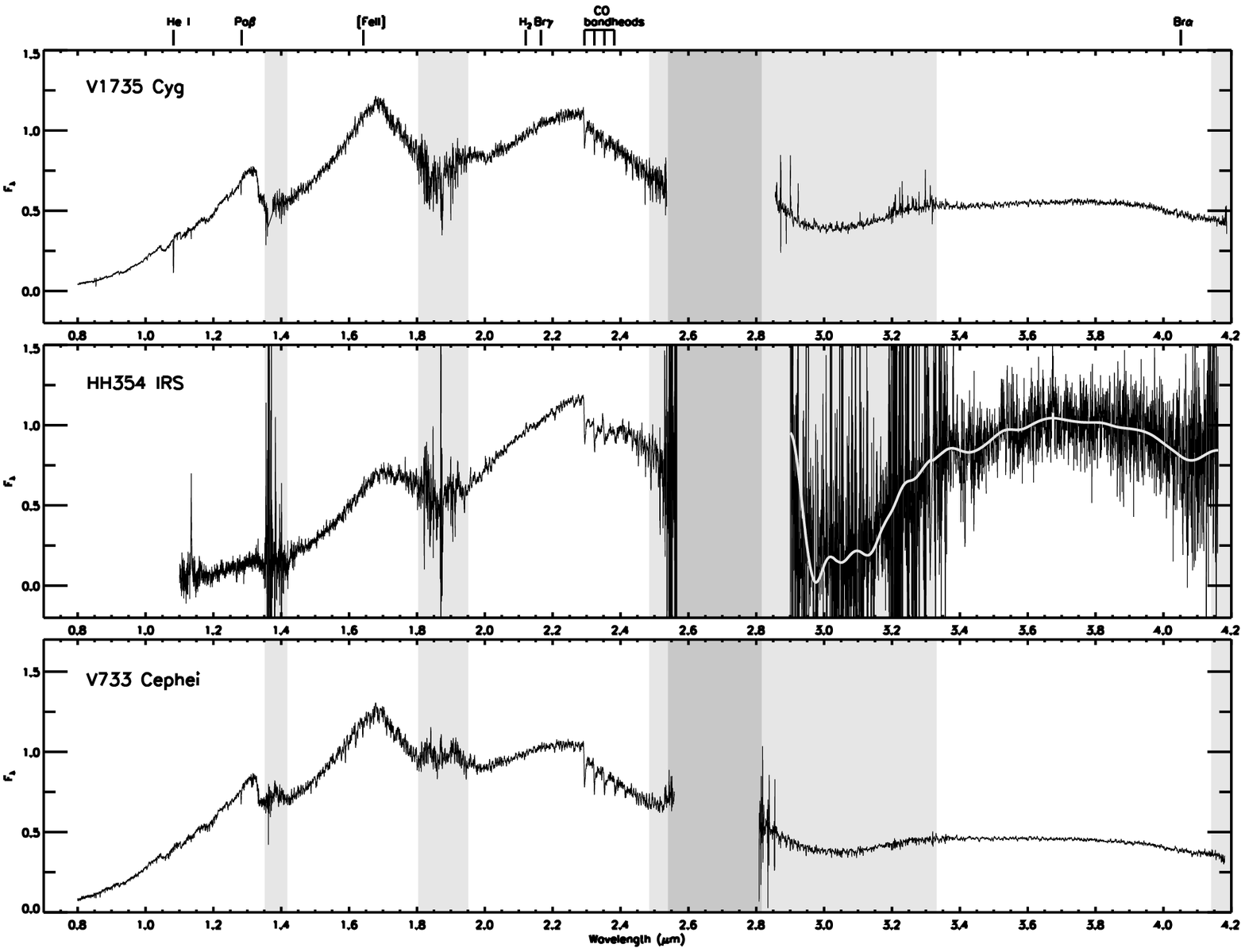}
 \caption{A smoothed spectrum of HH 354 IRS has been overlaid
   on the 3 to 4~$\mu$m region using a Gaussian kernel with a
   0.02~$\mu$m standard deviation. 
 \label{page7}}
 \end{figure}
\clearpage

 \begin{figure}
\epsscale{0.8}
 \plotone{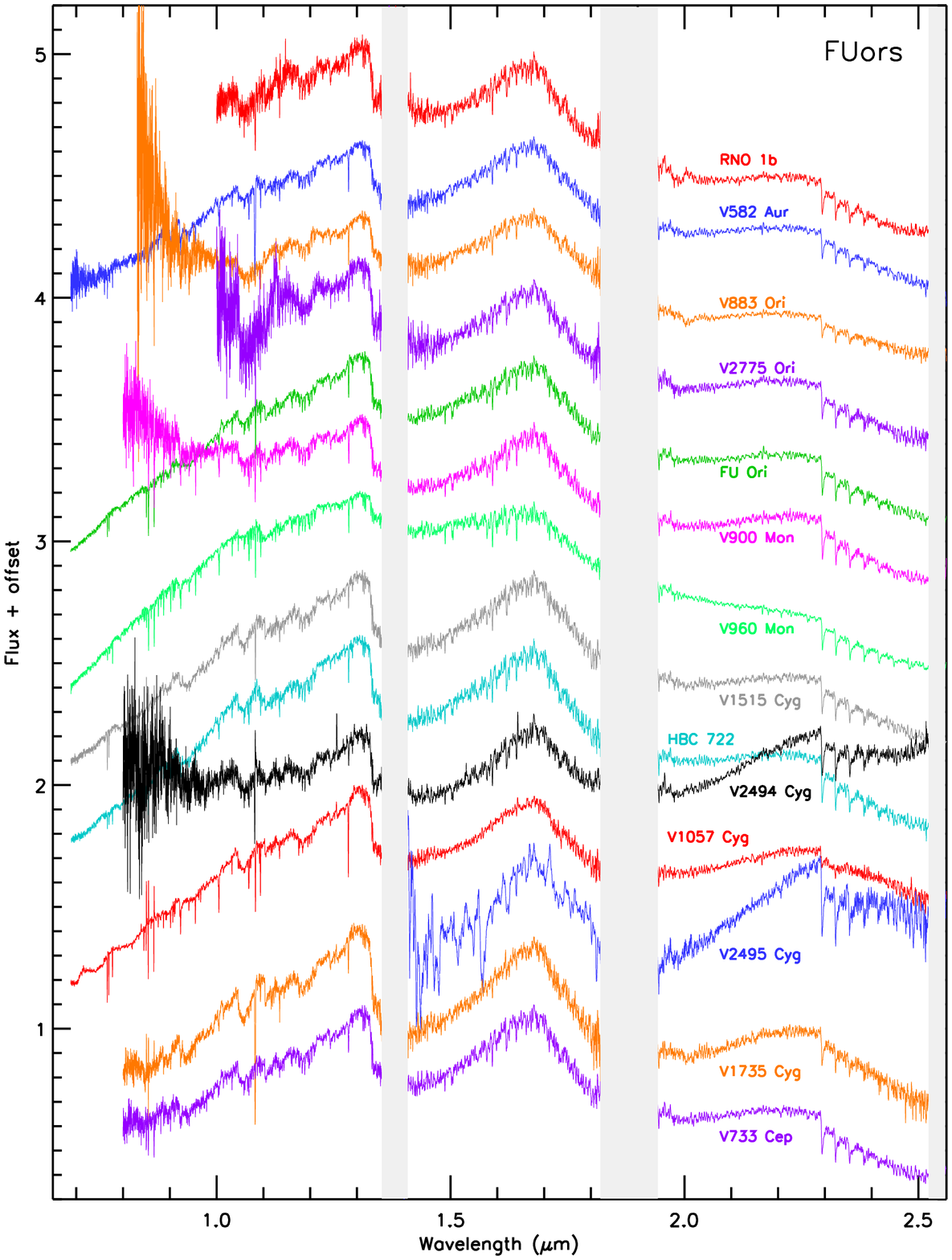}
 \caption{0.7~$\mu$m to 2.5~$\mu$m spectra of only those FUors whose outburst was observed, with extinction added or removed to minimize the overall spectral slope.  The common spectral features thus become apparent: strong He I absorption at 1.08~$\mu$m, a strong break at 1.32~$\mu$m and a 'triangular' H-band continuum due to water vapor absorption and low gravity, strong CO band heads starting at 2.29~$\mu$m, and weak metal absorption lines if any.  Pa$\beta$ at 1.28~$\mu$m is also seen in absorption.  
The flux of each spectrum has been normalized at 2.15~$\mu$m, then offset.
   \label{FUor_outburst}}
 \end{figure}
\clearpage

 \begin{figure}
 \plotone{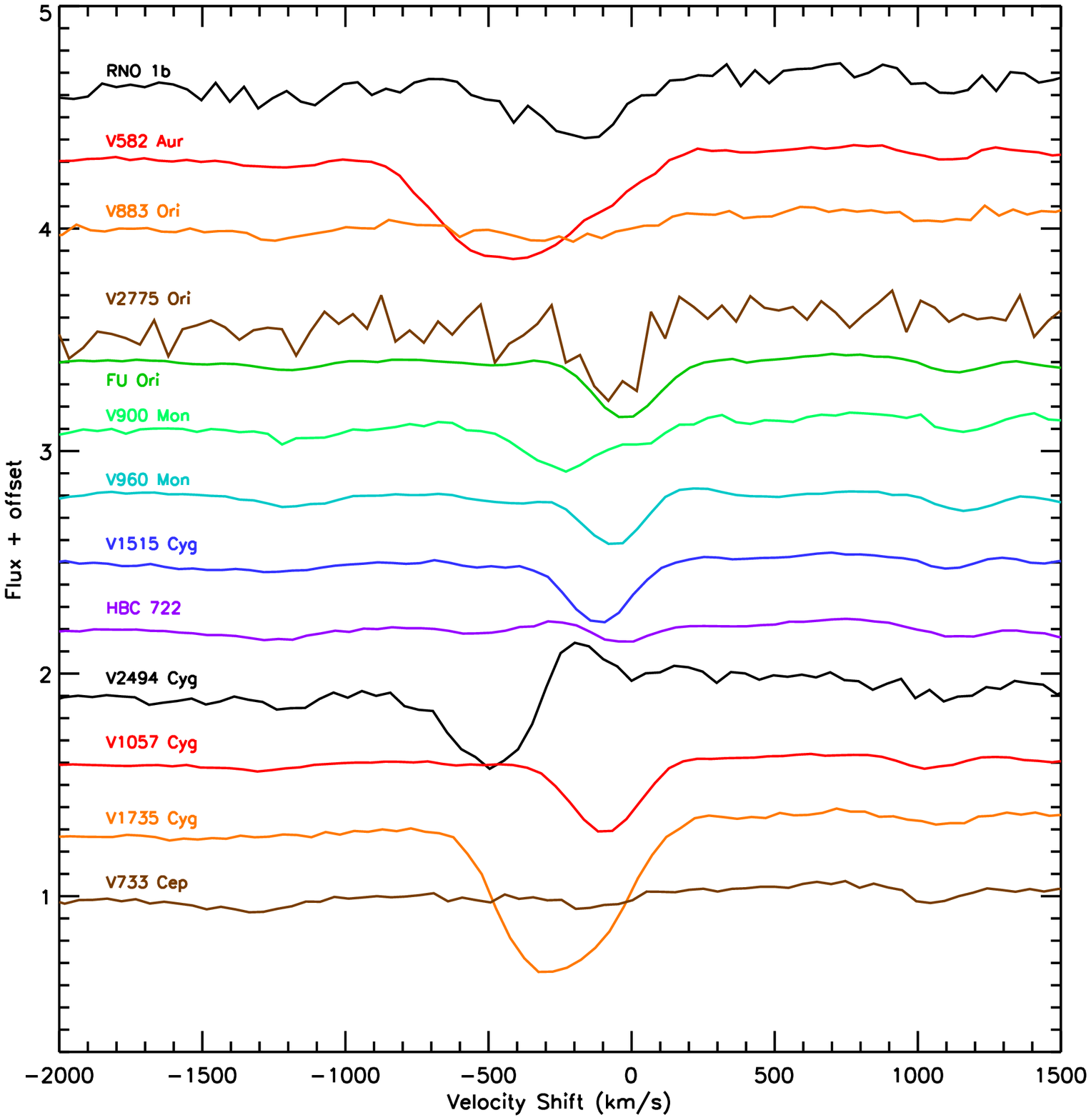}
   \caption{The He I $\lambda$~1.08~$\mu$m line profile, for only bona
     fide FUors.  Nearly all of these lines show blue shifted
     absorption, with a mean velocity of $\sim$350~kms$^{-1}$.
  \label{FUor_outburst_HeI}}
 \end{figure}
\clearpage

 \begin{figure}
 \plotone{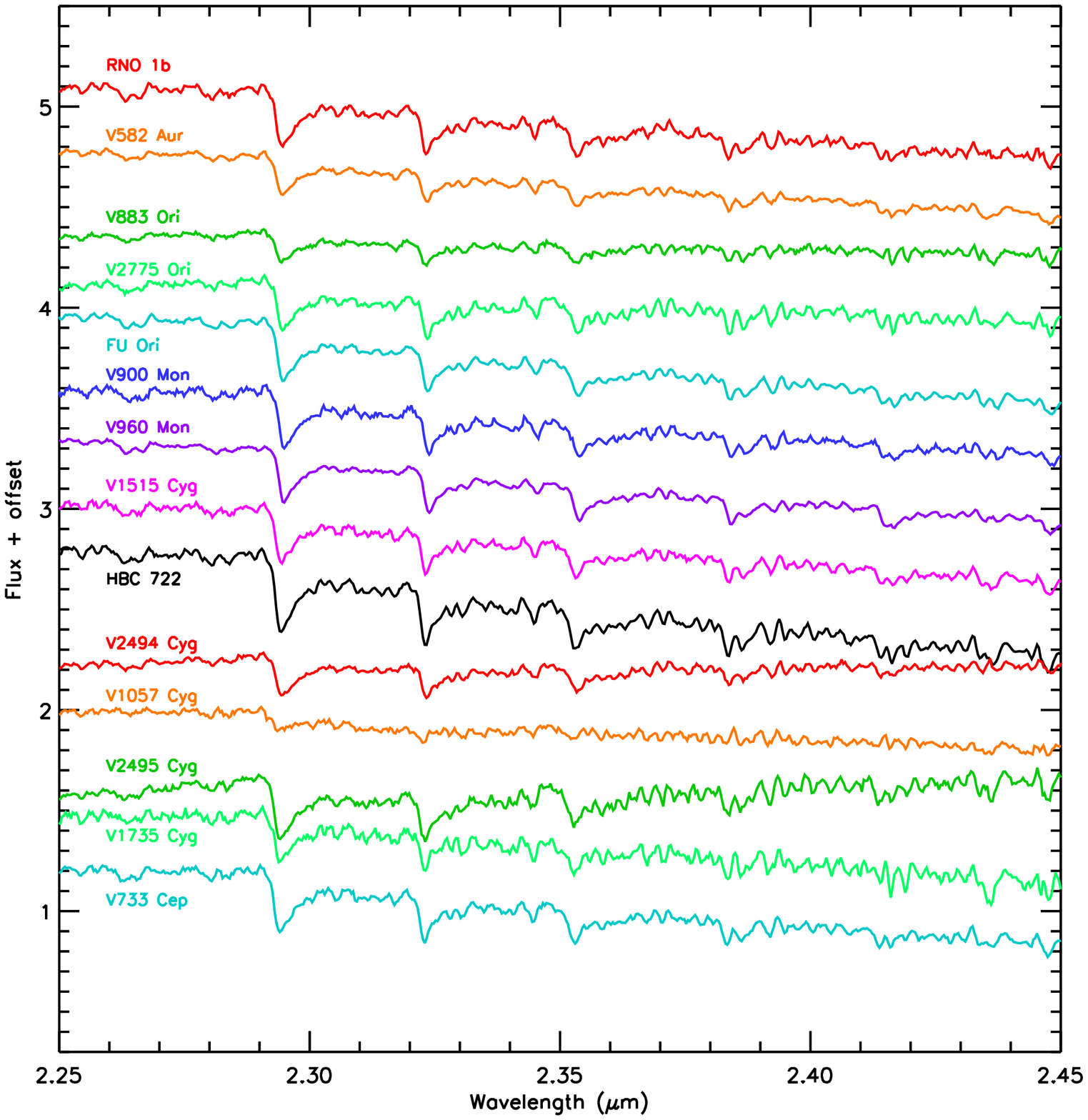}
 \caption{The CO band heads, for only bona fide FUors.   
  \label{FUor_outburst_CO}}
 \end{figure}
\clearpage

 \begin{figure}
\epsscale{1.0}
 \plotone{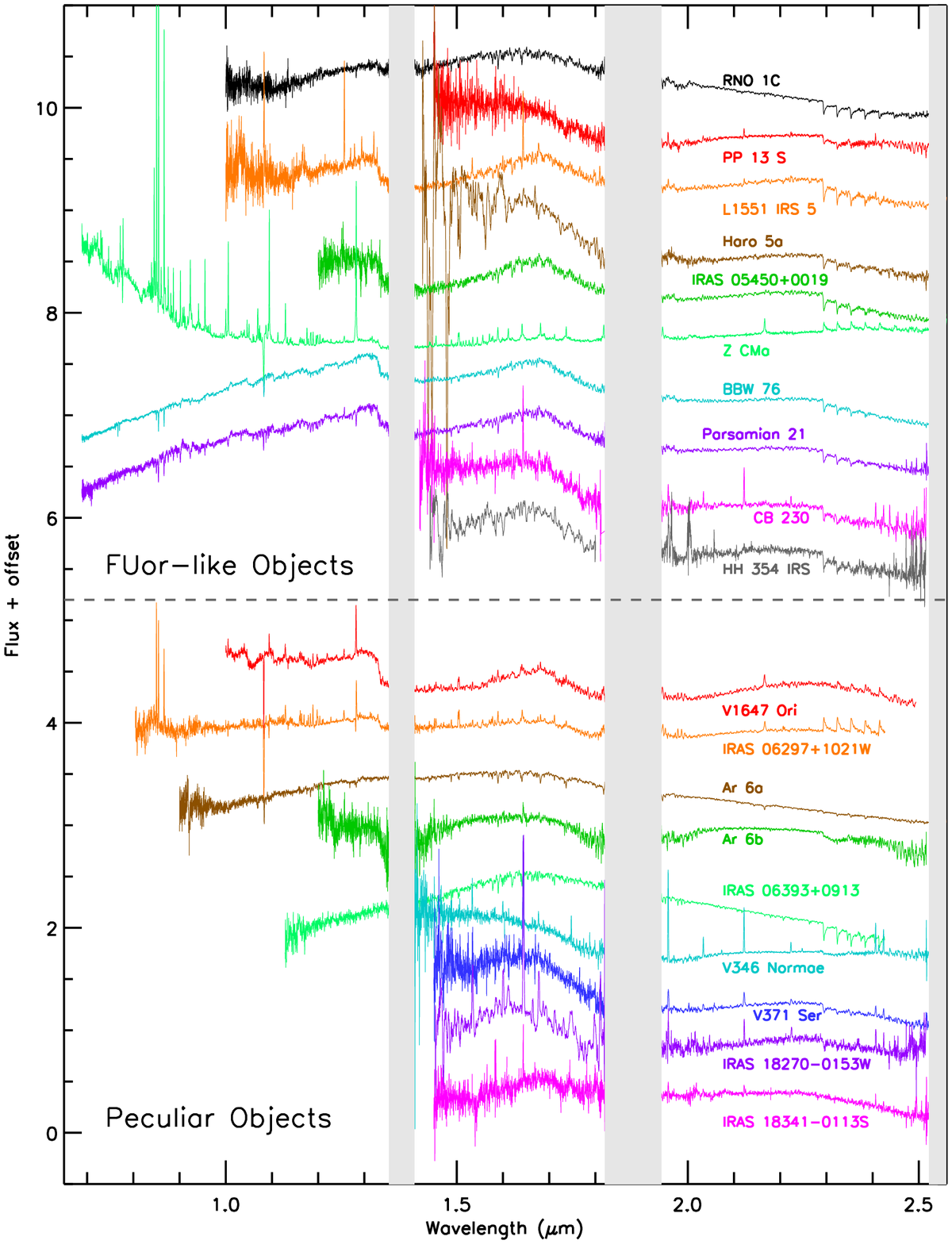}
 \caption{Same as for Figure 3, but for FUor-like objects (upper part) and peculiar objects with some FUor characteristics (lower part). 
   \label{FUor_like_flat}}
 \end{figure}
\clearpage

\begin{figure}
 \plotone{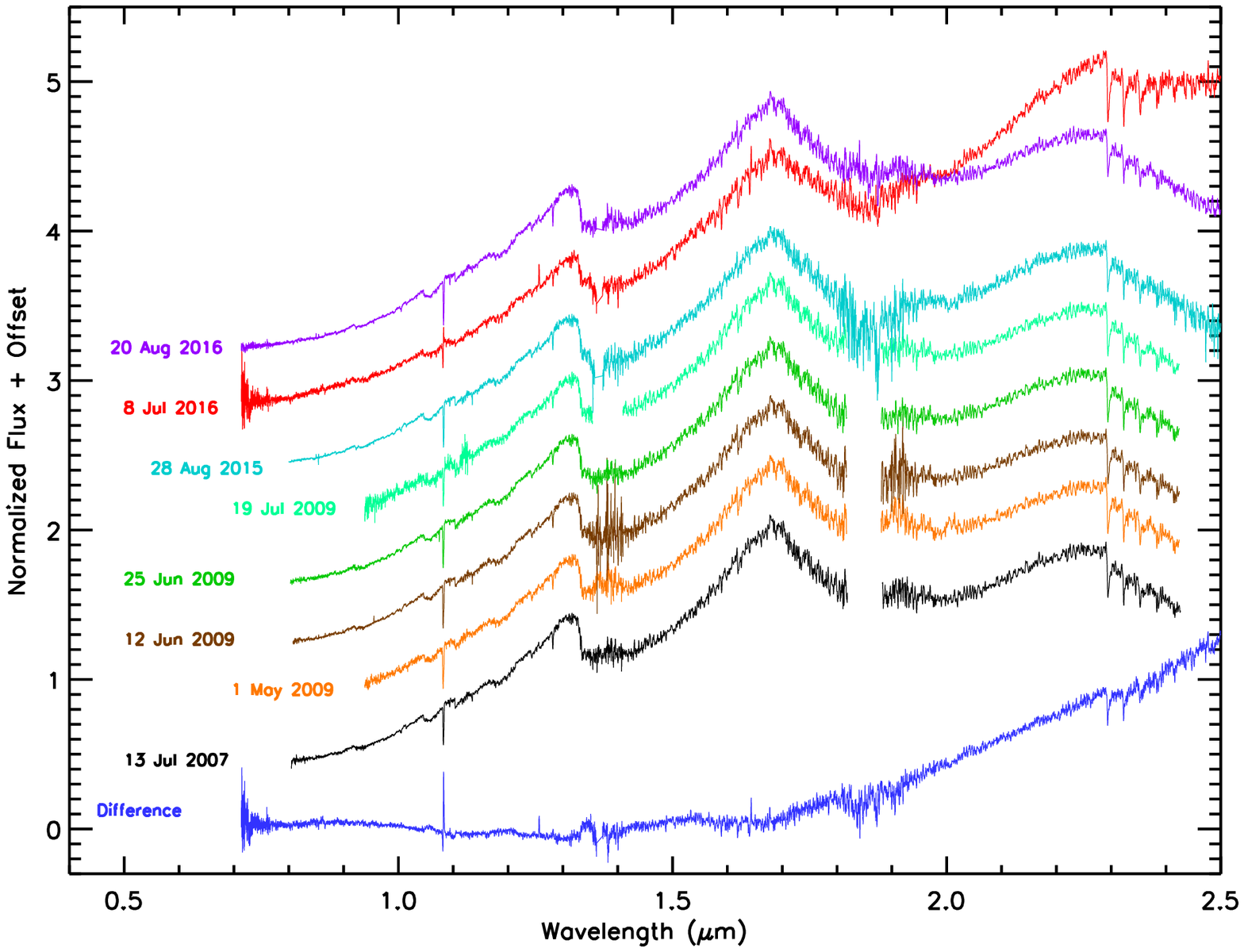}
\caption{V1735 Cyg briefly showed unexpected spectroscopic variability in July 2016, then returned to its nominal state when next observed the following month.  The difference between the July 2016 and August 2016 spectra is shown at the bottom.  We see the addition of a featureless red continuum, but with additional CO absorption.    
  \label{V1735Cyg_var}}
\end{figure}
\clearpage

\begin{figure}
 \plotone{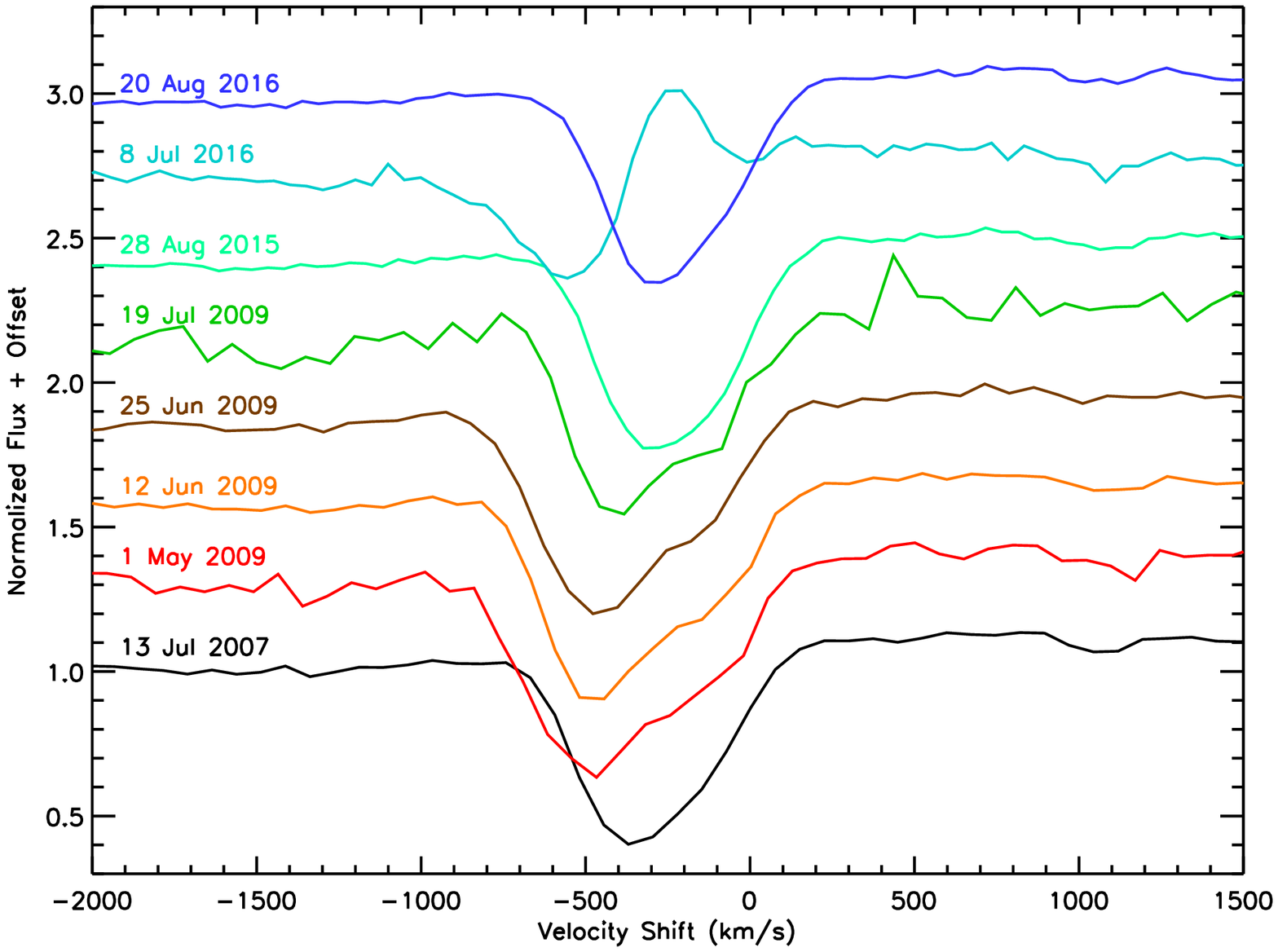}
\caption{The He I $\lambda$ 1.08~$\mu$m line for V1735~Cyg also showed blue shifted emission and faster blueshifted absorption with a velocity of $\sim$950~kms$^{-1}$ when its spectrum changed during the July 2016 event.  
  \label{V1735Cyg_HeI}}
\end{figure}
\clearpage

\begin{figure}
 \plotone{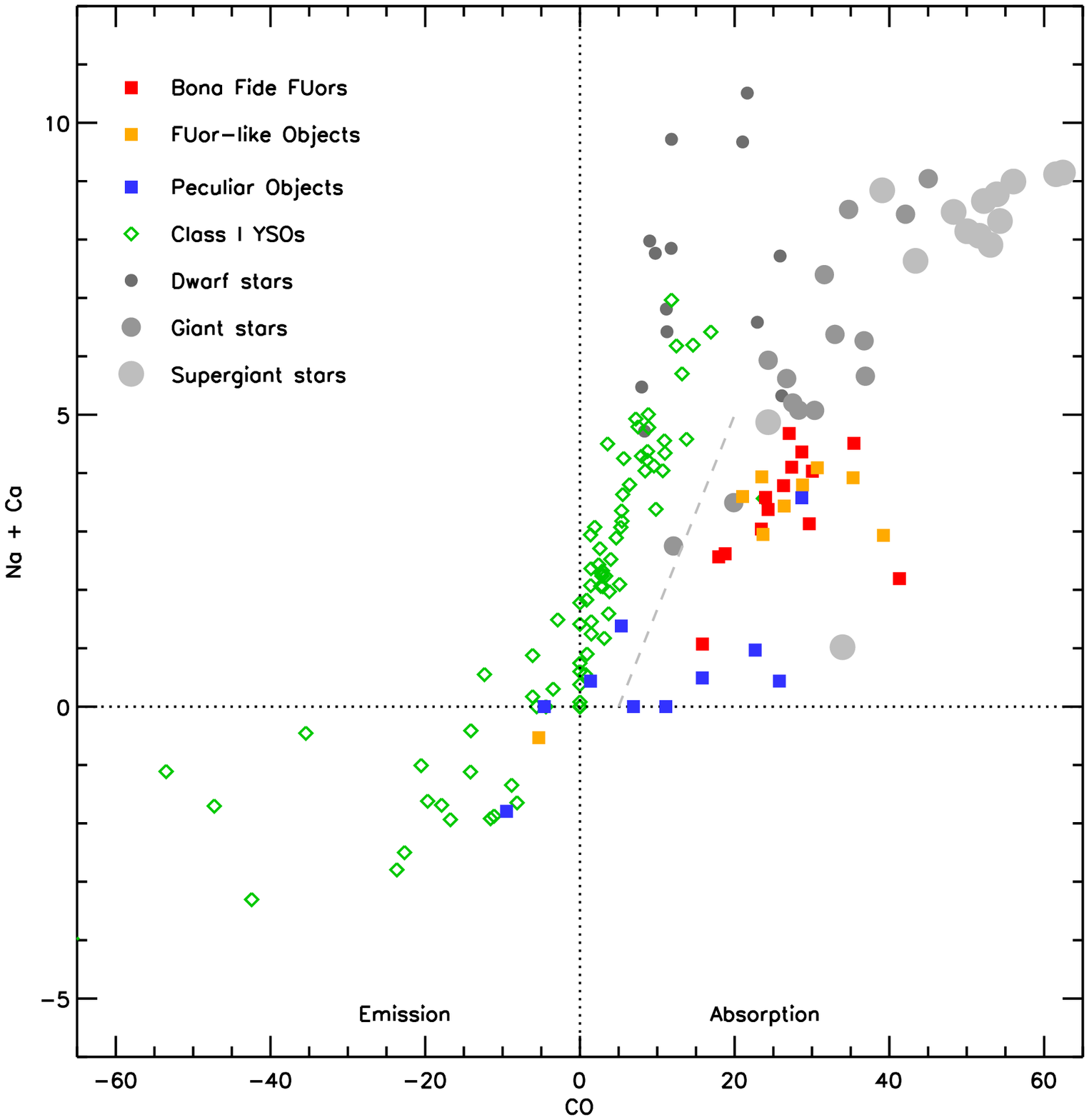}
\caption{In a plot of the equivalent width (EW) of CO vs the EW of Na+Ca, FUors and FUor-like objects lie in a unique part of this space without significant overlap with field stars or other YSOs.  The peculiar objects tend to be located closer to (0,0).  Field M-stars from the SpeX library, shown as gray circles, are to the upper right.  The dotted lines separate the regions of absorption and emission in the figure.  A gray dashed line separates the FUors from the rest of the YSOs.  If $x$ is the EW of Na+Ca, then the CO EW is greater than $3x + 5$.  
  \label{CO_vs_metals2}}
\end{figure}
\clearpage

\begin{figure}
 \plotone{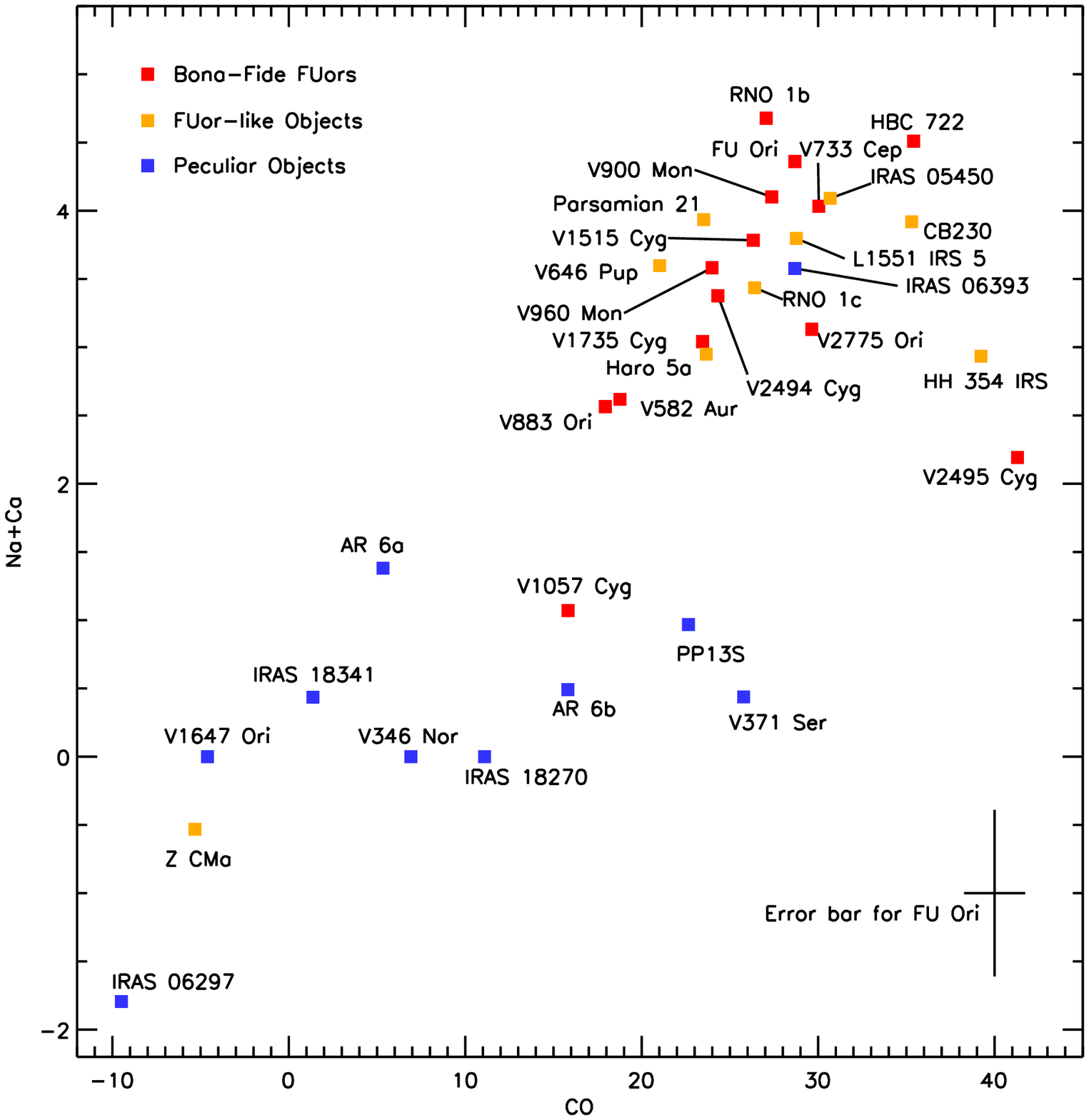}
\caption{An enlargement of the previous figure showing the region of FUors, FUor-like objects, and peculiar objects, with identifications. 
  \label{CO_vs_metals_crop}}
\end{figure}
\clearpage

 \begin{figure}
  \plotone{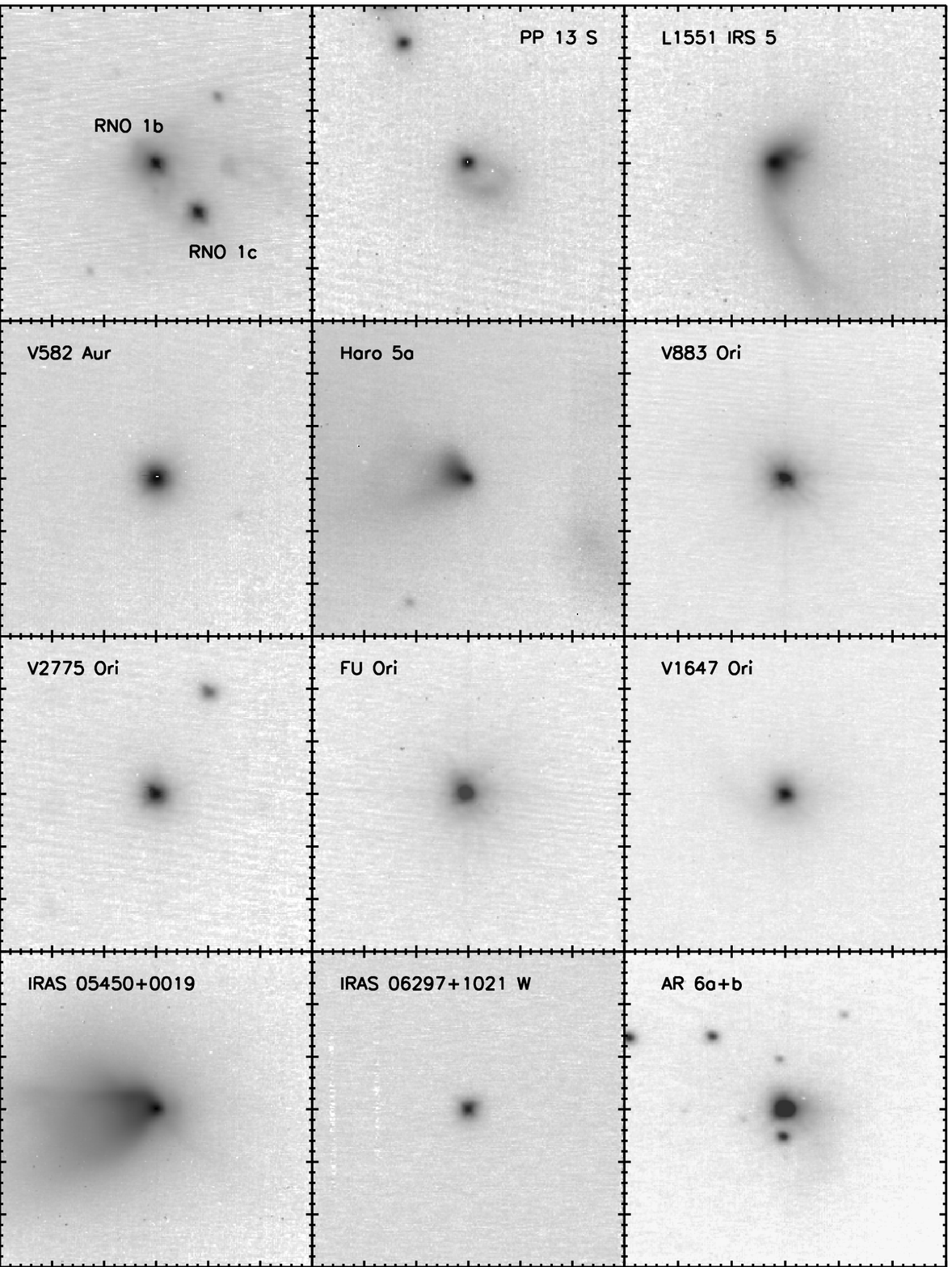}
 \caption{30\arcsec~K-band images of our targets. 
  \label{Images1}}
 \end{figure}
\clearpage

 \begin{figure}
  \addtocounter{figure}{-1}
 \plotone{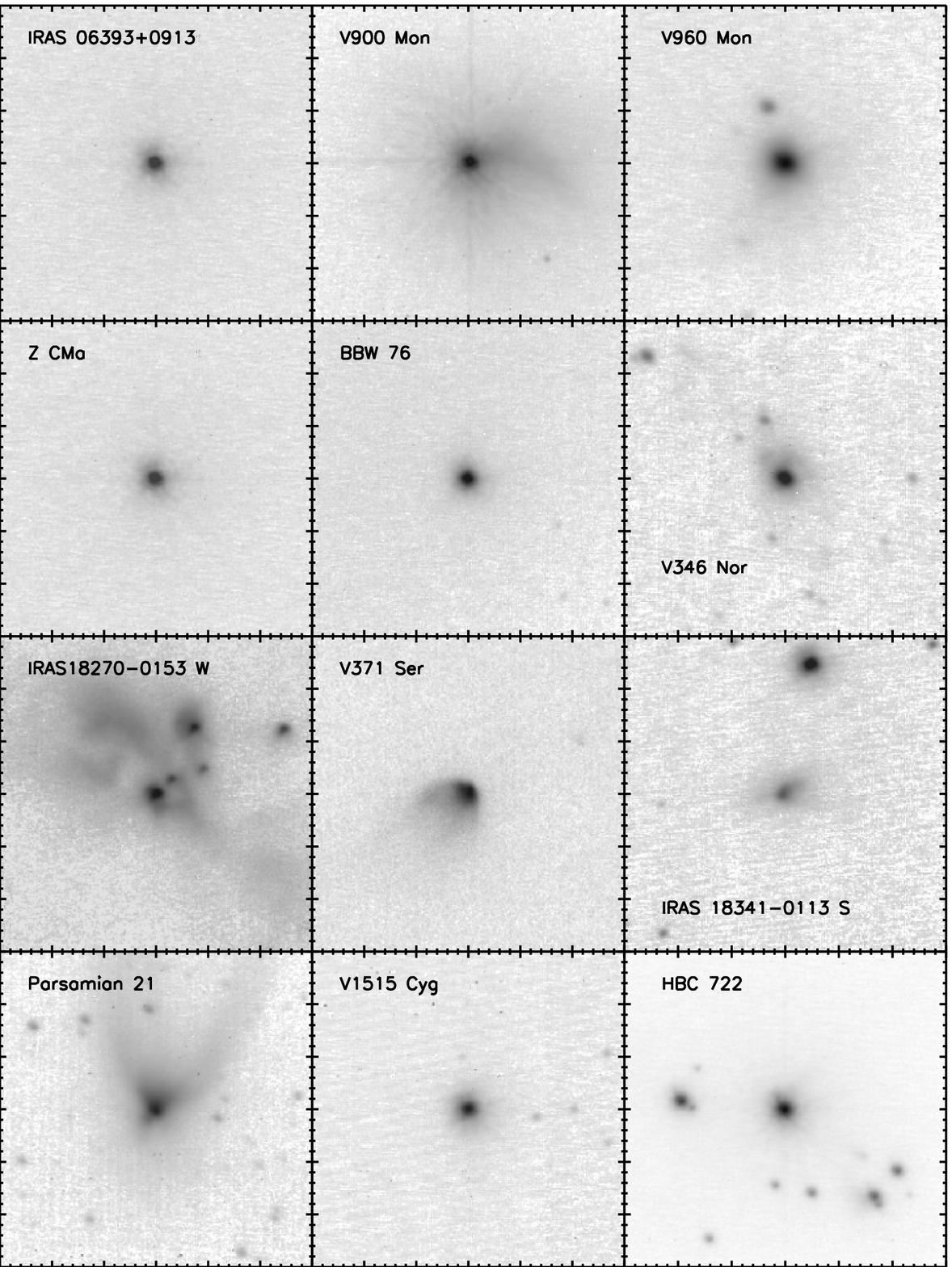}
 \caption{Continued 
  \label{Images2}}
 \end{figure}
\clearpage

 \begin{figure}
  \addtocounter{figure}{-1}
 \plotone{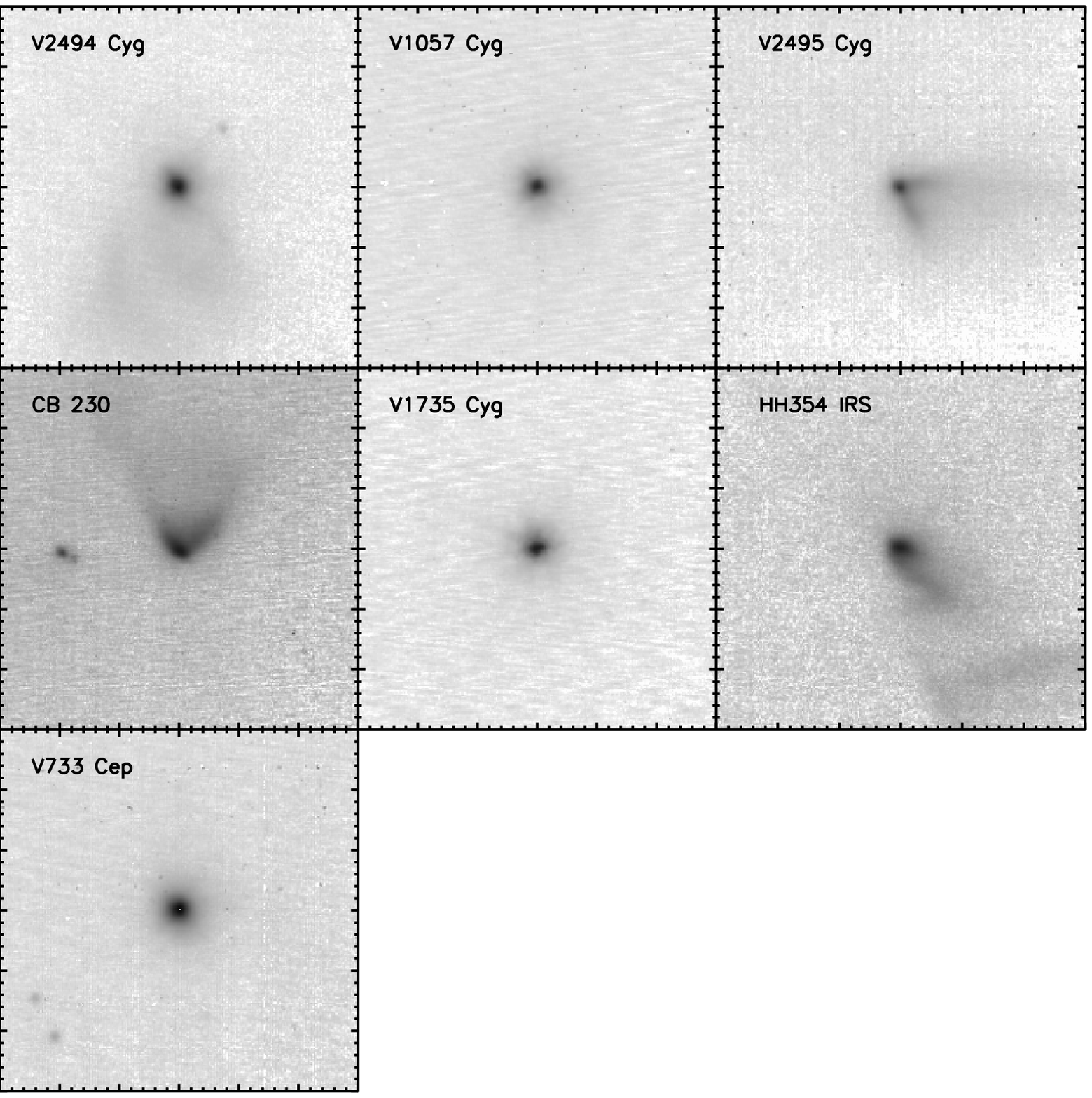}
 \caption{Continued
  \label{Images3}}
 \end{figure}
\clearpage

\begin{figure}
 \plotone{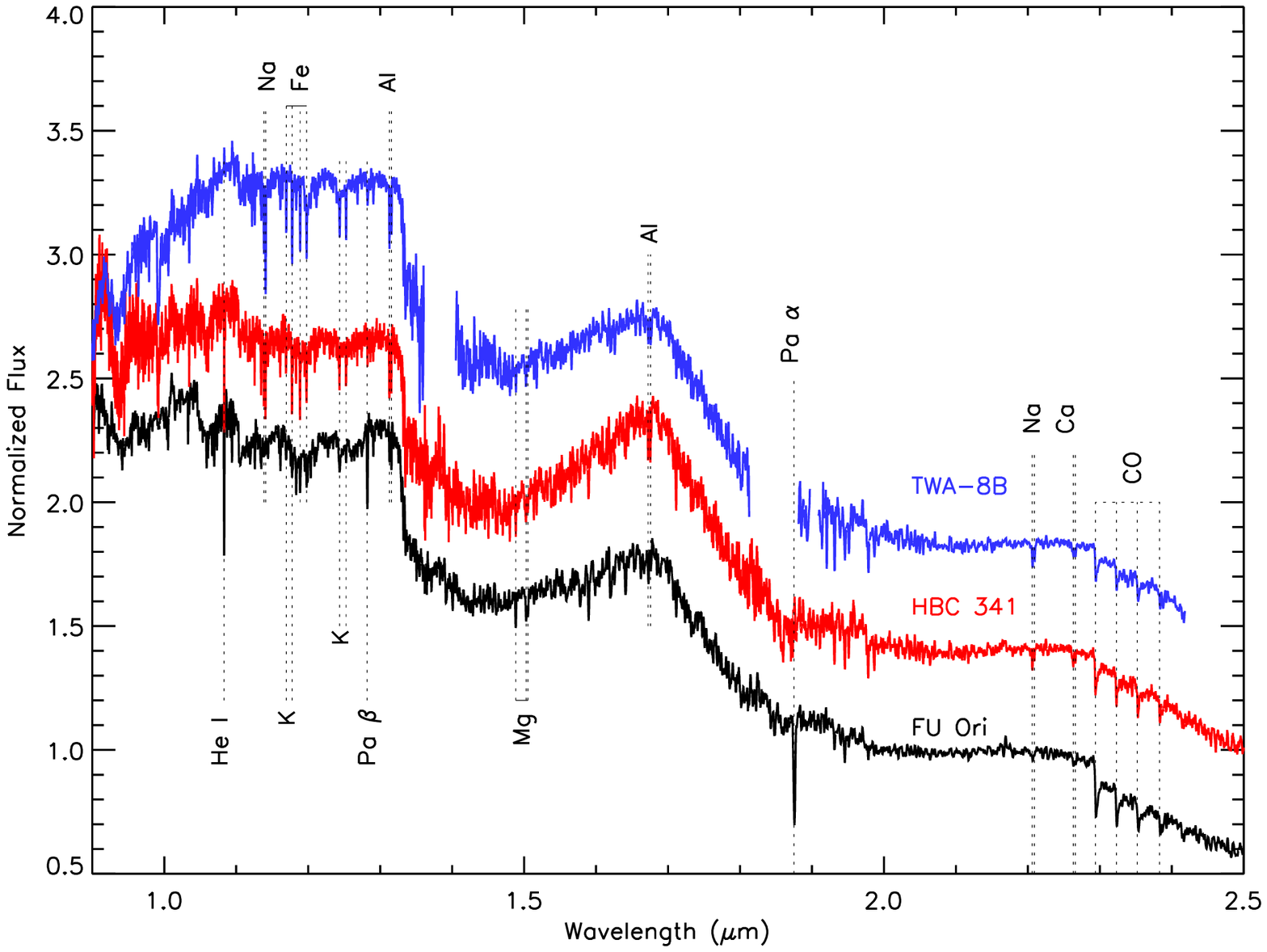}
\caption{ A comparison of the near-infrared spectrum of
FU~Ori with that of the newborn M5-type border-line brown dwarf HBC~341 (from
Dahm \& Hillenbrand  2017) and the 10~Myr old brown dwarf TWA-8B (from
Allers \& Liu 2013). Selected lines are identified.
  \label{FUOri+BD}}
\end{figure}
\clearpage

\begin{figure}
 \plotone{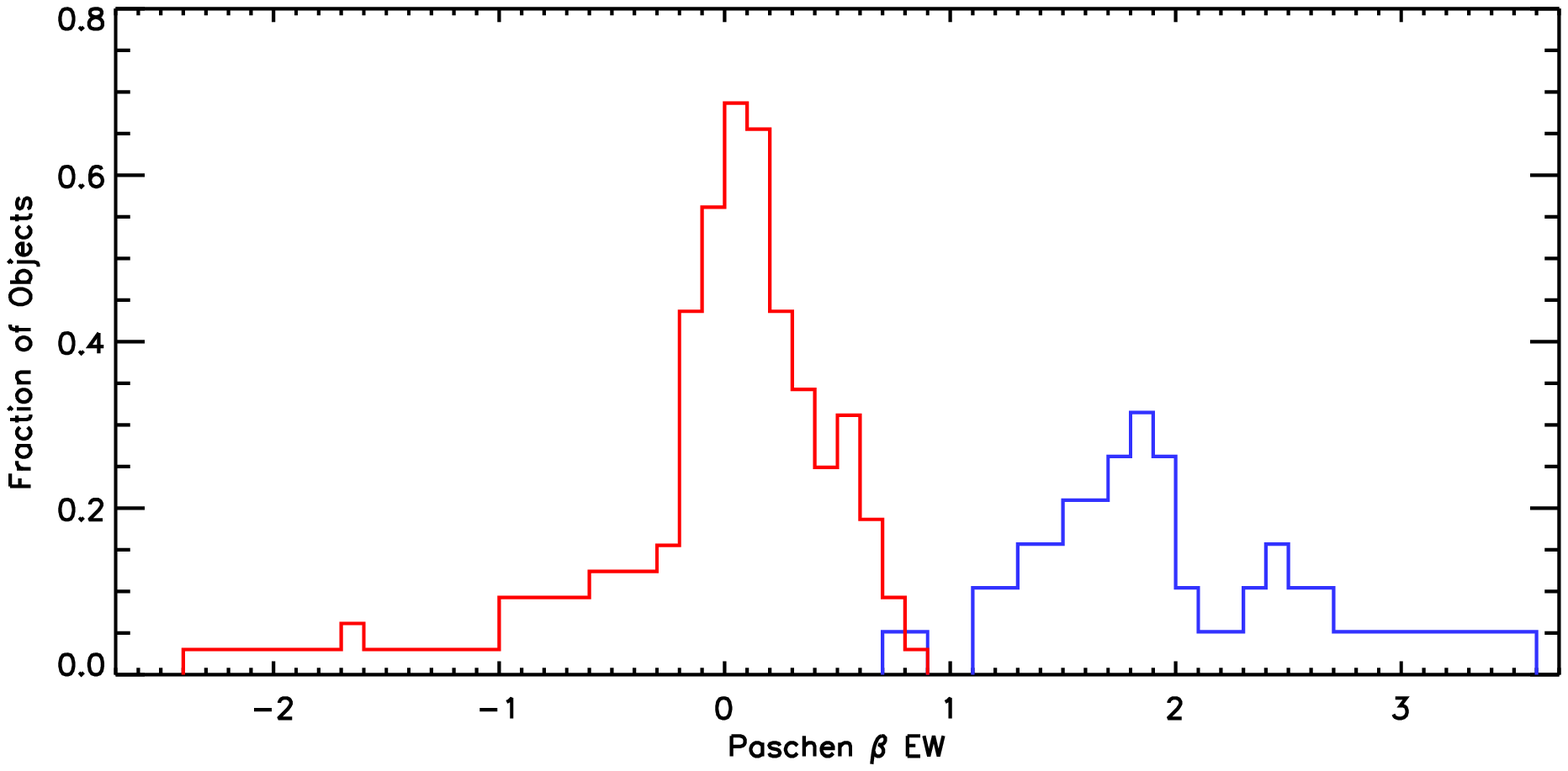}
\caption{ Distribution of equivalent widths of the Paschen-$\beta$
line in young brown dwarfs (red, from Allers \& Liu 2013) just overlaps with the distribution 
for FUors and FUor-like objects (blue, this paper), suggesting that Pa~$\beta$ is a useful spectroscopic
discriminator between the two types of objects.
  \label{FUOri+BD-EQW}}
\end{figure}
\clearpage

\begin{figure}
 \plotone{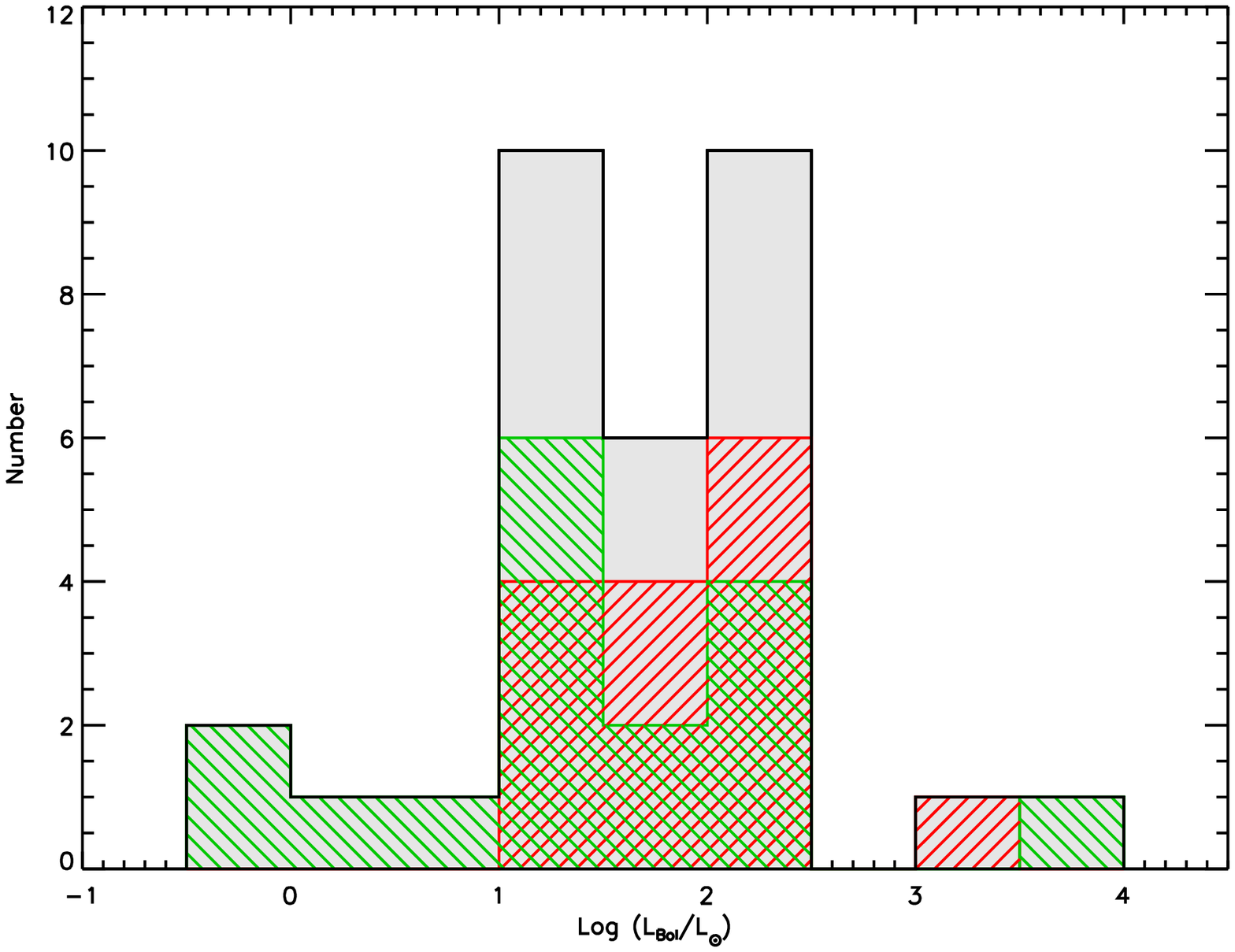}
\caption{The bolometric luminosity distributions for the FUors (red),
  FUor-like objects (green), and the combined sample (black).  Although
  the FUor-like objects have more low-luminosity objects, these two
  distributions are not significantly distinguishable.
  \label{Lbol_dist}}
\end{figure}
\clearpage

 \begin{figure}
 \plotone{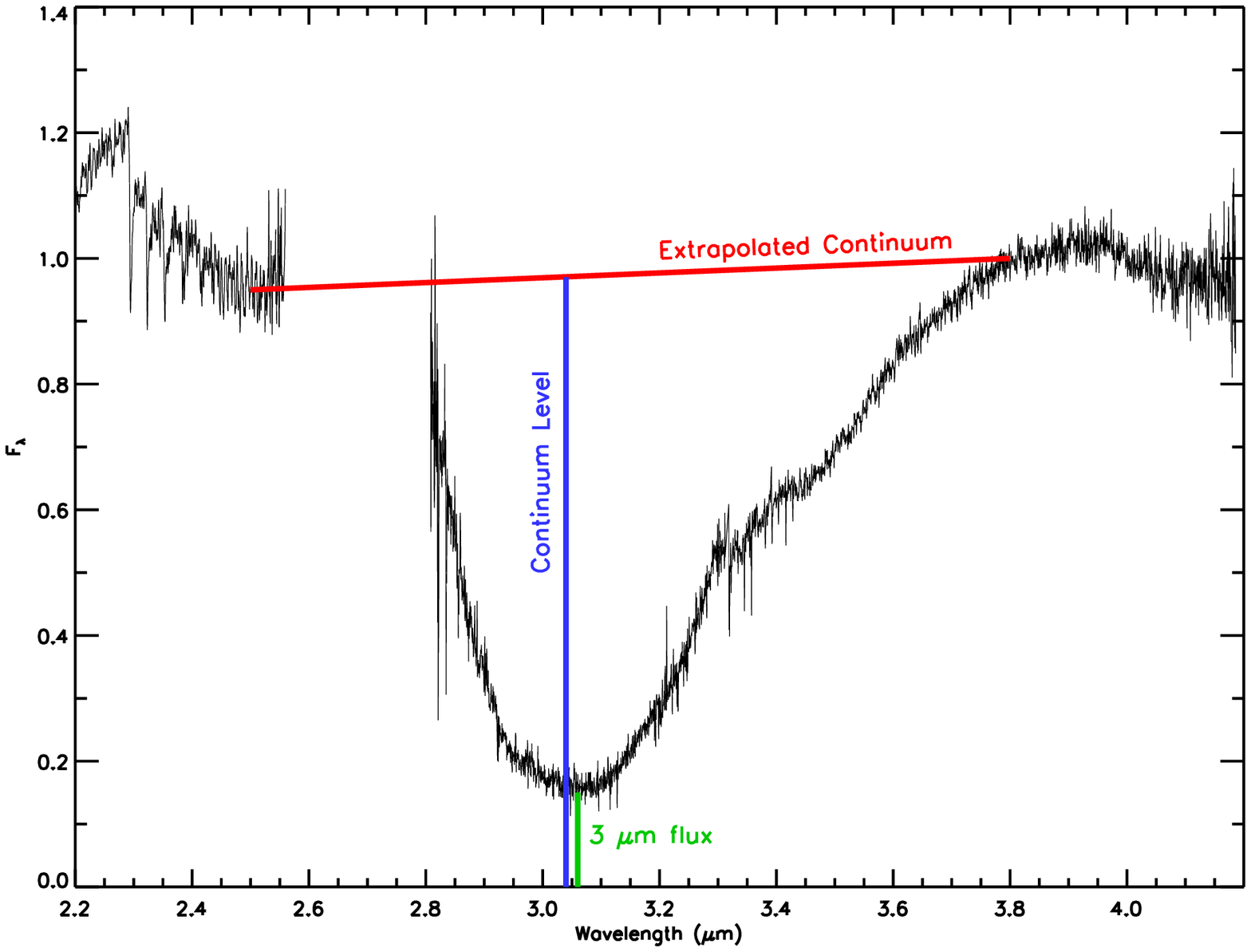}
 \caption{ To measure the depth of the ice band, we first extrapolated the 
  continuum from 2.5 to 3.8~$\mu$m.  The relative ice band optical depth 
  is calculated as -ln (continuum level/flux at 3~$\mu$m).  
  \label{Icedefinition}}
 \end{figure}
\clearpage

\begin{figure}
 \plotone{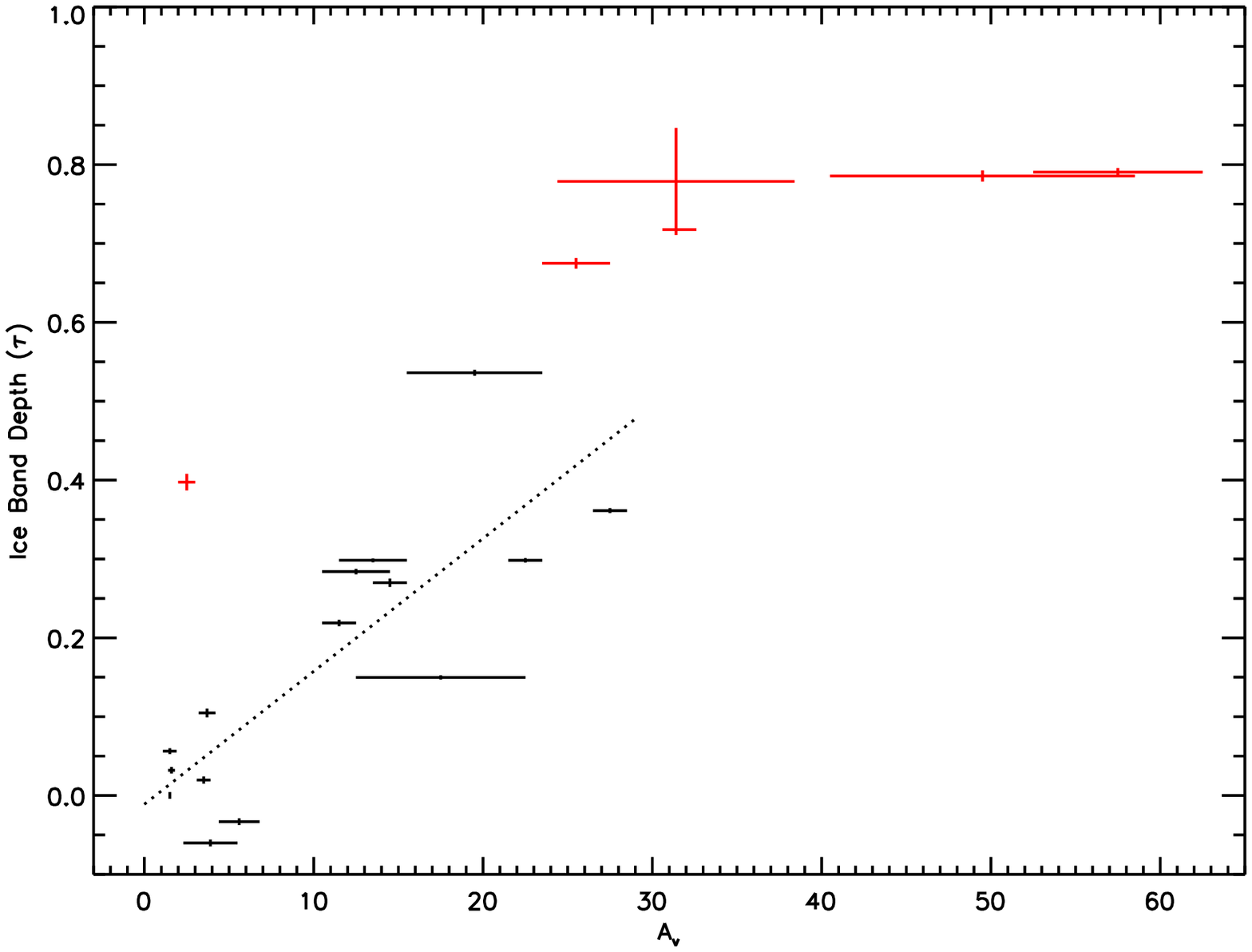}
  \caption{The 3.0~$\mu$m water ice absorption band optical depth ($\tau$) is well correlated with the
    extinction for FUors, but there are a few notable exceptions.  The black data points
    are objects not having a reflection nebula at K-band.  The red
    data points are those objects with an infrared reflection nebula,
    and thus mostly seen in scattered light at K-band, which were also
    bright enough at L-band (Parsamian 21, L1551 IRS 5, HH354 IRS, IRAS
    05450+0019, Haro 5a. and V2495 Cyg).  The dotted line is our
    linear fit to the data except for the two objects with the highest extinction, as the optical 
    depth appears to saturate when A$_v$ exceeds $\sim$30.  In most cases, the y-axis 
    error bars are too small to be seen.
\label{ice_depth}}
\end{figure}
\clearpage

\begin{deluxetable}{lccccccccccccc}
\tabletypesize{\scriptsize}
\tablecaption{Observing Log}
\tablewidth{0pt}
\tablecolumns{4}
\tablehead{
\colhead{} & \colhead{} & \multicolumn{2}{c}{SXD} & \multicolumn{2}{c}{LXD} \\
\colhead{Name} & \colhead{Photometry} & \colhead{Date} &  \colhead{Int. time\tablenotemark{a}} & \colhead{Date} &  \colhead{Int. time} 
}
\startdata   
RNO 1b              &  28 Aug 2015  &  28 Aug 2015  &    18.0         &  28 Aug 2015 &      4.0       \\
RNO 1c              &  28 Aug 2015  &  28 Aug 2015  &    12.0         &  28 Aug 2015 &      4.0       \\
PP 13 S             &  21 Nov 2015  &  16 Jan 2015  &    12.0         &  31 Oct 2015 &      8.0       \\
L1551 IRS 5         &  21 Nov 2015  &  04 Nov 2015  &    24.0         &  28 Aug 2015 &      4.0       \\
V582 Aur            &  30 Oct 2015  &  20 Jan 2015  &     8.0         &  30 Oct 2015 &      4.0       \\
Haro 5a/6a          &  30 Oct 2015  &  30 Oct 2015  &    24.0         &  05 Jan 2016 &     16.0       \\
V883 Ori            &  30 Oct 2015  &  30 Oct 2015  &     4.0         &  30 Oct 2015 &      1.2       \\
V2775 Ori           &  30 Oct 2015  &  30 Oct 2015  &    16.0         &  04 Nov 2015 &      4.0       \\
FU Ori              &  21 Nov 2015  &  08 Jan 2015  &     1.3         &  08 Jan 2015 &      2.7       \\
V1647 Ori           &  21 Nov 2015  &  04 Nov 2015  &    18.0         &  04 Nov 2015 &      4.0       \\
NGC2071 MM3         &  21 Nov 2015  &  04 Nov 2015  &    24.0         &  04 Nov 2015 &      4.0       \\
IRAS 06297+1021W    &  16 Nov 2006  &  16 Nov 2006  &     8.0         &              &                \\
AR 6a 	            &  21 Nov 2015  &  04 Nov 2015  &    24.0         &  04 Nov 2015 &      4.0       \\
AR 6b	            &  21 Nov 2015  &  21 Nov 2015  &    36.0         &              &                \\
IRAS 06393+0913     &  15 Oct 2007  &  15 Oct 2007  &    30.0         &              &                \\
V900 Mon            &  21 Nov 2015  &  20 Jan 2015  &    12.0         &  22 Jan 2015 &      5.3       \\
V960 Mon            &  21 Dec 2014  &  19 Dec 2014  &    16.0         &  19 Dec 2014 &      5.3       \\
Z CMa               &  21 Nov 2015  &  09 Jan 2015  &     1.6         &  09 Jan 2015 &      1.3       \\
BBW 76              &  21 Nov 2015  &  09 Jan 2015  &     8.0         &  09 Jan 2015 &      2.7       \\
V346 Nor            &  22 Jul 2015  &  22 Jul 2015  &    36.0         &  22 Jul 2015 &      8.0       \\
IRAS 18270-0153 W &               &  31 Aug 2015  &    24.0         &  31 Aug 2015 &      8.0       \\
V371 Ser            &  18 Jul 2016  &  18 Jul 2016  &   264.0\tablenotemark{b}   &              &                \\
IRAS 18341-0113 S    &  26 Jun 2015  &  26 Jun 2015  &    36.0         &              &                \\
Parsamian 21        &  27 May 2016  &  31 Aug 2015  &    16.0         &  31 Aug 2015 &      8.0      \\
V1515 Cyg           &  26 Jun 2015  &  26 Jun 2015  &    12.0         &  26 Jun 2015 &      5.3       \\
HBC 722           &  31 Oct 2015  &  31 Oct 2015  &    12.0         &  31 Oct 2015 &      4.0       \\
V2494 Cyg           &  30 Oct 2015  &  30 Oct 2015  &    24.0         &  30 Oct 2015 &      4.0       \\
V1057 Cyg           &  26 Jun 2015  &  26 Jun 2015  &     8.0         &  26 Jun 2015 &      2.7       \\
V2495 Cyg           &  10 Dec 2015  &  27 May 2016  &    64.0         &  18 Jul 2016 &     20.0       \\
CB230               &  28 Aug 2015  &  28 Aug 2015  &    24.0         &  28 Aug 2015 &      2.7       \\
V1735 Cyg           &  28 Aug 2015  &  28 Aug 2015  &    12.0         &  28 Aug 2015 &      2.7       \\
HH354 IRS           &  30 Oct 2015  &  08 Jul 2016  &    72.0\tablenotemark{b}   &  31 Oct 2015 &      8.0       \\
V733 Cep            &  26 Jun 2015  &  26 Jun 2015  &    18.0         &  26 Jun 2015 &      8.0       \\
\enddata

\tablenotetext{a}{The integration times are given in minutes}
\tablenotetext{b}{Observed with the $0\farcs8$ slit.  All others were observed with the $0\farcs5$ slit.}

\end{deluxetable}

\begin{deluxetable}{lccccccccccccc}
\tabletypesize{\scriptsize}
\tablecaption{Sample Characteristics}
\tablewidth{0pt}
\tablecolumns{9}
\tablehead{
\colhead{Name} &
\colhead{Alt. Name} &
\colhead{$\alpha$(J2000)\tablenotemark{a}} &
\colhead{$\delta$(J2000)\tablenotemark{a}} & 
\colhead{K\tablenotemark{b}} & 
\colhead{Group\tablenotemark{c}} &
\colhead{Dist (pc)\tablenotemark{d}} &
\colhead{L$_{bol}$\tablenotemark{e}} &
\colhead{A$_{v}$\tablenotemark{f}}
}  
\startdata   
RNO 1b             & V710 Cas                    & 00:36:46.3 & +63:28:54 &  8.34                  &   FUor     & 930$\pm$35 (1)  & 1652  &   14.5$\pm$1  \\
RNO 1c             & V710 Cas                    & 00:36:46.4 & +63:28:55 &  7.73                  &  FUor-like & 930$\pm$35 (1)  & 1652  &   19.5$\pm$4  \\
PP 13 S            & IRAS 04073+3800             & 04:10:41.1 & +38:07:53 & 10.82                  &  FUor-like & 450$\pm$23 (2)  &   51  &   56.5$\pm$5  \\
L1551 IRS 5        & IRAS 04287+1801             & 04:31:34.2 & +18:08:05 &  9.21                  &  FUor-like & 147$\pm$5 (3,4) &   29  &   25.5$\pm$2 \\
V582 Aur           &                             & 05:25:52.0 & +34:52:30 &  8.18                  &  FUor      & 1300 (5)        &  168  &  5.6$\pm$1.2  \\
Haro 5a/6a         & IRAS 05329-0505             & 05:35:26.6 & -05:03:56 &  9.85                  &  FUor-like & 388$\pm$5 (6)   &   18  &   57.5$\pm$5  \\
V883 Ori           & HBC 489                     & 05:38:18.1 & -07:02:26 &  5.53                  &  FUor      & 388$\pm$5 (6)   &  212  &   22.5$\pm$1  \\
V2775 Ori          & HOPS 223                    & 05:42:48.5 & -08:16:35 &  8.43                  &  FUor      & 428$\pm$10 (6)  &   29  &   27.5$\pm$1  \\
FU Ori             & IRAS 05426+0903             & 05:45:22.4 & +09:04:12 &  5.79                  &  FUor      & 400$\pm$40 (7)  &   66  &   1.5$\pm$0.2  \\
V1647 Ori          & McNeil's Nebula             & 05:46:13.1 & -00:06:05 &  7.90                  &  Peculiar  & 388$\pm$5 (6)   &   21  &   22.5$\pm$7  \\
IRAS 05450+0019    & NGC2071 MM3                 & 05:47:36.6 & +00:20:06 &  8.76                  &  FUor-like & 388$\pm$5 (6)   &   35  &   31.5$\pm$1  \\
IRAS 06297+1021W   &                             & 06:32:26.1 & +10:19:18 &  8.14                  &  Peculiar  & 738$\pm$57 (8)  &   31  &  10.3$\pm$3.8 \\
AR 6a 	           & V912 Mon                    & 06:40:59.3 & +09:35:52 &  7.88                  &  Peculiar  & 738$\pm$57 (8)  &  310  &   20.5$\pm$2  \\
AR 6b	           &                             & 06:40:59.3 & +09:35:52 & 10.89                  &  Peculiar  & 738$\pm$57 (8)  &  310  &   28.5$\pm$3  \\
IRAS 06393+0913    &                             & 06:42:08.1 & +09:10:30 & 10.45                  &  Peculiar  & 738$\pm$57 (8)  &  0.9  &   23.5$\pm$7  \\
V900 Mon           &                             & 06:57:22.2 & -08:23:18 &  7.51                  &  FUor      & 1100 (9)        &   99  &   13.5$\pm$2  \\
V960 Mon           &                             & 06:59:31.6 & -04:05:28 &  7.42                  &  FUor      & 1100 (9)        &   48  &  1.5$\pm$0.4  \\
Z CMa              & HD 53179                    & 07:03:43.2 & -11:33:06 &  3.77\tablenotemark{b} &  FUor-like & 990$\pm$50 (10) & 3548  &  7.1$\pm$8.0  \\
BBW 76             & V646 Pup                    & 07:50:35.6 & -33:06:24 &  8.60                  &  FUor-like & 1800 (11)       &  114  &  1.6$\pm$0.2  \\
V346 Nor           & HH57 IRS                    & 16:32:32.1 & -44:55:31 & 10.51                  &  Peculiar  & 700 (12)        &  176  &   46.5$\pm$9  \\
IRAS 18270-0153 W  &                             & 18:29:38.9 & -01:51:06 & 12.87\tablenotemark{b} &  Peculiar  & 436$\pm$9 (13)  &   30  &   41.5$\pm$11  \\
V371 Ser           & EC 53                       & 18:29:51.2 & +01:16:39 & 12.19                  &  Peculiar  & 429$\pm$2 (14)  &  1.6  &   47.5$\pm$9  \\
IRAS 18341-0113 S  &                             & 18:36:46.5 & -01:10:42 & 11.39                  &  Peculiar  & 259$\pm$37 (15) &  0.8  &  31.5$\pm$6  \\
Parsamian 21       & IRAS 19266+0932, HBC 687    & 19:29:00.8 & +09:38:43 &  9.55                  &  FUor-like & 500 (16)        &   16  &  2.5$\pm$0.5  \\
V1515 Cyg          & IRAS 20220+4202             & 20:23:48.0 & +42:12:26 &  7.95                  &  FUor      & 1050 (17)       &  103  &  3.5$\pm$0.4  \\
HBC 722            & V2493 Cyg                   & 20:58:17.0 & +43:53:43 &  6.31                  &  FUor      & 550$\pm$50 (18) &   17  &  3.7$\pm$0.5  \\
V2494 Cyg          & IRAS 20568+5217, HH381 IRS  & 20:58:21.4 & +52:29:27 &  8.37                  &  FUor      & 600 (19)        &  187  &   17.5$\pm$5 \\
V1057 Cyg          & LkH$\alpha$ 190             & 20:58:53.7 & +44:15:29 &  6.59                  &  FUor      & 550$\pm$50 (18) &  100  &  3.9$\pm$1.6  \\
V2495 Cyg          & Braid star                  & 21:00:25.4 & +52:30:16 & 11.79                  &  FUor      & 600 (19)        &   21  &   49.5$\pm$9  \\
CB230              & IRAS 21169+6804             & 21:17:39.4 & +68:17:32 & 10.27                  &  FUor-like & 300$\pm$30 (20) &  6.6  &   26.5$\pm$7  \\
V1735 Cyg          & IRAS 21454+4718, HBC 733    & 21:47:20.7 & +47:32:04 &  7.54                  &  FUor      & 950$\pm$80 (21) &  166  &   12.5$\pm$2  \\
HH354 IRS          & IRAS 22051+5848, L1165 IRS1 & 22:06:50.2 & +59:02:45 & 10.83                  &  FUor-like & 300$\pm$100 (22) &   16  &   31.5$\pm$7  \\
V733 Cep           & Persson's Star              & 22:53:33.3 & +62:32:24 &  8.29                  &  FUor      & 800 (23)        &   43  &   11.5$\pm$1  \\
\enddata

\tablenotetext{a}{ 2MASS coordinate for the target}
\tablenotetext{b}{ K-band magnitudes from our observations, and are in the MKO photometric system unless otherwise noted.  Z CMa was observed through the narrow K-continuum filter to avoid saturation. 2MASS photometry is given for IRAS 18270-0153 \#7.  }
\tablenotetext{c}{ FUor: Eruption observed; FUor-like: Eruption not observed, but spectrum similar to FUors; Peculiar: Has some spectroscopic or photometric similaries to FUors}
\tablenotetext{d}{
(1) Reid et al. 2014
(2) Lada et al. 2009
(3) Loinard et al. 2007
(4) This paper
(5) Kun et al. 2017
(6) Kounkel et al. 2017
(7) Murdin \& Penston 1977
(8) Kamezaki et al. 2014
(9) Kim et al. 2004
(10) Kaltcheva \& Hilditch 2000 
(11) Reipurth et al. 2002
(12) Graham \& Frogel 1985
(13) Ortiz-Le\'on et al. 2016
(14) Dzib et al. 2011
(15) Straizys et al. 1996 
(16) Dame \& Thaddeus 1985
(17) Racine 1968
(18) Laugalys et al. 2006
(19) Herbig \& Dahm 2006
(20) Kun et al. 1998
(21) Harvey et al. 2008
(22) Dobashi et al. 1994
(23) Reipurth et al. 2007
}
\tablenotetext{e}{The bolometric luminosity was calculated using photometry from 2MASS (1.2 to 2.2~$\mu$m), WISE (3.4 to 22~$\mu$m), and Akari (65 to 160~$\mu$m).  An 80 K black body was added to reddest data point. For RNO 1b/c and AR 6a/b the binary was not resolved.}
\tablenotetext{f}{Visual extinction in magnitudes, derived from fitting the spectrum to FU Ori and adding FU Ori's 1.5 magnitudes of extinction.}
\end{deluxetable}

\begin{deluxetable}{lccccccccccccc}
\tabletypesize{\scriptsize}
\tablecaption{Basis of FUor Classification}
\tablewidth{0pt}
\tablecolumns{9}
\tablehead{
\colhead{Name} &\multicolumn{1}{c}{Eruption} &\multicolumn{1}{c}{CO} &\multicolumn{1}{c}{Water} &\multicolumn{1}{c}{VO or\tablenotemark{a}} &\colhead{Pa $\beta$} &\multicolumn{1}{c}{Emission} &\multicolumn{1}{c}{weak} &\colhead{He I} \\
\colhead{}          &\colhead{Observed?}              &\colhead{abs.}               &\colhead{abs.}                          &\colhead{TiO}                                                & \colhead{abs.}                     & \colhead{Lines}                 &\colhead{metals}        & \colhead{abs.}
}  
\startdata   
\textbf{Bona fide FUors} & &        &         &         &           &         &         &           \\
RNO 1b            &   Y   &    Y    &    Y    &    Y    &     Y     &    N    &    Y    &    Y      \\
V582 Aur          &   Y   &    Y    &    Y    &    Y    &     Y     &    N    &    Y    &    Y      \\
V883 Ori          &   Y   &    Y    &    Y    &    Y    &     Y     &    N    &    Y    &    N      \\
V2775 Ori         &   Y   &    Y    &    Y    &    Y    &     Y     &    N    &    Y    &    Y      \\
FU Ori            &   Y   &    Y    &    Y    &    Y    &     Y     &    N    &    Y    &    Y      \\
V900 Mon          &   Y   &    Y    &    Y    &    Y    &     Y     &    N    &    Y    &    Y      \\
V960 Mon          &   Y   &    Y    &    Y    &    Y    &     Y     &    N    &    Y    &    Y      \\
V1515 Cyg         &   Y   &    Y    &    Y    &    Y    &     Y     &    N    &    Y    &    Y      \\
HBC 722           &   Y   &    Y    &    Y    &    Y    &     Y     &    N    &    Y    &    Y      \\
V2494 Cyg         &   Y   &    Y    &    Y    &    Y    &     Y     &    Y    &    Y    &    Y      \\
V1057 Cyg         &   Y   &    Y    &    Y    &    Y    &     Y     &    N    &    Y    &    Y      \\
V2495 Cyg         &   Y   &    Y    &    Y    & no data &  no data  &    N    &    Y    &  no data  \\
V1735 Cyg         &   Y   &    Y    &    Y    &    Y    &     Y     &    N    &    Y    &    Y      \\
V733 Cep          &   Y   &    Y    &    Y    &    Y    &     Y     &    N    &    Y    &    N      \\

                   &         &         &         &           &         &         &           \\

\textbf{FUor-like}&        &         &         &         &           &         &         & \\
RNO 1c           &    N    &    Y    &    N    &    N    &     Y     &    N    &    Y    &    N      \\
PP 13 S          &    N    &    Y    &    Y    & no data &  no data  &    Y    &    Y    & no data   \\
L1551 IRS 5      &    N    &    Y    &    Y    &    Y    &     N     &    Y    &    Y    &    N      \\
Haro 5a/6a       &    N    &    Y    &    Y    & no data &  no data  &    N    &    Y    & no data   \\
IRAS 05450+0019  &    N    &    Y    &    Y    & no data &     N     &    N    &    Y    & no data   \\
Z CMa            &    N    &    N    &    N    &    N    &     N     &    Y    &    Y    &    Y      \\
BBW 76           &    N    &    Y    &    Y    &    Y    &     Y     &    N    &    Y    &    Y      \\
Parsamian 21     &    N    &    Y    &    Y    &    Y    &     Y     &    N    &    Y    &    Y      \\
CB230 IRS1       &    N    &    Y    &    Y    & no data &  no data  &    Y    &    Y    & no data   \\
HH354 IRS        &    N    &    Y    &    Y    & no data &  no data  &    N    &    Y    & no data   \\

                   &         &         &         &           &         &         &           \\

\textbf{Peculiar}  &        &         &         &         &           &         &         & \\
V1647 Ori          &    Y   &    N    &    Y    &    Y    &     N     &    Y    &    Y    &    Y      \\
IRAS 06297+1021W   &    N   &    N    &    Y    &    N    &     N     &    Y    &    Y    &    Y      \\
AR 6a 	           &    N   &    N    &    N    &    N    &     Y     &    N    &    Y    &    Y      \\
AR 6b	           &    N   &    Y    &    N    & no data &     N     &    N    &    Y    & no data   \\
IRAS 06393+0913    &    N   &    Y    &    N    &    N    &     N     &    N    &    Y    & no data   \\
V346 Nor           &    Y   &    N    &    N    & no data &  no data  &    Y    &    Y    & no data   \\
V371 Ser           &    N   &    Y    &    Y    & no data &  no data  &    Y    &    Y    & no data   \\
IRAS 18270-0153W   &    N   &    N    &    Y    & no data &  no data  &    Y    &    Y    & no data   \\
IRAS 18341-0113S   &    N   &    N    &    Y    & no data &  no data  &    Y    &    Y    & no data   \\
\enddata

\tablenotetext{a}{ See Figure 1 or location of VO and TiO absorption bands}

\end{deluxetable}


\clearpage

\begin{thebibliography}{}

\bibitem[Abraham et al.(2004)]{Abr2004} \'Abrah\'am, P., K\'osp\'al, \'A., Csizmadia, S. et al., 2004, \aap, 428, 89

\bibitem[Allers \& Liu(2013)]{All2013} Allers, K., \& Liu, M., 2013, ApJ, 772, 79

\bibitem[Antoniucci et al.(2016)]{Ant2016} Antoniucci, S., Podio, L., Nisini, B. et al., 2016, A\&A, 593, L13

\bibitem[Ambartsumian(1971)]{Amb1971} Ambartsumian, V.A., 1971, Astrophysics, 7, 331

\bibitem[Armitage et al.(2001)]{Arm2001} Armitage, P., Livio, M., Pringle, J., 2001, MNRAS, 324, 705

\bibitem[Armond et al.(2011)]{Arm2011} Armond, T., Reipurth, B, Bally, J., \& Aspin, C., 2011, A\&A, 528, 125

\bibitem[Aspin \& Repurth(2000)]{Asp2000} Aspin, C. \& Reipurth, B., 2000, MNRAS, 311, 522

\bibitem[Aspin \& Reipurth(2003)]{Asp2003}Aspin, C. \& Reipurth, B., 2003, AJ, 126, 2936

\bibitem[Aspin \& Sandell(2001)]{Asp2001} Aspin, C. \& Sandell, G., 2001, MNRAS, 328, 751

\bibitem[Aspin et al.(2009a)]{Asp2009a} Aspin, C., Reipurth, B., Beck, T., et al., 2009a, ApJ, 692, L67

\bibitem[Aspin et al.(2009b)]{Asp2009b} Aspin, C., Beck, T.L., Pyo, T.-S. et al., 2009b, AJ, 137, 431 
 

\bibitem[Aspin et al.(2006)]{Asp2006} Aspin, C., Barbieri, C., Boschi, F., et al., 2006, AJ, 132, 1298

\bibitem[Audard et al.(2014)]{Aur2014} Audard, M., Abraham, P., Dunham, M.M. et al., 2014, in {\em Protostars and Planets VI}, eds. H. Beuther et al., Univ. Arizona Press, p.387

\bibitem[Bally(2008)]{Bal2008} Bally, J. 2008, in {\em Handbook of Star Forming Regions Vol. I}, ed. Bo Reipurth, ASP, p. 459

\bibitem[Baraffe et al.(2012)]{Bar2012} Baraffe, I., Vorobyov, E., Chabrier, G. 2012, ApJ, 756, A118

\bibitem[Barman et al.(2011)]{Bar2011} Barman, T., Macintosh, B., Konopacky, Q., \& Marois, C., 2011, ApJ, 733, 65

\bibitem[Baxter et al.(2009)]{Bax2009} Baxter, E.J., Covey, K.R., Muench, A.A. et al., 2009, AJ, 138, 963

\bibitem[Beck \& Aspin(2012)]{Bec2012} Beck, T.L. \& Aspin, C. 2012, AJ, 143, A55

\bibitem[Beck (2007)]{Beck2007} Beck, T., 2007, AJ, 133, 1673

\bibitem[Beck et al.(2001)]{Beck2001} Beck, T., Prato, L., \& Simon, M., 2001, ApJ, 551, 1031

\bibitem[xyz(000)]{xyz001} Bell, K.R. \& Lin, D. 1994, ApJ, 427, 987

\bibitem[xyz(002)]{} Bieging, J.H. \& Cohen, M. 1985, ApJ, 289, L5
 

\bibitem[xyz(003)]{xyz003} Bonnefoy, M., Chauvin, G., Dougados, C. et al., 2017, A\&A, 597, A91

\bibitem[xyz(004)]{xyz004} Bonnell, I. \& Bastien, P. 1992, ApJ, 401, L31

\bibitem[Borysow et al. (1997)]{Bor1997} Borysow, A., J\o rgensen, U., \& Zheng, C., 1997, \aap, 324, 185

\bibitem[xyz(005)]{xyz005} Brice\~no, C., Vivas, A.K., Hern\'andez, J. et al., 2004, ApJ, 606, L123

\bibitem[Calvet et al.(1991)]{Cal1991} Calvet, N., Pati\~{n}o, A., Magris, G., \& D'Alessio, P., 1991, \apj, 380, 617

\bibitem[Caratti o Garatti et al.(2011)]{Car2011} Caratti o Garatti, A., Garcia-Lopez, R., Scholz, A. et al., 2011, \aap, 526, L1

\bibitem[xyz(006)]{xyz006} Caratti o Garatti, A., Garcia Lopez, R., Ray, T.P. et al., 2015, ApJ, 806, L4

\bibitem[xyz(007)]{xyz007} Carr, J.S. 1989, ApJ, 345, 522

\bibitem[xyz(008)]{xyz008} Chiang, H.-F., Reipurth, B., Walawender, J. et al., 2015, ApJ, 805, L54

\bibitem[xyz(009)]{xyz009} Chiar, J., Pendleton, Y., Allamandola, L., et al., 2011, ApJ, 731, 9

\bibitem[xyz(010)]{xyz010} Cieza, L.A., Casassus, S., Tobin, J. et al., 2016, Nature, 535, 258

\bibitem[Clarke et al.(2005)]{Cla2005} Clark, C., Lodato, G., Melnikov, S., \& Ibrahimov, M., 2005, MNRAS, 361, 942

\bibitem[Cohen \& Kuhi(1979)]{Coh1979} Cohen, M., \& Kuhi, L., 1979, ApJS, 41, 743

\bibitem[xyz(011)]{xyz011} Cohen M., Aitken D. K., Roche P. F., Williams P. M., 1983, ApJ, 273, 624

\bibitem[Connelley et al.(2007)]{Con2007} Connelley, M., Reipurth, B., \& Tokunaga, A., 2007, \aj, 133, 1528 

\bibitem[Connelley et al.(2008a)]{Con2008a} Connelley, M., Reipurth, B., \& Tokunaga, A., 2008a, \aj, 135, 2496 

\bibitem[Connelley et al.(2008b)]{Con2008b} Connelley, M., Reipurth, B., \& Tokunaga, A., 2008b, \aj, 135, 2526 
 

\bibitem[Connelley \& Greene(2014)]{Con2014} Connelley, M., \& Greene, T., 2014, \aj, 147, 125 

\bibitem[Connelley \& Greene(2010)]{Con2010} Connelley, M.S., \& Greene, T., 2010, \aj, 140, 1214 

\bibitem[xyz(012)]{xyz012} Connelley, M.S., Reipurth, B., Tokunaga, A. 2007, AJ, 133, 1528

\bibitem[xyz(013)]{xyz013} Contreras, M.E., Sicilia-Aguilar, A., Muzerolle, J. et al., 2002, AJ, 124, 1585

\bibitem[xyz(014)]{xyz014} Contreras Pe\~na, C., Lucas, P.W., Minniti, D. et al., 2017, MNRAS, 465, 3039

\bibitem[Covino et al.(1984)]{Cov1984} Covino, E., Terranegra, A., Vittone, A., \& Russo, G., 1984, AJ, 89, 1868

\bibitem[Cushing et al.(2004)]{Cus2004} Cushing, M., Vacca, W., \& Rayner, J., 2004, \pasp, 116, 362  

 \bibitem[xyz(015)]{xyz015} Dahm, S.E. 2008, in {\em Handbook of Star Forming Regions Vol. I}, Ed. Bo Reipurth, ASP, p. 966
 

\bibitem[xyz(016)]{xyz016} Dahm, S. \& Hillenbrand, L.A. 2017, AJ, 154, A177

\bibitem[xyz(017)]{xyz017} Dame, T.M. \& Thaddeus, P. 1985, ApJ, 297, 751

\bibitem[xyz(018)]{xyz018} Das, A., Das, H.S., Senorita Devi, A. 2015, MNRAS, 452, 389

\bibitem[xyz(019)]{xyz019} Devine, D., Reipurth, B., Bally, J. 1997, in {\em Low Mass Star Formation - from Infall to Outflow}, Poster proceedings of IAU Symposium No. 182, eds. F. Malbet \& A. Castets, p. 91.

\bibitem[xyz(020)]{xyz020} Dobashi, K., Bernard, J.-P., Yonekura, Y., Fukui, Y. 1994, ApJS, 95, 419

\bibitem[Doppmann et al.(2005)]{Dop2005} Doppmann, G., Greene, T., Covey, K., \& Lada, C., 2005, AJ, 130, 1145

\bibitem[Dunham et al.(2012)]{Dun2012} Dunham, M., Arce, H., Bourke, T., et al., 2012, \apj, 755, 157

\bibitem[xyz(021)]{xyz021} Dzib, S., Loinard, L., Mioduszewski, A.J. et al., 2010,  ApJ, 718, 610

\bibitem[xyz(022)]{xyz022} Dzib, S., Loinard, L., Mioduszewski, A.J. et al., 2011, Rev. Mex. AA Serie Conf. 40, 231

\bibitem[xyz(023)]{xyz023} Eiroa, C. \& Casali, M.M. 1992, A\&A, 262, 468
 

\bibitem[Elias(1978)]{Eli1978} Elias, J., 1978, \apj, 223, 859

\bibitem[xyz(024)]{xyz024} Evans, N.J., Balkum, S., Levreault, R.M., Hartmann, L., Kenyon, S.J., 1994, ApJ, 424, 793
 
\bibitem[Fischer et al.(2012)]{Fis2012} Fischer, W., Megeath, S. T., Tobin, J. J., et al., 2012, \apj, 756, 99

 \bibitem[xyz(025]{xyz025} Fridlund, C.V.M. \& Liseau, R. 1998, ApJ, 499, L75

\bibitem[xyz(026)]{xyz026} Fridlund, C.V.M., Nordh, H.L., van Duinen, R.J. et al., 1980, A\&A, 91, L1

\bibitem[xyz(027)]{xyz027} Friesen, R.K., Bourke, T.L., Di Francesco, J. et al., 2016, ApJ, 833, A204

\bibitem[xyz(028)]{xyz028} Garrison, R.F. \& Kormendy, J. 1976, PASP, 88, 865

\bibitem[xyz(029)]{xyz029} Goodrich, R.W. 1987, PASP, 99, 116

\bibitem[Graham\& Frogel(1985)]{Gra1985} Graham, J., \& Frogel, J., 1985, \apj, 289, 331

\bibitem[Gramajo et al.(2014)]{Gra2014} Gramajo, L., Rodon, J., \& Gomez, M., 2014, AJ, 147, 140

 \bibitem[xyz(030)]{xyz030} Green, J.D., Kraus, A.L., Rizzuto, A.C. et al., 2016, ApJ, 830, A29

\bibitem[Greene \& Lada(1996)]{Gre1996} Greene, T., \& Lada, C., 1996, AJ, 112, 2184

\bibitem[xyz(031)]{xyz031} Greene, T.P., Aspin, C., Reipurth, B. 2008, AJ, 135, 1421

\bibitem[Haas et al.(1993)]{Haa1993} Haas, M., Christou, J., Zinnecker, H., Ridgway, S., \& Leinert, Ch., 1993, \aap, 269, 282

\bibitem[Hackstein et al.(2015)]{Hac2015} Hackstein, M., Haas, M., K\'osp\'al, A. et al., 2015, A\&A, 582, L12 

\bibitem[Haro(1953)]{Har1953} Haro, G., 1953, ApJ, 117, 73

\bibitem[xyz(032)]{xyz032} Hartmann, L. \& Kenyon, S.J. 1985, ApJ, 299, 462

\bibitem[xyz(033)]{xyz033} Hartmann, L. \& Kenyon, S.J. 1987, ApJ, 322, 393

 \bibitem[Hartmann \& Kenyon(1996)]{Har1996} Hartmann, L., \& Kenyon, S., 1996, ARA\&A, 34, 207

\bibitem[Hartmann et al.(1989)]{Har1989} Hartmann, L., Kenyon, S.J., Hewett, R. et al., 1989, ApJ, 338, 1001

\bibitem[xyz(034)]{xyz034} Harvey, P.M., Huard, T.L., J{\o}rgensen, J.K. et al., 2008, ApJ, 680, 495

\bibitem[Herbig(1960)]{Her1960} Herbig, G., 1960, ApJS, 4, 337

\bibitem[Herbig(1966)]{Her1966} Herbig, G., 1966, Vistas in Astronomy, 8, 109

\bibitem[Herbig(1977)]{Her1977} Herbig, G., 1977, \apj, 217, 693

\bibitem[xyz(035)]{xyz035} Herbig, G.H. 2007, AJ, 133, 2679

\bibitem[xyz(036)]{xyz036} Herbig, G.H. 2008, AJ, 135, 637

\bibitem[xyz(037)]{xyz037} Herbig, G.H. 2009, AJ, 138, 448

\bibitem[xyz(038)]{xyz038} Herbig, G.H., Petrov, P.P., Duemmler, R. 2003, ApJ, 595, 384

\bibitem[xyz(039)]{xyz039} Herbig, G.H. \& Dahm, S.E. 2006, AJ, 131, 1530

\bibitem[Herbig \& Harlan(1971)]{Her1971} Herbig, G.H., \& Harlan, E., 1971,  Information Bulletin of Variable Stars, No. 543, Konkoly Obs., Budapest

\bibitem[xyz(040)]{xyz040} Herczeg, G.J., Dong, S., Shappee, B.J. et al., 2016, ApJ, 831, 133

\bibitem[xyz(041)]{xyz041} Hessman, F.V., Eisl\"offel, J., Mundt, R. et al., 1991, ApJ, 370, 384

\bibitem[xyz(042)]{xyz042} Hillenbrand, L., \& Findeisen, K., 2015, ApJ, 808, 68

\bibitem[Hodapp (1999)]{Hod1999} Hodapp, 1999, AJ, 118, 1338

\bibitem[Hodapp et al.(2012)]{Hod2012} Hodapp, K., Chini, R., Waterman, R., Lemke, R., 2012, ApJ, 744, 56

\bibitem[xyz(043)]{xyz043} Hosokawa, T., Offner, S.R., Krumholz, M.R. 2011, ApJ, 738, A140

\bibitem[xyz(044)]{xyz044} Jeffries, R.D. 2007, MNRAS, 376, 1109 

\bibitem[xyz(045)]{xyz045} Jurdana-Sepic, R. \& Munari, U. 2016, New Astron., 43, 873

\bibitem[xyz(046)]{xyz046} Kaltcheva, N., Hilditch, R., 2000, MNRAS, 312, 753

\bibitem[xyz(047)]{xyz047} Kamezaki, T., Imura, K., Omodaka, T. et al., 2014, ApJS, 211, A18

\bibitem[xyz(048)]{xyz048} Kenyon, S.J., Hartmann, L., Hewett, R. 1988, ApJ, 325, 231

\bibitem[Kenyon et al.(1991)]{Ken1991} Kenyon, S., Hartmann, L., \& Kolotilov, E., 1991, PASP, 103, 1069

 
\bibitem[xyz(049)]{xyz049} Kenyon, S.J., Hartmann, L., Gomez, M., Carr, J.S. 1993, AJ, 105, 1505

\bibitem[Khanzadyan et al.(2012)]{Kha2012} Khanzadyan, T., Davis, C.J., Aspin, C. et al., 2012, A\&A, 542, A111

\bibitem[xyz(050)]{xyz050} Kim, B.G., Kawamura, A., Yonekura, Y., Fukui, Y. 2004, PASJ, 56, 313

\bibitem[xyz(051)]{xyz051} Kopatskaya, E.N., Kolotilov, E.A., Arkharov, A.A. 2013, MNRAS, 434, 38

\bibitem[K\'osp\'al et al.(2007)]{Kos2007} K\'{o}sp\'{a}l, A., \'Abrah\'am, P., Prusti, T. et al., 2007, A\&A, 470, 211

\bibitem[Kospal et al.(2008)]{Kos2008} K\'{o}sp\'{a}l, A., \'Abrah\'am, P., Apai, D., et al., 2008, MNRAS, 383, 1015

\bibitem[xyz(052)]{xyz052} K\'osp\'al, A., \'Abrah\'am, P., Mo\'or, A. et al., 2015, ApJ, 801, L5

\bibitem[xyz(053)]{xyz053} K\'osp\'al, A., \'Abrah\'am, P., Acosta-Pulido, J.A. et al. 2016, A\&A, 596, A52

\bibitem[xyz(054)]{xyz054} K\'osp\'al, A., \'Abrah\'am, P., Westhues, Ch., Haas, M., 2017, A\&A, 597, L10

\bibitem[xyz(055)]{xyz055} Kounkel, M., Hartmann, L., Loinard, L. et al., 2017, ApJ, 834, A142

\bibitem[xyz(056)]{xyz056} Kraus, S., Caratti o Garatti, A., Garcia-Lopez, R., Kreplin, A., Aarnio, A., Monnier, J., Naylor, T., Weigelt, G., 2016, MNRAS, 462L, 61K

\bibitem[xyz(057)]{xyz057} Kun, M. 1998, ApJS, 115, 59
\bibitem[xyz(058)]{xyz058} Kun, M. 2008, in {\em Handbook of Star Forming Regions Vol. I}, ed. Bo Reipurth, ASP, p.240

\bibitem[xyz(059)]{xyz059} Kun, M., Kiss, Z.T., Balog, Z. 2008, in {\em Handbook of Star Forming Regions Vol. I}, ed. Bo Reipurth, ASP, p.136

\bibitem[xyz(060)]{xyz060} Kun, M., Szegedi-Elek, E., Reipurth, B. 2017, MNRAS, 468, 2325

\bibitem[xyz(061)]{xyz061} Lada, C.J., Alves, J., Lada, E.A. 1999, ApJ, 512, 250

\bibitem[xyz(062)]{xyz062} Lada, C.J., Lombardi, M., Alves, J.F. 2009, ApJ, 703, 52

 \bibitem[xyz(063)]{xyz063} Larson, R.B. 1980, MNRAS, 190, 321

\bibitem[xyz(064]{xyz064} Laugalys, V., Straizys, V., Vrba, F.J. et al., 2006, Baltic Astron., 15, 483

\bibitem[xyz(065)]{xyz065} Lim, J., Yeung, P.K.H., Hanawa, T. et al., 2016, ApJ, 826, A153

\bibitem[xyz(066)]{xyz066} Loinard, L., Torres, R.M., Mioduszewski, A.J. et al., 2007, ApJ, 671, 546

\bibitem[Maehara et al.(2014)]{Mar2014} Maehara, H., Kojima, T., \& Fujii, M., 2014, ATel, 6770

\bibitem[xyz(067)]{xyz067} Magakian, T.Yu., Nikogossian, E.H., Aspin, C. et al., 2010, AJ, 139, 969

\bibitem[Magakian et al.(2013)]{Mag2013} Magakian, T., Nikogossian, E. H., Movsessian, T., et al., 2013, MNRAS, 432, 2685
\bibitem[Massi et al.(2008)]{Mas2008} Massi, F., Codella, C., Brand, J., di Fabrizio, L., \& Wouterloot, J., 2008, \aap, 490, 107
\bibitem[xyz(068)]{xyz068} Mayne, N.J. \& Naylor, T., 2008, MNRAS, 386, 261
\bibitem[McNeil et al.(2004)]{McN2004} McNeil, J., Reipurth, B., \& Meech, K., 2004, IAUC 8284
\bibitem[Miller et al.(2011)]{Mil2011} Miller, A., Hillenbrand, L.A., Covey, K.R. et al., 2011, \apj, 730, 80
\bibitem[Mould et al.(1978)]{Mou1978} Mould, J.R., Hall, D.N.B., Ridgway, S.T. et al., 1978, ApJ, 222, L123

\bibitem[Movsessian et al.(2006)]{Mov2006} Movsessian, T., Khanzadyan, T., Aspin, C. et al., 2006, \aap, 455, 1001


\bibitem[Munari et al.(2001)]{Mun2010} Munari, U., Milani, A., Valisa, P., \& Semkov, E., 2010, Atel, 2808

\bibitem[Mundt et al.(1985)]{Mun1985} Mundt, R., Stocke, J., Strom, S., Strom, K., \& Anderson, E., 1985, ApJ, 297, 41

\bibitem[xyz(069)]{xyz069} Murdin, P. \& Penston, M.V. 1977, MNRAS, 181, 657

\bibitem[xyz(070)]{xyz070} Ortiz-Le\'on, G.N., Dzib, S.A., Kounkel, M.A. et al., 2017, ApJ, 834, A143

\bibitem[stuff(2004)]{stuff2004} Padoan, P. \& Nordlund, A. 2004, ApJ, 617, 559


\bibitem[xyz(071)]{xyz071} Parsamian E. S., Petrossian V. M., 1979, Akad. Nauk.
  Armenian SSR,Soobschenia, No. 135, Yerevan

\bibitem[Peneva et al.(2010)]{Pen2010} Peneva, S., Semkov, E., Munari, U., \& Birkle, K., 2010, \aap, 515, 24

\bibitem[xyz(072)]{xyz072} Perez, L.M., Lamb, J.W., Woody, D.P. et al., 2010, ApJ, 724, 493

\bibitem[Persi et al.(1988)]{Per1988} Persi, P., Ferrari-Toniolo, M., Busso, M., et al., 1988, AJ, 95, 1167

\bibitem[Persson(2004)]{Per2004} Persson, R., 2004, IAU Circ, 8441

\bibitem[xyz(073)]{xyz073} Petrov, P.P. \& Herbig, G.H. 1992, ApJ, 392, 209

\bibitem[xyz(074)]{xyz074} Petrov, P.P. \& Herbig, G.H. 2008, 136, 676

\bibitem[xyz(075)]{xyz075} Petrov, P.P., Kurosawa, R., Romanova, M.M. et al., 2014, MNRAS, 442,3643

\bibitem[xyz(076)]{xyz076} Pfalzner, S. 2008, A\&A, 492, 735


\bibitem[xyz(077)]{xyz077} Pickering, E., 1890, Ann. Harvard College Obs, 18, 113

\bibitem[Poetzel et al.(1989)]{Poe1989} Poetzel, R., Mundt, R., \& Ray, T., 1989, \aap, 224, L13

\bibitem[xyz(078)]{xyz078} Principe, D.A., Cieza, L., Hales, A. et al., 2018, MNRAS, 473, 879

\bibitem[xyz(079)]{xyz079} Prusti, T., Bontekoe, Tj., Chiar, J., Kester, D., \& Whittet, D., 1993, A\&A, 279, 163

\bibitem[Pueyo et al.(2012)]{Pue2012} Pueyo, L., Hillenbrand, L., Vasisht, G., et al., 2012, \apj, 757, 57

\bibitem[xyz(080)]{xyz080} Racine, R. 1968, AJ, 73, 233

\bibitem[Rayner et al.(2003)]{Ray2003} Rayner, J., Toomey, D., Onaka, P., et al., 2003, \pasp, 15, 362

\bibitem[Rayner et al.(2009)]{Ray2009} Rayner, J., Cushing, M., \& Vacca, W., 2009, ApJS, 185, 289

\bibitem[xyz(081)]{xyz081} M.J. Reid et al., 2014, ApJ, 783, 130 

\bibitem[xyz(082)]{xyz082} Reipurth, B. 1985a, A\&A, 143, 435

\bibitem[Reipurth(1985b)]{Rei1985b} Reipurth, B., 1985b, in {\em ESO/IRAM/ONSALA Workshop on (Sub)-Millimeter Astronomy}, ed. P. Shaver \& K. Kj\"ar (Garching: ESO), p. 459

\bibitem[Reipurth(1990)]{Rei1990} Reipurth, B. 1990, in IAU Symp. No. 137 {\em Flare stars in star clusters, associations and the solar vicinity}, Kluwer, p. 229

\bibitem[xyz(083)]{xyz083} Reipurth, B. 2016, {\em George Herbig and Early Stellar Evolution}, http://ifa.hawaii.edu/SP1

\bibitem[Reipurth \& Aspin(1997)]{Rei97_1} Reipurth, B., \& Aspin, C., 1997, AJ, 114, 2700

\bibitem[xyz(084)]{xyz084} Reipurth, B. \& Aspin, C. 2004a, ApJ, 608, L65

\bibitem[Reipurth \& Aspin(2004b)]{Rei2004b} Reipurth, B. \& Aspin, C., 2004b, ApJ, 606, L119

\bibitem[xyz(085)]{xyz085} Reipurth, B. \& Aspin, C. 2010, in Victor Ambartsumian Centennial Volume "Evolution of Cosmic Objects through their Physical Activity", eds. Hayk Harutyunyan, Areg Mickaelian and Yervant Terzian, Gitutyun Publishing House, Yerevan, Armenia, p. 19

\bibitem[xyz(086)]{xyz086} Reipurth, B. \& Bally, J. 1986, Nature, 320, 336

\bibitem[stuff(2001)]{stuff2001} Reipurth, B. \& Clarke, C.J. 2001, AJ, 122, 432


\bibitem[Reipurth, Bally, \& Devine(1997a)]{Rei97_2} Reipurth, B., Bally, J., \& Devine, D., 1997a, AJ, 114, 2708


\bibitem[Reipurth et al.(1997b)]{Rei97_3} Reipurth, B., Olberg, M., Gredel, R., \& Booth, R., 1997b, \aap, 327, 1164


\bibitem[Reipurth et al.(2002)]{Rei2002} Reipurth, B., Hartmann, L., Kenyon, S., Smette, A., \& Bouchet, P., 2002, AJ, 124, 2194

\bibitem[Reipurth et al.(2007)]{Rei2007} Reipurth, B., Aspin, C., Beck, T., et al., 2007, AJ, 133, 1000

\bibitem[Reipurth et al.(2012)]{Rei2012} Reipurth, B., Aspin, C., \& Herbig, G., 2012, ApJ, 748, L5

\bibitem[xyz(087)]{xyz087} Rodriguez, L.F., Porras, A., Claussen, M.J. et al., 2003, ApJ, 586, L137


\bibitem[xyz(088)]{xyz088} Samus, N., 2009, CBET, 1896, 1

\bibitem[Sandell \& Aspin(1998)]{San1998} Sandell, G., \& Aspin, C., 1998, \aap, 333, 1016

\bibitem[xyz(089)]{xyz089} Sandell, G. \& Weintraub, D.A. 2001, ApJS, 134, 115

\bibitem[xyz(0000)]{xyz000} Sandstrom, K.M., Peek, J.E.G., Bower, G.C. et al., 2007,
  ApJ, 667, 1161


\bibitem[xyz(090)]{xyz090} Schoonenberg, D., Okuzumi, S., Ormel, C.W. 2017, A\&A, 605, L2



\bibitem[Semkov \& Peneva(2010)]{Sem2010} Semkov, E., \& Peneva, S., 2010, IAU Inform. Bull. of Variable Stars, No. 5939

\bibitem[xyz(091)]{xyz091} Semkov, E., Peneva, S., Munari, U., Milani, A., \& Valisa, P., 2010, A\&A, 523, L3 

\bibitem[Semkov et al.(2012)]{Sem2012} Semkov, E., Peneva, S.P., Munari, U., 2012, \aap, 542, 43

\bibitem[Semkov et al.(2013)]{Sem2013} Semkov, E., Peneva, S.P., Munari, U., 2013, \aap, 556, 60

\bibitem[Simons \& Tokunaga(2002)]{Sim2002} Simons, D., \& Tokunaga, A., 2002, PASP, 114, 169

\bibitem[xyz(092)]{xyz092} Skinner, S.L., Sokal, K.R., G\"udel, M., Briggs, K.R.
  2009, ApJ, 696, 766

\bibitem[xyz(093)]{xyz093} Smith R. G., 1993, MNRAS, 264, 587

\bibitem[Staude \& Neckel(1991)]{Sta1991} Staude, H., \& Neckel, Th., 1991, \aap, 244, L13

\bibitem[Staude \& Neckel(1992)]{Sta1992} Staude, H., \& Neckel, Th., 1992, \apj, 400, 556

\bibitem[Stocke et al.(1988)]{Sto1988} Stocke, J., Hartigan, P., Strom, S., et al., 1988, ApJS, 68, 229

\bibitem[xyz(094)]{xyz094} Straizys, V., Cernis, K., Bartasiute, S. 1996, Balt.
  Astr. 5, 125

\bibitem[Strom \& Strom(1993)]{Str1993} Strom, K., \& Strom, S., 1993, \apj, 412, L63

\bibitem[stuff(2016)]{stuff2016} Testi, L., Natta, A., Scholz, A. et al., 2016, A\&A, 593, A111


\bibitem[Thommes et al.(2011)]{Tho2011} Thommes, J., Reipurth, B., Aspin, C., \& Herbig, G., 2011, CBET, No. 2795

\bibitem[xyz(095)]{xyz095} Tobin, J.J., Chandler, C.J., Wilner, D.J. et al., 2013,
  ApJ, 779, A93

\bibitem[Tokunaga \& Simons(2002)]{Tok2002} Tokunaga, A., \& Simons, D., 2002, PASP, 114, 180

\bibitem[xyz(096)]{xyz096} Turner, D.G. 2012, Astron. Nach. 333, 174

\bibitem[Vacca et al.(2004)]{Vac2004} Vacca, W., Cushing, M., \& Simon, T., 2004, ApJ, 609, L29


\bibitem[xyz(097)]{xyz097} Varricatt, W.P., Kerr, T.H., Carroll, T., Moore, E. 2015, ATel 8174

\bibitem[xyz(098)]{xyz098} Velazquez, P.F. \& Rodriguez, L.F. 2001, Rev. Mex.
  Astron. Astrofis., 37, 261

\bibitem[xyz(099)]{xyz099} Vorobyov, E.I. \& Basu, S. 2015, ApJ, 805, A115

\bibitem[Wachmann(1954)]{Wac1954} Wachmann, A., 1954, ZfA, 35, 74

\bibitem[Wang(2004)]{Wan2004} Wang, H., Apai, D., Henning, T., \& Pascucci, I., ApJ, 601, L83

\bibitem[xyz(100)]{xyz100} Walker, M., 1959, ApJ, 130, 57

\bibitem[Welin(1971a)]{Wel1971a} Welin, G., 1971a, Information Bulletin on Variable Stars, No. 581

\bibitem[Welin(1971b)]{Wel1971b} Welin, G., 1971b, A\&A, 12, 312
 
\bibitem[Welin et al. (1976)]{Wel1976} Welin, G. 1976, A\&A, 49, 145
 
\bibitem[Welty et al. (1992)]{Wel1992} Welty, A.D., Strom, S.E., Edwards, S. et al., 1992, ApJ, 397, 260
 
 
\bibitem[Whelan et al. (2010)]{Whe2010} Whelan, E.T., Dougados, C., Perrin, M.D. et al., 2010, ApJ, 720, L119
 
\bibitem[stuff(2004)]{stuff2004} Whitworth, A.P. \& Zinnecker, H. 2004, A\&A, 427, 299



 \bibitem[Wolstencroft et al. (1986)]{Wol1986} Wolstencroft, R.D., Scarrott, S.M., Warren-Smith, R.F. et al., 1986, MNRAS, 218, 1P
 
 
 \bibitem[Zhu et al. (2007)]{Zhu2007} Zhu, Z., Hartmann, L., Calvet, N. et al., 2007, ApJ, 669, 483
 
 \bibitem[Zhu et al. (2008)]{Zhu2008} Zhu, Z., Hartmann, L., Calvet, N., et al., 2008, ApJ, 684, 1281

\bibitem[Zhu et al. (2009)]{Zhu2009} Zhu, Z., Hartmann, L., Gammie, C., McKinney, J. 2009, ApJ, 701, 620



\end{thebibliography}
\end{document}